\documentclass[12pt]{article}
\pdfoutput=1
\usepackage{epsfig,amsfonts,amsthm}
\usepackage{amsmath,amssymb}
\usepackage{array}
\usepackage{lipsum}
\usepackage[includefoot,top=80pt,bottom=85pt,left=40pt,right=40pt]{geometry}
\usepackage{cite}
\newcommand{\be}{\begin{equation}}
\newcommand{\ee}{\end{equation}}
\newcommand{\bea}{\begin{eqnarray}}
\newcommand{\eea}{\end{eqnarray}}

\newcommand{\doublet}[2]{ \left( \begin{array}{c}#1 \\ #2 \end{array}\right) }
\usepackage[normalem]{ulem}
\usepackage{array}
\newcolumntype{P}[1]{>{\centering\arraybackslash}p{#1}}
\newcolumntype{M}[1]{>{\centering\arraybackslash}m{#1}}
\usepackage{soul}

\newcommand{\GeV}{{\ensuremath\rm \; GeV}}

\usepackage{color}
\definecolor{orange}{rgb}{1,0.5,0}

\numberwithin{equation}{section}

\usepackage[compat=1.0.0]{tikz-feynman}
 \tikzfeynmanset{compat=1.0.0} 
\usepackage{tikz}

\begin{document}
\bibliographystyle{unsrt}

\title{\hfill ~\\[-30mm]
\begin{footnotesize}
\hspace{90mm}
HIP-2020-35/TH,\,
IIPDM-20\\
\end{footnotesize}
\vspace{3mm}
\textbf{\large Complementary Probes of Two-component Dark Matter}        
}
\date{}

\author{\\[-7mm]
\normalsize
J. Hern\'andez-S\'anchez\footnote{E-mail: {\tt jaime.hernandez@correo.buap.mx}}$^{~1}$,\ 
V.  Keus\footnote{E-mail: {\tt Venus.Keus@helsinki.fi}}$^{~2,3}$,
S.  Moretti\footnote{E-mail: {\tt S.Moretti@soton.ac.uk}} $^{3,4}$,
D.  Rojas-Ciofalo\footnote{E-mail: {\tt Diana.Rojas-Ciofalo@ncbj.gov.pl}} $^{3,5}$,\  
D.  Soko\l{}owska\footnote{E-mail: {\tt dsokolowska@iip.ufrn.br}} $^{6,7}$
\\ \\
\emph{\small $^1$ Instituto de F\'isica and Facultad de Ciencias de la Electr\'onica,}\\  
\emph{\small 
Benem\'erita Universidad Aut\'onoma de Puebla,
Apdo. Postal 542, C.P. 72570 Puebla, M\'exico,}\\
  \emph{\small $^2$ Department of Physics and Helsinki Institute of Physics,}\\
 \emph{\small Gustaf Hallstromin katu 2, FIN-00014 University of Helsinki, Finland}\\
  \emph{\small $^3$ School of Physics and Astronomy, University of Southampton,}\\
  \emph{\small Southampton, SO17 1BJ, United Kingdom}\\
  \emph{\small  $^4$ Particle Physics Department, Rutherford Appleton Laboratory,}\\
 \emph{\small Chilton, Didcot, Oxon OX11 0QX, United Kingdom}\\
 \emph{\small $^5$ National Centre for Nuclear Research, Pasteura 7, 02-093 Warsaw, Poland}\\
  \emph{\small  $^6$ University of Warsaw, Faculty of Physics, Pasteura 5, 02-093 Warsaw, Poland.}\\
  \emph{\small  $^7$ International Institute of Physics, Universidade Federal do Rio Grande do Norte,}\\
\emph{\small Campus Universitario, Lagoa Nova, Natal-RN 59078-970, Brazil}
}

\maketitle

\vspace{-3mm}

\begin{abstract}
\noindent\small
{We study a $Z_2 \times Z'_2$ symmetric 3-Higgs Doublet Model (3HDM), wherein two of the doublets are inert and one is active (thus denoted in literature as I(2+1)HDM), yielding a two-component Dark Matter (DM) sector. The two DM candidates emerge as the lightest scalar component of a different inert doublet, each with a different odd discrete parity, and cooperate to achieve the correct relic density. When a sufficient mass difference exists between the two DM candidates, it is possible to test the presence of both in present and/or forthcoming facilities, as the corresponding masses are typically at the electroweak scale. Specifically, 
the light DM component can be probed by the nuclear recoil energy in direct detection experiments while the heavy DM  component appears  through the photon flux in indirect detection experiments. In fact, the DM mass sensitivity that the two experimental set-ups can achieve should be adequate to establish the presence of two different DM signals.  This result has been obtained in the presence of a thorough theoretical analysis of the stability conditions of the vacuum structure emerging from our I(2+1)HDM construct, ensuring that the model configurations adopted are physical, and of up-to-date  constraints coming from data collected by both space and ground experiments, ensuring that the coupling and mass spectra investigated are viable phenomenologically.} 
 \end{abstract}
\thispagestyle{empty}
\vfill
\newpage
\setcounter{page}{1}

%\tableofcontents

\section{Introduction}

The Standard Model (SM) of particle physics has been extensively tested in the recent decades with the search for its  last missing particle – the SM Higgs boson – ending in 2012 with the discovery of a scalar boson with a mass of $\approx$ 125 GeV by ATLAS and CMS experiments at the CERN Large Hadron Collider (LHC) \cite{Aad:2012tfa,Chatrchyan:2012ufa}.  
Although the properties of the observed scalar are in agreement with those of the SM-Higgs boson, it remains an intriguing possibility that it may just be one member of an extended scalar sector.

Even though so far no signs of detection of
physics Beyond SM (BSM) have been reported, it is well understood that the SM of particle physics is incomplete.
One of the issues that the SM fails to address is providing a viable Dark Matter (DM) candidate. 
According to the standard cosmological $\Lambda$CDM Model \cite{Ade:2015xua}, DM should be a particle which is stable on cosmological time scales, cold (i.e., non-relativistic at the onset of galaxy formation), non-baryonic, neutral and weakly interacting: such a state does not exist in the SM. 
Various such candidates exist in the literature with the most well-studied being the Weakly Interacting Massive Particles (WIMPs) \cite{Jungman:1995df, Bertone:2004pz, Bergstrom:2000pn} with masses between a few GeV and a few TeV.
However, the exact nature of such a particle is still unknown.

A well-known example of an extended scalar sector that can accommodate a particle with such characteristics is the Inert Doublet Model (IDM), a 2-Higgs Doublet Model (2HDM) with an unbroken discrete $Z_2$ symmetry \cite{Deshpande:1977rw}. 
This model involves 1 \textit{inert} doublet, a $Z_2$-odd doublet which does not develop a vacuum expectation value (VEV) and does not couple to fermions, plus 1 \textit{active} doublet, a $Z_2$-even doublet with a non-zero VEV which couples to fermions. 
We shall thus refer to the IDM as the I(1+1)HDM to explicitly show the number of inert (I) and active Higgs (H)doublets. 
An essential feature of this model is that, due to the unbroken $Z_2$ symmetry, the lightest neutral $Z_2$-odd particle, coming from the inert doublet, is stable DM candidate.

The I(1+1)HDM remains a viable model for a scalar DM candidate, however, the accessible parameter space in agreement with current experimental constraints has significantly been reduced 
\cite{Arina:2009um,Nezri:2009jd,Miao:2010rg,Gustafsson:2012aj,Arhrib:2012ia,Krawczyk:2013pea,Goudelis:2013uca,Arhrib:2013ela,Krawczyk:2015vka,Ilnicka:2015jba,Diaz:2015pyv,Modak:2015uda,Queiroz:2015utg,Garcia-Cely:2015khw,Hashemi:2016wup,Poulose:2016lvz,Alves:2016bib,Datta:2016nfz,Belyaev:2016lok,Belyaev:2018ext,Sokolowska:2019xhe,Kalinowski:2019cxe}.
As of now, there are two regions of DM mass that remain viable: a low mass region, 53 GeV $\lesssim m_{\text{DM}} \lesssim m_W$, and a heavy mass region $m_{\text{DM}}\lesssim$ 525 GeV. 
In all the region between $m_W$ and $~ 525$ GeV, the DM candidate annihilates very efficiently, giving a total relic density below the observations.
In order to revive this DM region and give a comprehensive model of DM, one needs to invoke the presence of a least a second DM candidate.
A simple way to accomplish this is by the addition of an extra inert singlet \cite{Belanger:2012vp,Yaguna:2019cvp,Belanger:2020hyh}, or  doublet scalar\cite{Ivanov:2012hc, Aranda:2019vda}, that is, within the framework of the 3-Higgs doublet model (3HDM), a.k.a the I(2+1)HDM \cite{Keus:2014jha,Keus:2014isa,Keus:2015xya,Cordero-Cid:2016krd,Cordero:2017owj,Cordero-Cid:2018man,Keus:2019szx,Cordero-Cid:2020yba}.

Here, we propose a two-component DM in the context of an I(2+1)HDM framework symmetric under a $Z_2\times Z'_2$ group with one inert doublet odd under $Z_2$ and even under the $Z'_2$ and the other vice versa. 
The lightest particle in from each inert doublet is a viable DM candidate, each with a different odd discrete parity which cooperate to achieve the correct relic density. 
We show that other dark particles from both doublets have a significant impact on the final relic abundances of these two stable particles, as they influence the thermal evolution and decoupling rate of DM particles. A similar analysis was performed in the context of a supersymmetric model in \cite{Khalil:2020syr}.

When there is a sufficient mass difference between the two DM candidates, we show that it is possible to test the presence of both in current and upcoming experiments/observations, as the corresponding masses are typically at the electroweak (EW) scale.
Specifically, the light DM component can be probed by the nuclear recoil energy in direct detection experiments while the heavy DM  component appears  through its contribution to the photon flux in indirect detection experiments.
We have also performed a thorough theoretical analysis of the stability conditions of the vacuum structure emerging from our I(2+1)HDM construct.
Specifically, when running random scans over the parameter space, one can easily cover regions where the chosen minimum is not a global one.
Therefore, establishing the minimum as a global is a complicated but necessary task while dealing with multi-scalar models, and extrapolating from properties of other models may not be correct.

The outline of the paper is as follows. In Section \ref{sec-general}, we present the scalar potential and discuss the properties of different possible minima. In Section \ref{sec-coexistence}, we derive conditions which ensure that the minimum with two potential DM candidates is a true vacuum of the potential. Experimental and theoretical constraints are presented in Section \ref{sec-constraints}. In Section \ref{sec-DM} we discuss properties of DM particles in the model and in Section \ref{sec-num} we present numerical analysis of chosen benchmarks. Additional information, including Feynman rules, mass formulas for different minima and further discussion regarding the coexistence of various minima, are collected in the appendices.

\section{The $Z_2\times Z'_2$ symmetric I(2+1)HDM \label{sec-general}}

\subsection{Scalar potential}
The most general $Z_2 \times Z'_2$ symmetric 3HDM potential has the following form \cite{Ivanov:2011ae,Keus:2013hya}:
\bea
\label{potential}
V &=& V_0+V_{Z_2 \times Z'_2},\\[1mm]
V_0 &=& - \mu^2_{1} (\phi_1^\dagger \phi_1) -\mu^2_2 (\phi_2^\dagger \phi_2) - \mu^2_3(\phi_3^\dagger \phi_3) \nonumber
+ \lambda_{11} (\phi_1^\dagger \phi_1)^2+ \lambda_{22} (\phi_2^\dagger \phi_2)^2  + \lambda_{33} (\phi_3^\dagger \phi_3)^2 \nonumber\\
&& + \lambda_{12}  (\phi_1^\dagger \phi_1)(\phi_2^\dagger \phi_2)
 + \lambda_{23}  (\phi_2^\dagger \phi_2)(\phi_3^\dagger \phi_3) + \lambda_{31} (\phi_3^\dagger \phi_3)(\phi_1^\dagger \phi_1) \nonumber\\
&& + \lambda'_{12} (\phi_1^\dagger \phi_2)(\phi_2^\dagger \phi_1) 
 + \lambda'_{23} (\phi_2^\dagger \phi_3)(\phi_3^\dagger \phi_2) + \lambda'_{31} (\phi_3^\dagger \phi_1)(\phi_1^\dagger \phi_3),  \nonumber\\[1mm]
V_{Z_2 \times Z'_2}&=&  \lambda_1 (\phi_1^\dagger \phi_2)^2 + \lambda_2(\phi_2^\dagger \phi_3)^2 + \lambda_3 (\phi_3^\dagger \phi_1)^2 + h.c.. \nonumber 
\eea
where $V_0$ is invariant under any phase rotation, while $V_{Z_2 \times Z'_2}$ ensures symmetry under the $Z_2 \times Z'_2$ group generated by
\be 
g_{Z_2} = \mathrm{diag}(-1,1,1)\, , \qquad
g_{Z'_2} = \mathrm{diag}(1,-1,1) \,.
\ee
With this charge assignment, the only doublet odd under the $Z_2$ symmetry is $\phi_1$, the only doublet odd under the $Z'_2$ symmetry is $\phi_2$, and $\phi_3$ is even under both $Z_2$ and $Z'_2$.
In this paper, we assume that all parameters in the potential are real; therefore, we do not introduce any explicit CP-violation in the scalar sector. In particular, we do not consider the possible effects of dark CP-violation, a feature that was introduced for the first time in \cite{Cordero-Cid:2016krd} and further studied in \cite{Keus:2016orl,Cordero:2017owj,Cordero-Cid:2018man,Cordero-Cid:2020yba} which could arise in an extended dark sector. 
However, there is still the possibility of spontaneous breaking of the CP symmetry in the active sector for particular choices of parameters. 

We extend the $Z_2 \times Z'_2$ to the full Lagrangian by assigning an even $Z_2 \times Z'_2$ charge to all SM gauge bosons and fermions. 
Following this choice, we  the Yukawa interactions are set to ''Type-I'' interactions, i.e. only the third doublet, $\phi_3$, will couple to fermions:
\bea
\mathcal{L}_{Y} = \Gamma^u_{mn} \bar{q}_{m,L} \tilde{\phi}_3 u_{n,R} + \Gamma^d_{mn} \bar{q}_{m,L} \phi_3 d_{n,R}
 +  \Gamma^e_{mn} \bar{l}_{m,L} \phi_3 e_{n,R} + \Gamma^{\nu}_{mn} \bar{l}_{m,L} \tilde{\phi}_3 {\nu}_{n,R} + h.c.  \label{yukawa}
\eea
This ensures that there are no Flavour Changing Neutral Currents (FCNCs) and, in cases when doublets $\phi_{1,2}$ do not develop Vacuum Expectation Values (VEVs), fields from these doublets will not couple to fermions.

\paragraph{Stability of the potential:} Parameters of the potential $V$ are subject to a number of various theoretical and experimental constraints (described in details in Section \ref{sec-constraints}). Here, we want to focus on the stability of the potential. For this potential to have a stable vacuum (i.e. for the potential to be bounded from below) the following conditions are required \cite{Grzadkowski:2009bt}:
\bea
&& \lambda_{ii}>0, \quad i =1,2,3, \label{positivity1} \\
&& \lambda_x > - 2 \sqrt{\lambda_{11} \lambda_{22}}, \quad \lambda_y > - 2 \sqrt{\lambda_{11} \lambda_{33}}, \quad \lambda_z > - 2 \sqrt{\lambda_{22} \lambda_{33}}, \label{positivity2}\\
&&\left\lbrace  \begin{array}{l} 
\sqrt{\lambda_{33}} \lambda_x + \sqrt{\lambda_{11}} \lambda_z+\sqrt{\lambda_{22}} \lambda_y \geq 0\\
\quad \textrm{or}\\[1mm]
\lambda_{33} \lambda_x^2 + \lambda_{11} \lambda_z^2+\lambda_{22} \lambda_y^2 -\lambda_{11} \lambda_{22} \lambda_{33} - 2 \lambda_x \lambda_y \lambda_z <0.
\end{array}\right.
\label{positivity3}
\eea
where 
\bea
\lambda_x = \lambda_{12}+\textrm{min}(0,\lambda_{12}'-2|\lambda_1|),\\
\lambda_y = \lambda_{31}+\textrm{min}(0,\lambda_{31}'-2|\lambda_3|),\\
\lambda_z = \lambda_{23}+\textrm{min}(0,\lambda_{23}'-2|\lambda_2|).
\eea
As noted in \cite{Faro:2019vcd} these conditions are in fact sufficient but
not necessary, as it is possible to construct examples of this model in which the potential is bounded from below, but which violate conditions (\ref{positivity1}-\ref{positivity3}). We do not explore this region of parameter space in this work and remain agnostic as to how big the impact of these ``neglected'' regions would be on the phenomenology of the model. In particular, positivity conditions can have a direct impact on the (co)existence of different possible minima, and combined with other constraints they significantly limit the allowed parameter space. It would be interesting to see if a change in imposed positivity constraints modifies this picture. %\orange{VK: Is this more constraining or less constraining?}

\subsection{Possible neutral vacuum states}

In principle, every doublet can develop a VEV and different choices of $(\langle \phi_1 \rangle,\langle \phi_2 \rangle, \langle \phi_3 \rangle)$ can lead to the realisation of different scenarios. In case of neutral vacuum states\footnote{Discussion of potential non-physical charge-breaking minima with massive photons, when at least one doublet develops a VEV in the upper component, is presented in Section \ref{ap-charged}.}, the most general decomposition of the three scalar doublets can be written as:
\be 
\phi_1= \doublet{$\begin{scriptsize}$ H^+_1 $\end{scriptsize}$}{\frac{v_1e^{i\xi_1}+H^0_1+iA^0_1}{\sqrt{2}}} ,\quad 
\phi_2= \doublet{$\begin{scriptsize}$ H^+_2 $\end{scriptsize}$}{\frac{v_2e^{i\xi_2}+H^0_2+iA^0_2}{\sqrt{2}}} , \quad 
\phi_3= \doublet{$\begin{scriptsize}$ H^+_3 $\end{scriptsize}$}{\frac{v_3+H^0_3+iA^0_3}{\sqrt{2}}}  .
\label{explicit-fields}
\ee
All possible neutral extrema are listed in Table \ref{extrema-table}. Depending on the VEV alignment we observe significant differences between the properties of the model, both in the scalar and in the fermionic sector. 
The aim of the paper is to study the phenomenology of the \texttt{2-Inert} minimum configuration, which corresponds to having two \textit{inert} doublets ($\phi_1$ and $\phi_2$) and one \textit{active} doublet ($\phi_3$).  The label ``inert'' refers not only to zero VEV of the doublet but also to the absence of any couplings to fermions, following the choice of Yukawa interactions in the form of Eq. (\ref{yukawa}). 
With this choice, we can interpret the lightest particles from doublets $\phi_1$ and $\phi_2$, which are stable due to the unbroken $Z_2\times Z'_2$ symmetry, as DM candidates. 
However, this desirable minimum is not the only possible vacuum state which can be realised in the model. 
In this work we do not attempt to study the entire phase space, but rather to highlight the differences and similarities of the model to other multi-scalar models, in particular the 2HDM, which our limited analysis presented here already does. We aim to identify the viable parameter space where the \texttt{2-Inert} configuration is the {\sl global} minimum of the model, while other possible minima either do not exist or have higher energy. 
To do so, below we discuss the general properties of neutral vacuum states and in section \ref{sec-coexistence} we
identify the conditions required for the \texttt{2-Inert} configuration to be the global minimum.  
%Case of the charge breaking vacua is presented in Section \ref{ap-charged}.

\begin{table} [h!]
\begin{footnotesize}
\begin{center}
\begin{tabular}{|c || c |P{2cm}| c|c|} \hline
%\begin{tabular}{|m{2cm} || m{2cm} | m{2cm}| m{2cm}|m{4cm}|} \hline
%& & & & \\[-2ex]
& & & &  \\[-2ex]
 \textbf{Name} &\textbf{VEV}    & 
\mbox{\textbf{Remnant}} \mbox{\textbf{symmetry}} &
\mbox{\textbf{Generators}} &
\mbox{\textbf{Properties}} \\[2ex]
   \hline  \hline
& & & &  \\[-2ex]
\texttt{EWs} & $(0,0,0)$ & $Z_2 \times Z'_2$   & $(-1,1,1)$, $(1,-1,1)$   &EW symmetry  \\[1ex] \hline  

& & & &  \\[-2ex]
\texttt{2-Inert} & $(0,0,v_3)$ & $Z_2 \times Z'_2$   & $(-1,1,1)$, $(1,-1,1)$   & \mbox{SM+ 2 DM candidates}  \\[1ex] \hline

& & & &  \\[-2ex]
\texttt{DM1} & $(0,v_2,v_3)$ & $Z_2$   & $(-1,1,1)$ & \mbox{2HDM +1 DM candidate}  \\[1ex] \hline  

& & & &  \\[-2ex]
\texttt{DM2} & $(v_1,0,v_3)$ & $Z'_2$   & $(1,-1,1)$ & \mbox{2HDM +1 DM candidate}  \\[1ex] \hline  

& & & &  \\[-2ex]
\texttt{F0DM1} & $(0,v_2,0)$ & $Z_2$   & $(-1,1,1)$ & \mbox{1 DM candidate}, \mbox{massless fermions}  \\[1ex] \hline  

& & & &  \\[-2ex]
\texttt{F0DM2} & $(v_1,0,0)$ & $Z'_2$   & $(1,-1,1)$ & \mbox{1 DM candidate}, \mbox{massless fermions}  \\[1ex] \hline  

& & & &  \\[-2ex]
\texttt{F0DM0} & $(v_1,v_2,0)$ & None & -- & \mbox{no DM candidate}, \mbox{massless fermions}  \\[1ex] \hline  

& & & &  \\[-2ex]
\texttt{N} & $(v_1,v_2,v_3)$ & None & -- & \mbox{no DM candidate} \\[1ex] \hline \hline 
& & & &  \\[-2ex]
\texttt{sCPv} & $(v_1e^{i\xi_1},v_2e^{i\xi_2},v_3)$ & None & -- & \mbox{spontaneous CP-violation} \\[1ex] \hline 
\end{tabular}
\end{center}
\end{footnotesize}
\caption{\footnotesize Possible CP-conserving (first eight entries) and CP-violating (last entry) neutral extrema in the $Z_2 \times Z'_2$ symmetric 3HDM.}
\label{extrema-table}
\end{table}

\paragraph{EWs:} This is a trivial non-physical minimum in which none of the doublets develops VEVs, leading to an unbroken $SU(2)_L \times U(1)_Y$ gauge symmetry. Fermions and all four gauge bosons are massless. This minimum is realised only if all quadratic parameters in Eq.~(\ref{potential}) are negative, $\mu_i <0, i=1,2,3$. At tree-level, no other minimum can be realised in this region of parameter space.

\paragraph{DM1:}

For the \texttt{DM1} minimum the scalar doublets can be defined as:
\be 
\phi_1= \doublet{$\begin{scriptsize}$ H^+_1 $\end{scriptsize}$}{\frac{H_1+iA_1}{\sqrt{2}}} ,\quad 
\phi_2= \doublet{$\begin{scriptsize}$ H^+_2 $\end{scriptsize}$}{\frac{v_2+H^0_2+iA^0_2}{\sqrt{2}}} , \quad 
\phi_3= \doublet{$\begin{scriptsize}$ H^+_3 $\end{scriptsize}$}{\frac{v_3+H^0_3+iA^0_3}{\sqrt{2}}}  .
\label{vac-DM1}
\ee
In this case, two doublets develop VEVs, while the first doublet remains inert. Due to the non-zero $v_2$ the original discrete symmetry of the potential is spontaneously broken into a remnant symmetry $Z_2$ with generator $(-1,1,1)$. In the classification of inert-active doublets this extremum would correspond to a CP-conserving version of I(1+2)HDM.

Spontaneous breaking of a discrete symmetry is often avoided by adding a $Z_2 \times Z'_2$ soft-breaking term $\mu_{23}^2 \phi_2^\dagger \phi_3 + h.c.$ to the potential,
%\orange{(VK: meaning that the potential is not $Z_2 \times Z'_2$ symmetric from the start)} 
see e.g. \cite{Grzadkowski:2009bt,Grzadkowski:2010au,Osland:2013sla,Merchand:2019bod}.
Notice also that without introducing a $Z_2 \times Z'_2$ soft-breaking term it is not possible to induce spontaneous CP-violation for this vacuum choice. An apparent CP-violation appears if VEVs are chosen\footnote{Due to the gauge freedom we can choose the VEVs such that $\langle \phi_3 \rangle = v_3 \in \mathbf{R}$.} to be $(0,v_2 e^{i\xi},v_3)$, however the extremum conditions are satisfied only for the two possible choices of $\xi=0, \pi$. The first one is a CP-conserving extremum, while $\xi=\pi$ is its equivalent by rephrasing of the potential by $\lambda_2 \to -\lambda_2$. 

The physical spectrum is separated into inert and active/Higgs scalars. The inert sector consists of four inert particles, $H_1, A_1, H_1^\pm$, the lightest of which is stable and can serve as a DM candidate. In short, the dark sector is precisely like the one from the I(1+1)HDM. However, mass relations and the DM evolution are different since the inert doublet couples to two active doublets. Among five physical Higgs particles, there are two CP-even Higgs bosons, $h$ and $H$, a pseudoscalar $A$ and two charged particles, $H^\pm$, which come from the mixing between $\phi_2$ and $\phi_3$. Relevant mass formulas are presented in Appendix \ref{ap-masses}.

\paragraph{DM2:} This type of vacuum is similar to the one discussed above, with roles of $\phi_1$ and $\phi_2$ reversed:
\be 
\phi_1= \doublet{$\begin{scriptsize}$ H^+_1 $\end{scriptsize}$}{\frac{v_1+H^0_1+iA^0_1}{\sqrt{2}}} ,\quad 
\phi_2= \doublet{$\begin{scriptsize}$ H^+_2 $\end{scriptsize}$}{\frac{H_2+iA_2}{\sqrt{2}}} , \quad 
\phi_3= \doublet{$\begin{scriptsize}$ H^+_3 $\end{scriptsize}$}{\frac{v_3+H^0_3+iA^0_3}{\sqrt{2}}} .
\label{vac-DM2}
\ee

Extremum conditions and mass formulae can be obtained from the \texttt{DM1} case by replacements: $\lambda_{11} \leftrightarrow \lambda_{22}, \, \mu_1^2 \leftrightarrow  \mu_2^2,\, \Lambda_2 \leftrightarrow \Lambda_3$ and are listed in Appendix \ref{ap-masses}.

One may wonder if these two extrema, \texttt{DM1} and \texttt{DM2}, are in fact equivalent to each other through a basis rotation. However, in Appendix \ref{ap-DM1DM2} we show that in fact minima \texttt{DM1} and \texttt{DM2} in principle have different depths and can coexist. 
%We have also found several numerical solutions where both minima coexist and depending on the parameters; one is deeper than the other. 
It is known that the 2HDM allows for the coexistence of similar symmetry-breaking minima \cite{Ivanov:2007de}. In our model, of course, symmetry is not the same, as in each case it is a different $Z_2$ that is broken by either $\langle \phi_1 \rangle$ or $\langle \phi_2 \rangle$. However, in both cases, one $Z_2$ symmetry remains unbroken, which may be the reason why these two distinct states can be realised at the same time. The same situation was found e.g. in N2HDM \cite{Engeln:2020fld} and in the Higgs complex singlet-doublet model \cite{Ferreira:2016tcu}, where phases with symmetry breaking patterns $Z_2 \times Z'_2 \to Z_2$ and $Z_2 \times Z'_2 \to Z'_2$ are distinct and can simultaneously exist.

\paragraph{F0DM1:}
In this case, only the second doublet acquires a VEV, leading to a field decomposition of the form:
\be 
\phi_1= \doublet{$\begin{scriptsize}$ H^+_1 $\end{scriptsize}$}{\frac{H_1+iA_1}{\sqrt{2}}} ,\quad 
\phi_2= \doublet{$\begin{scriptsize}$ H^+_2 $\end{scriptsize}$}{\frac{v_2+H^0_2+iA^0_2}{\sqrt{2}}} , \quad 
\phi_3= \doublet{$\begin{scriptsize}$ H^+_3 $\end{scriptsize}$}{\frac{H_3+iA_3}{\sqrt{2}}}  .
\label{vac-F0DM1}
\ee
This type of extremum, if realised as a vacuum, has very devastating consequences for the phenomenology of the model. Since only the third doublet, $\phi_3$, has interactions with fermions, its zero VEV leads to tree-level fermion masses to be zero (hence \texttt{F0} in our notation). We would like to mention that there is a possibility of generating fermion masses at loop level, however achieving correct masses at least for the third generation of fermions seems unlikely. Notice, that unlike the \texttt{EWs}, in this case gauge bosons are still massive. 

The model would still have a DM candidate since the $Z_2$ symmetry remains unbroken, and the I(1+1)HDM-like dark sector is generated from doublet $\phi_1$. Notice that this is only due to the existence of three doublets. The massless fermion phase, called the \textit{inert-like} minimum in \cite{Ginzburg:2010wa}, can coexist with the inert minimum in the I(1+1)HDM. However, in the I(1+1)HDM, there is no DM candidate in the inert-like vacuum, as there is no remaining discrete symmetry to stabilize any particles.  

\paragraph{F0DM2:} As in the case of \texttt{DM1} and \texttt{DM2} extrema, similar to \texttt{F0DM1} there is another phase where fermions are massless and one doublet provides a DM candidate where the doublets can be written as:
\be 
\phi_1= \doublet{$\begin{scriptsize}$ H^+_1 $\end{scriptsize}$}{\frac{v_1+H^0_1+iA^0_1}{\sqrt{2}}} ,\quad 
\phi_2= \doublet{$\begin{scriptsize}$ H^+_2 $\end{scriptsize}$}{\frac{H_2+iA_2}{\sqrt{2}}} , \quad 
\phi_3= \doublet{$\begin{scriptsize}$ H^+_3 $\end{scriptsize}$}{\frac{H_3+iA_3}{\sqrt{2}}}  .
\label{vac-F0DM2}
\ee
Again, as in the discussion of the \texttt{DM1}--\texttt{DM2} pair, the difference between the \texttt{F0DM1} and \texttt{F0DM2} case
seems to be related only to the interchanges of $\phi_1 \leftrightarrow \phi_2, \, \lambda_{11} \leftrightarrow \lambda_{22},  \,\mu_1^2 \leftrightarrow  \mu_2^2, \, \Lambda_2 \leftrightarrow \Lambda_3$. This is yet another situation, where two extrema of the same level of symmetry breaking can coexist, even if the actual remaining symmetry is different (see Appendix \ref{ap-DM1DM2}), and therefore cannot be dismissed as equivalent through a basis change.

\paragraph{F0DM0:} The last possibility of having a vacuum with tree-level massless fermions arises with the following vacuum structure: 
\be 
\phi_1= \doublet{$\begin{scriptsize}$ H^+_1 $\end{scriptsize}$}{\frac{v_1+H^0_1+iA^0_1}{\sqrt{2}}} ,\quad 
\phi_2= \doublet{$\begin{scriptsize}$ H^+_2 $\end{scriptsize}$}{\frac{v_2+H^0_2+iA^0_2}{\sqrt{2}}} , \quad 
\phi_3= \doublet{$\begin{scriptsize}$ H^+_3 $\end{scriptsize}$}{\frac{H_3+iA_3}{\sqrt{2}}}  .
\label{vac-F0DM0}
\ee
Here both discrete symmetries $Z_2$ and $Z'_2$ are broken, therefore there is no stable particle that could serve as a DM candidate. 

\paragraph{N:} This is an analogue to the \textit{normal} extremum from the 2HDMs, where every doublet acquires a (real) VEV and the discrete symmetry is broken completely:
\be 
\phi_1= \doublet{$\begin{scriptsize}$ H^+_1 $\end{scriptsize}$}{\frac{v_1+H^0_1+iA^0_1}{\sqrt{2}}} ,\quad 
\phi_2= \doublet{$\begin{scriptsize}$ H^+_2 $\end{scriptsize}$}{\frac{v_2+H^0_2+iA^0_2}{\sqrt{2}}} , \quad 
\phi_3= \doublet{$\begin{scriptsize}$ H^+_3 $\end{scriptsize}$}{\frac{v_3+H^0_3+iA^0_3}{\sqrt{2}}}  .
\label{vac-mixed}
\ee
There is no DM candidate in this state, and the scalar sector consists solely of the active Higgs particles: three CP-even Higgses, $h_1,h_2,h_3$, two CP-odd Higgses, $a_1,a_2$ and two pairs of charged scalars, $h_1^\pm, h_2^\pm$.

\paragraph{sCPv:} Spontaneous CP-violation may arise in the model if the extremum is of the following form:
\be 
\phi_1= \doublet{$\begin{scriptsize}$ H^+_1 $\end{scriptsize}$}{\frac{v_1e^{i\xi_1}+H^0_1+iA^0_1}{\sqrt{2}}} ,\quad 
\phi_2= \doublet{$\begin{scriptsize}$ H^+_2 $\end{scriptsize}$}{\frac{v_2e^{i\xi_2}+H^0_2+iA^0_2}{\sqrt{2}}} , \quad 
\phi_3= \doublet{$\begin{scriptsize}$ H^+_3 $\end{scriptsize}$}{\frac{v_3+H^0_3+iA^0_3}{\sqrt{2}}}  .
\label{vac-sCPV}
\ee
Here, CP symmetry is spotnaneously broken by two non-zero phases, $\xi_1,\xi_2$. As the potential is $Z_2\times Z'_2$ symmetric, both phases have to be non-zero for the CP-violation to appear. This type of extremum was introduced as the Weinberg's 3HDM \cite{Weinberg:1976hu} and expanded on in \cite{Branco:1980sz}. As with the normal minimum, here there are five neutral Higgs particles (with mixed CP charges) and two charged Higgses. Due to spontaneous violation of both $Z_2$ symmetries there is no DM candidate.

\subsection{2-Inert vacuum \label{sec-inertvac}}
The \texttt{2-Inert} vacuum state has the following form:
\be 
\phi_1= \doublet{$\begin{scriptsize}$ H^+_1 $\end{scriptsize}$}{\frac{H_1+iA_1}{\sqrt{2}}} ,\quad 
\phi_2= \doublet{$\begin{scriptsize}$ H^+_2 $\end{scriptsize}$}{\frac{H_2+iA_2}{\sqrt{2}}} , \quad 
\phi_3= \doublet{$\begin{scriptsize}$ H^+_3 $\end{scriptsize}$}{\frac{v+h+iA^0_3}{\sqrt{2}}} \, ,
\label{vac-inert}
\ee
and the extremum condition for this state reads:
\be 
v^2=\frac{\mu_3^2}{\lambda_{33}}\,.
 \label{inertex}
\ee
The third doublet, $\phi_3$ plays the role of the SM-like Higgs doublet, with the Higgs particle $h$ having, by construction, tree-level interactions with gauge bosons and fermions identical to those of the SM-Higgs boson. Its mass is fixed through the tadpole conditions to be:
\be
m^2_h= 2\mu_3^2 = 2 v^2 \lambda_{33},
\ee
while ${A^0}_3$ and ${H}^\pm_3$ are the Goldstone bosons. 

Two inert doublets, $\phi_1$ and $\phi_2$, provide DM candidates. Each doublet consists of two neutral particles\footnote{As usual, inert scalars $H_i$ and $A_i$ have opposite CP parity, as evident from the gauge interactions, however it is not possible to establish their definite CP properties, as they do not couple to fermions.}, $H_i$ and $A_i$, and two charged $H^\pm_i$, $i=1,2$. The mass spectrum is as follows:
\bea
&& m^2_{H_1}= -\mu^2_1  +\frac{1}{2}(\lambda_{31}+\lambda'_{31} +2\lambda_3 )v^2 \equiv -\mu_1^2 + \Lambda_3 v^2, \label{massmh1}\\
&& m^2_{A_1}= -\mu^2_1  +\frac{1}{2}(\lambda_{31}+\lambda'_{31} -2\lambda_3)v^2 \equiv -\mu_1^2 + \bar{\Lambda}_3 v^2,\\
&& m^2_{H^\pm_1}= -\mu^2_1 +\frac{1}{2}\lambda_{31}v^2. 
\eea
\bea 
&& m^2_{H_2}= -\mu^2_2  +\frac{1}{2}(\lambda_{23}+\lambda'_{23} +2\lambda_2)v^2\equiv -\mu_2^2 + \Lambda_2 v^2, \label{massmh2}\\
&& m^2_{A_2}= -\mu^2_2  +\frac{1}{2}(\lambda_{23}+\lambda'_{23} -2\lambda_2)v^2\equiv -\mu_2^2 + \bar{\Lambda}_2 v^2,\\
&& m^2_{H^\pm_2}= -\mu^2_2 +\frac{1}{2}\lambda_{23} v^2. \label{massmhc2}
\eea
In principle any particle among $(H_i, A_i, H^\pm_i)$ can be the lightest. We dismiss the possibility of $H^\pm_i$ being the lightest, as it would mean that DM candidate is a charged particle. Choosing between $H_1$ and $A_1$ (or $H_2$ and $A_2$) is related only to a change of the sign of the quartic parameter $\lambda_{3}$ ($\lambda_2$) and has no impact on the phenomenology. Therefore, from the demand that $m_{H_i} < m_{A_i}, m_{H^\pm_i}$ we obtain the following relations between parameters:
\be
\lambda_2<0, \;\; \lambda_3<0, \;\; \lambda_{31}'+ 2\lambda_3<0, \;\; \lambda_{23}'+ 2\lambda_2<0. \label{Hlightest}
\ee
Notice, that unlike in  many $Z_N$ symmetric models, two lightest states from two doublets are automatically stable, regardless of their mass relation, as they are stabilized by different $Z_2$ symmetries.

For $(0,0,v)$ to be a local minimum, all mass-squared values have to be positive, and for it to be a global minimum, i.e. the true vacuum, its energy, $\mathcal{V}_{\texttt{2Inert}}$ has to be lower than energy of any other possible minima, $\mathcal{V}_{\texttt{X}}$, that exists at the same time. Following the chosen mass order and resulting relations in Eq.(\ref{Hlightest}) we arrive at the following conditions:
\bea
 \textrm{Local minimum if:}&& \left\lbrace \begin{array}{l}
v^2={\mu_3^2}/{\lambda_{33}} > 0\\[1mm]
\Lambda_2 > \mu^2_2/v^2  \\[1mm]
\Lambda_3 > \mu^2_1/v^2 \end{array} \right. 
\hspace{3cm}
\label{inert-loc}
\\
\textrm{Global minimum if in addition:}&& \mathcal{V}_{\texttt{2Inert}}= -\frac{\mu_3^4}{4\lambda_{33}} < \mathcal{V}_{\texttt{X}} 
\hspace{3cm}
\label{inert-glob}
\eea

\paragraph{Physical parameters:} Parameters of the potential can be rephrased in terms of the physical observables, such as masses and couplings. For our studies we have chosen masses of inert particles and the Higgs-DM couplings to describe parameters from the visible sector, and self-interaction parameters to describe the dark sector:
\be
v^2, m_h^2, m^2_{H_1}, m^2_{H_2}, m^2_{A_1}, m^2_{A_2}, m^2_{H^\pm_1}, m^2_{H^\pm_2}, \Lambda_2, \Lambda_3, \Lambda_1, \lambda_{11}, \lambda_{22}, \lambda'_{12}, \lambda_{12} \label{physpar}.
\ee
Self-couplings $\lambda_{11}, \lambda_{22}, \lambda'_{12}, \lambda_{12}$ correspond exactly to the terms in Eq.~(\ref{potential}), while relations between remaining parameters and our chosen physical basis is as follows:
\bea
&& \mu_1^2=-m_{H_1}^2+\Lambda_{3}v^2, \\
&& \lambda_{3}=(m_{H_1}^2-m_{A_1}^2)/(2v^2), \\
&& \lambda_{31}'=(m_{H_1}^2+m_{A_1}^2-2m_{H^\pm_1}^2)/v^2, \\
&& \lambda_{31}=2\Lambda_{3}-2\lambda_{3}-\lambda_{31}', 
\eea
\bea
&& \mu_2^2=-m_{H_2}^2+\Lambda_{2}v^2, \\
&& \lambda_{2}=(m_{H_2}^2-m_{A_2}^2)/(2v^2),\\
&& \lambda_{23}'=(m_{H_2}^2+m_{A_2}^2-2m_{H^\pm_2}^2)/v^2, \\
&& \lambda_{23}=2\Lambda_{2}-2\lambda_{2}-\lambda_{23}',\\
&& \lambda_1 = 2\Lambda_1 - (\lambda_{12}+\lambda'_{12}).
\eea

\paragraph{Note about the ``dark democracy'' limit:} Due to the large number of parameters in 3HDM potentials, it is often convenient to study the model in the so-called ``dark democracy'' limit, by assuming that two inert doublets have identical interactions with the active doublet. This can be translated into imposing additional relations on the parameters of the potential where:
\be
\mu_1^2 = \mu_2^2, \, \lambda_2 = \lambda_3, \,\lambda_{23} = \lambda_{31},\, \lambda'_{23} = \lambda'_{31}\,. 
\label{darkdem}
\ee
In the $Z_2$ symmetric I(2+1)HDM imposing this limit does not render both doublets degenerate due to the presence of the quadratic term $\mu_{12}^2 \phi_1^\dagger  \phi_2 +h.c.$, which introduces a mass splitting between doublets. 
However, in the $Z_2 \times Z'_2$ symmetric I(2+1)HDM this term is forbidden by the symmetry of the potential. Applying Eq.~(\ref{darkdem}) to the mass relations in Eqs.~(\ref{massmh1}-\ref{massmhc2}), leads to $m^2_{H_1} = m^2_{H_2}, \, m^2_{A_1} = m^2_{A_2}$ and $m^2_{H^\pm_1} = m^2_{H^\pm_2} $. Therefore, imposing the dark democracy limit here has stronger consequences  than e.g. in the $Z_2$ symmetric I(2+1)HDM. In this paper we do not consider this limit and any possible degeneracy in numerical analysis will be purely coincidental.

\section{Vacuum stability in the I(2+1)HDM \label{sec-coexistence}}

To identify the parameter space where the \texttt{2-Inert} configuration is the global minimum, i.e. the vacuum, we use the following procedure:

\begin{enumerate}
\item We assume that the \texttt{2-Inert} extremum exists and it is always a minimum $\Rightarrow$ conditions (\ref{inert-loc}) are always satisfied.
\item If stationary point \texttt{X} exists, then its energy has to be larger than the energy of \texttt{2-Inert} state, $\mathcal{V}_{\texttt{X}} - \mathcal{V}_{\texttt{2-Inert}} > 0$. We are not concerned whether in this situation \texttt{X} is a minimum, or a saddle point, as the proper energy difference is enough to guarantee the stability of the \texttt{2-Inert} state. The analysis is performed at tree-level only, and we comment on possible issues arising from higher-order corrections in Section \ref{sec-stabsummary}.
%\item If $E_{\texttt{2-Inert}} - E_x > 0$ then \texttt{X} cannot be a minimum.
\end{enumerate}

\subsection{Bilinear formalism \label{sec-bilinear}}

We will use a a bilinear formalism similar to the one developed for the 2HDM \cite{Ferreira:2004yd,Barroso:2007rr}, and which has been since adapted to study the vacuum
structure of other multi-scalar models, with additional doublets and singlets, see e.g. \cite{Degee:2012sk,Engeln:2020fld,Ferreira:2016tcu,Ivanov:2010ww,Ivanov:2014doa,Ivanov:2018jmz,Ferreira:2019iqb}.

The potential of the model can be written in terms of gauge-invariant iso-scalar
combinations of the field operators that are quadratic in the fields, therefore greatly simplifying the minimisation procedure. In general, the potential in Eq.~(\ref{potential}) can be rewritten as:
\be
V = a_1 x_1 + a_2 x_2 + a_3 x_3 + \sum\limits_{i} b_{ii} x_i^2 + \sum\limits_{j} \sum\limits_{i<j}  b_{ij} x_i x_j, \quad i,j=1,...,9, \label{pot-bil}
\ee
where $x_i$ are defined as:
\bea
&& x_1 = |\phi_1|^2, ~
x_2 = |\phi_2|^2, ~
x_3 = |\phi_3|^2, \\ 
&& x_4 = \textrm{Re}(\phi_1^\dagger \phi_2),~ 
x_5 = \textrm{Re}(\phi_1^\dagger \phi_3), ~
x_6 = \textrm{Re}(\phi_2^\dagger \phi_3),\\ 
&& x_7 = \textrm{Im}(\phi_1^\dagger \phi_2), ~
x_8 = \textrm{Im}(\phi_1^\dagger \phi_3), ~
x_9 = \textrm{Im}(\phi_2^\dagger \phi_3)\, .
\eea
Some of the $b_{ij}$ coefficients in Eq.~(\ref{pot-bil}) are zero in the $Z_2 \times Z'_2$ symmetric 3HDM. In fact, every entry with an odd number of $\phi_i$ is forbidden. Therefore,  Eq.~(\ref{pot-bil}) can be explicitly written as:
\be
V = \sum\limits_{i=1}^3 a_i x_i + \sum\limits_{i=1}^6 b_{ii} x_i^2 +  b_{12} x_1 x_2 + b_{13} x_1 x_3 + b_{23} x_2 x_3 + \sum\limits_{i=7}^9 b_{ii} x_i^2. \label{pot-bilin2}
\ee
Further simplification arises when considering the value of the potential in Eq.~(\ref{pot-bil}) at a given stationary point. As long as there is no spontaneous CP-violation, $\langle x_7 \rangle = \langle x_8 \rangle = \langle x_9 \rangle =0 $, and thus the last sum in above equation will not contribute to the value of the potential at a given CP-conserving minimum. 

The potential in Eq.~(\ref{pot-bilin2}) is equivalent to the potential in Eq.~(\ref{potential}) and can be expressed as:
\be
V = A^T X + \frac{1}{2} X^T B X,
\ee
where
\be
X^T = (x_1, x_2, x_3, x_4, x_5, x_6, x_7, x_8, x_9), \quad A^T = (-\mu_1^2, -\mu_2^2, -\mu_3^2,0,0,0,0,0,0), \label{bilin-vec}
\ee
\bea
B = \left( \begin{array}{ccc} B_{13} & 0_{3\times 3} & 0_{3\times 3} \\
0_{3\times 3} & B_{46} & 0_{3\times 3} \\
0_{3\times 3} & 0_{3\times 3} & B_{79}
\end{array} \right), \label{bilin-B}
\eea
and 
\bea
&& B_{13}=\left(\begin{array}{ccc}
2 \lambda_{11} & \lambda_{12} & \lambda_{31}\\
\lambda_{12}& 2 \lambda_{22}& \lambda_{23} \\
\lambda_{31}& \lambda_{23}& 2 \lambda_{33} \end{array} \right), 
\quad 
B_{46}= 2\left(\begin{array}{ccc}
2 \lambda_{1} + \lambda'_{12}& 0& 0\\
0& 2 \lambda_{3} + \lambda'_{31}& 0\\
0& 0& 2 \lambda_{2} + \lambda'_{23}
\end{array} \right), \label{bilin-BCPc}\\[4mm]
&& B_{79}= 2\left(\begin{array}{ccc}
-2 \lambda_{1} + \lambda'_{12}& 0& 0\\
0& -2 \lambda_{3} + \lambda'_{31}& 0\\
0& 0& -2 \lambda_{2} + \lambda'_{23}
\end{array} \right). \label{bilin-BCPv}
\eea
The texture of the $B$ matrix shows the impact of the imposed $Z_2 \times Z'_2$ symmetry on the model, as well as the distinction between CP-conserving and potentially CP-violating contributions.

Using the identities in Eqs.~(\ref{bilin-vec}-\ref{bilin-BCPv}) we can express the potential in Eq.~(\ref{potential}) at a stationary point $a$ through:
\be
\mathcal{V}_a = A^T X_a + \frac{1}{2} X^T_a B X_a,
\ee
where 
\be
X_a^T = (\langle x_1 \rangle_a, \langle x_2 \rangle_a, \langle x_3 \rangle_a , \langle x_4 \rangle_a, \langle x_5 \rangle_a , \langle x_6 \rangle_a , \langle x_7 \rangle_a , \langle x_8 \rangle_a , \langle x_9 \rangle_a).
\ee
Notice, that the values of VEVs are in principle different at each stationary point, and are hence denoted by appropriate label $\langle x_i \rangle_a$. Furthermore, in general we cannot expect that every possible minimum respects the observed value of $v^2$, i.e. it is possible that $\sum_i v_i^2|_a \neq \sum_i v_i^2|_b$. For certain stationary points there appears the so called ``panic vacuum'', i.e. another deeper minimum with the same characteristics (i.e. same remnant symmetry and field content), but where the value of VEV, and hence the fermion and boson masses, does not match the observed values. Fortunately, there is no ``panic vacuum'' associated with \texttt{2-Inert}, as Eq~(\ref{inertex}) has a unique solution.

%
%First, we define:
%\bea
%&& X = \left(\begin{array}{c} |\phi_1|^2\\ |\phi_2|^2\\ |\phi_3|^2\\ (\phi_1^\dagger \phi_2)\\ (\phi_1^\dagger \phi_3) \\ (\phi_2^\dagger \phi_3)\end{array}\right), \quad X_a = X\Big\lvert_{\langle\phi_i\rangle} = \frac{1}{2} \left(\begin{array}{c} (v_1^2)^a\\ (v_2^2)^a\\ (v_3^2)^a\\ (v_1 v_2)^a\\ (v_1 v_3)^a \\ (v_2 v_3)^a\end{array}\right), \quad A = \left(\begin{array}{c} -\mu_1^2\\ -\mu_2^2\\ -\mu_3^2\\ 0\\ 0\\0 \end{array}\right), \label{bilin-vec}\\
%&& B=\left(\begin{array}{cccccc}
%2 \lambda_{11} & \lambda_{12} & \lambda_{31}& 0& 0& 0\\
%\lambda_{12}& 2 \lambda_{22}& \lambda_{23}& 0& 0& 0 \\
%\lambda_{31}& \lambda_{23}& 2 \lambda_{33}& 0& 0& 0 \\
%0& 0& 0& 2 (2 \lambda_{1} + \lambda'_{12})& 0& 0\\
%0& 0& 0& 0& 2 (2 \lambda_{3} + \lambda'_{31})& 0\\
%0& 0& 0& 0& 0& 2 (2 \lambda_{2} + \lambda'_{23})
%\end{array} \right). \label{bilin-B}
%\eea

At a given stationary point quadratic and quartic parts of the potential are related through  minimisation conditions and it can be explicitly checked that: 
\be
\mathcal{V}_a = \frac{1}{2} A^T X_a = - \frac{1}{2} X_a^T B X_a. \label{bilin-extr}
\ee
In the language of bilinears we can now define the derivative of the potential at a stationary point as:
\be
\mathcal{V}'_a = \frac{\partial V}{\partial X^T}\Bigg\lvert_{X=X_a} = A + B X_a. \label{bilin-der}
\ee
We can then derive the energy difference between two stationary points $V_1 - V_2$ by using Eq.~(\ref{bilin-extr}) and defining:
\bea
X_1^T \mathcal{V}'_2 = X^T_1 A + X^T_1 B X_2 = 2 \mathcal{V}_1 + X^T_1 B X_2,\\
X_2^T \mathcal{V}'_1 = X^T_2 A + X^T_2 B X_1 = 2 \mathcal{V}_2 + X^T_2 B X_1.
\eea
From here, and because $X^T_2 B X_1 = X^T_1 B X_2$ as $B$ is symmetric, we arrive at the relation:
\be
\mathcal{V}_1 - \mathcal{V}_2 = \frac{1}{2} (X_1^T \mathcal{V}'_2 - X_2^T \mathcal{V}'_1). \label{energydiff}
\ee
This quantity can inform us about the relative depths of different stationary points, however, on its own it cannot establish whether they are both minima or whether  one is e.g. a saddle point (i.e. some of the masses-squared are negative). 
In this paper, we are interested in establishing the region where \texttt{2-Inert} is the global minimum, therefore we aim to find parts of parameter space where $ \mathcal{V}_{X}- \mathcal{V}_{\texttt{2-Inert}} > 0$ for all stationary points listed in Table \ref{extrema-table}, regardless of their nature. For the convenience of the reader, we will present detailed analysis for the \texttt{DM1} stationary point, and provide final result for other combinations with details in Appendix~\ref{ap-masses}.

\subsection{Stability against \texttt{DM1}}  
The relevant vectors in Eqs.~(\ref{bilin-vec}) and (\ref{bilin-der}) for the \texttt{DM1} and \texttt{2-Inert} configurations are defined as
\be 
X_{\texttt{2-Inert}} = \frac{1}{2} \left(\begin{array}{c}0\\ 0\\ (v_3^2)^{\texttt{2-Inert}}\\ 0\\ 0 \\ 0\\ 0\\ 0 \\ 0\end{array}\right),
~ 
X_{\texttt{DM1}} = \frac{1}{2} \left(\begin{array}{c} 0\\ (v_2^2)^\texttt{DM1}\\ (v_3^2)^\texttt{DM1}\\ 0\\ 0\\ (v_2 v_3)^\texttt{DM1}\\ 0\\ 0 \\ 0\end{array}\right),
\ee
\be 
\mathcal{V}'_{\texttt{2-Inert}} = \left(\begin{array}{c}((v_3^2)^{\texttt{2-Inert}} \lambda_{31})/2 - \mu_1^2\\ ((v_3^2)^{\texttt{2-Inert}} \lambda_{23})/2 - \mu_2^2\\ 0\\ 0\\ 0\\0\\ 0\\ 0 \\ 0\end{array}\right), 
~
\mathcal{V}'_{\texttt{DM1}} = \left(\begin{array}{c}((v_2^2)^{\texttt{DM1}} \lambda_{12} + (v_3^2)^{\texttt{DM1}} \lambda_{31} - 2 \mu_1^2)/2\\ -
    (v_3^2)^{\texttt{DM1}} (2 \lambda_{2} + \lambda'_{23})/2\\ -
    (v_2^2)^{\texttt{DM1}} (2 \lambda_{2} + \lambda'_{23})/2\\ 0\\ 0\\ 
 (v_2)^{\texttt{DM1}} (v_3)^{\texttt{DM1}} (2 \lambda_{2} + \lambda'_{23})\\ 0\\ 0 \\ 0\end{array}\right).\label{VprimeInert}
\ee
We have explicitly included entries related to $x_{7}-x_{9}$ and as expected, they are all zero.

Using the relations above and following Eq.~(\ref{energydiff}) the energy difference between points \texttt{2-Inert} and \texttt{DM1} is:
\be
\mathcal{V}_{\texttt{DM1}} - \mathcal{V}_{\texttt{2-Inert}} = \frac{1}{4} (v_2^2)^{\texttt{DM1}} ((v_3^2)^{\texttt{2-Inert}} \Lambda_2 - \mu_2^2) = \frac{(v_2^2)^{\texttt{DM1}}}{4} (m_{H_2}^2)^{\texttt{2-Inert}}, \label{InertDM1}
\ee
where $(m_{H_2}^2)^{\texttt{2-Inert}}$ is the mass-squared of particle $H_2$ in the \texttt{2-Inert} minimum (Eq.~(\ref{massmh2})) which is always positive. 
From this we can conclude that if the \texttt{2-Inert} configuration \textit{is a minimum, then it will be deeper than the stationary point} \texttt{DM1}, regardless of the values of $(v_1^2)^{\texttt{DM1}}, (v_2^2)^{\texttt{DM1}}$. 
It is also worth pointing out that we can think of the inert particle $H_2$ as ``associated'' with the imposed $Z'_2$ symmetry with generator $(1,-1,0)$. This is precisely the symmetry that is broken by the VEV of \texttt{DM1}. As we will see, this feature is present in all cases.

As mentioned earlier, Eq.~\ref{InertDM1} does not tell us anything about the nature of the higher stationary point. 
As we are not focusing on the full analysis of the parameter space, rather on establishing the \texttt{2-Inert} configuration as the global minimum of the potential, we are not concerned with the properties of \texttt{DM1}. However, in this case it is easy to show that Eq.~(\ref{InertDM1}) can be written as (see Eqs.(\ref{inert-glob}) and (\ref{DM1energy})):
\be
\mathcal{V}_{\texttt{DM1}} - \mathcal{V}_{\texttt{2-Inert}} = - \frac{(\lambda_{33} \mu_2^2 - \Lambda_2 \mu_3^2)^2}{4 \lambda_{33} (\lambda_{22} \lambda_{33} - \Lambda_2^2)}\, ,
\ee
 which in turn requires $(\lambda_{22} \lambda_{33} - \Lambda_2^2)<0$. 
 As $\lambda_{33} >0$ from positivity conditions, this contradicts the condition in Eq.~(\ref{DM1global1}) for positive masses in \texttt{DM1}. We can then state that if \texttt{2-Inert} is a minimum then \texttt{DM1} can only be a saddle point.
 
 \subsection{Stability against other neutral minima}
 
\paragraph{2-Inert and DM2} Following the same steps as for \texttt{DM1} we arrive at the condition:
\be
\mathcal{V}_{\texttt{DM2}} - \mathcal{V}_{\texttt{2-Inert}} = \frac{(v_1^2)^{\texttt{DM2}}}{4} (m_{H_1}^2)^{\texttt{2-Inert}}. \label{InertDM2}
\ee
As the difference between \texttt{DM1} and \texttt{DM2} is related to switching roles of $\phi_1$ and $\phi_2$, it is not surprising that the conclusion is the same: if \texttt{2-Inert} \textit{is a minimum, then it will be deeper than the stationary point} \texttt{DM2}. Furthermore, we can show in the same way as above that in this case \texttt{DM2} will be a saddle point.

\paragraph{2-Inert and F0DM1} In contrast to the previous two cases, here the relation between the energies does not clearly indicate which point is the deeper one:
\be
\mathcal{V}_{\texttt{F0DM1}} - \mathcal{V}_{\texttt{2-Inert}}  =  \frac{1}{4} \left(-\frac{\mu_2^4}{\lambda_{22}} + \frac{\mu_3^4}{\lambda_{33}}  \right) = \frac{1}{16} \left( \frac{(m_h^4)^{\texttt{2-Inert}}}{\lambda_{33}} -\frac{(m_{H_2}^4)^{\texttt{F0DM1}}}{\lambda_{22}} \right).
\ee

\begin{figure}[h]
\begin{center}
\includegraphics[width=120mm]{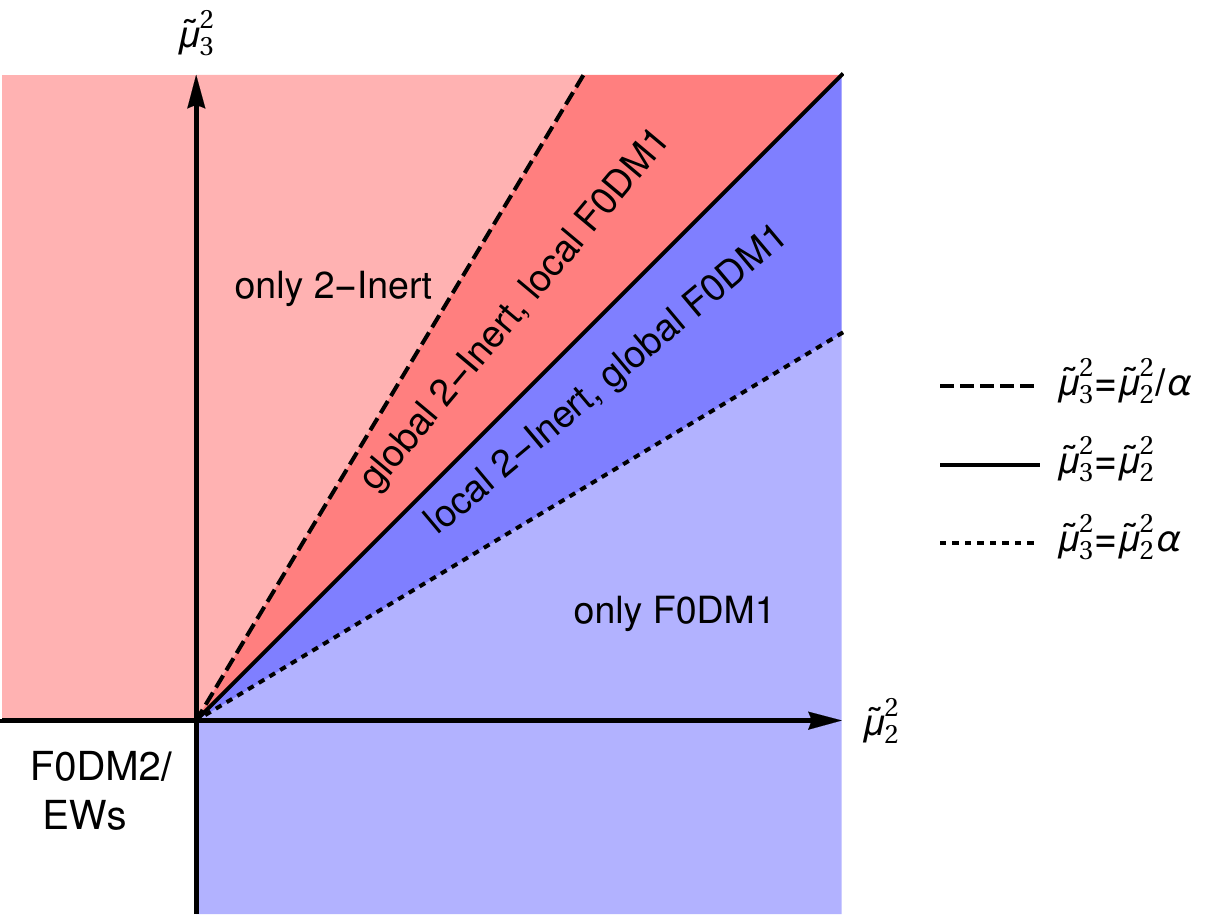}
\caption{Coexistence of the \texttt{2-Inert} and \texttt{F0DM1} states in the $(\tilde{\mu}^2_2,\tilde{\mu}^3_2)$ plane where $\tilde{\mu}^2_2 = \mu_2^2/ \sqrt{\lambda_{22}}$, $\tilde{\mu}^3_2 = \mu_3^2/ \sqrt{\lambda_{33}}$ and $\alpha = \sqrt{\lambda_{22} \lambda_{33}}/\Lambda_2<1$. 
In the light red region only \texttt{2-Inert} is a minimum, therefore a vacuum. In the dark red region \texttt{2-Inert} and \texttt{F0DM1} coexist while \texttt{2-Inert} is the global minimum. In the dark blue region, the \texttt{2-Inert} and \texttt{F0DM1} coexist with the \texttt{F0DM1} as the global minimum. In the light blue region only the \texttt{F0DM1} exists. The white region shows the region where none of the states \texttt{F0DM1} and \texttt{2-Inert} are minima of the potential.}
 \label{fig-coex}
\end{center}
\end{figure}

In fact, this is one of the cases where there is the possibility of coexistence of two minima at the same time. This is not an unexpected situation, as the coexistence of the \textit{inert} and  the so-called \textit{inert-like} minima also occurs in the I(1+1)HDM~\cite{Ginzburg:2010wa}. We arrive at a very similar condition to the one from the I(1+1)HDM, i.e. for \texttt{2-Inert} to be a deeper minimum the following relation must hold:
\be
\frac{\mu_3^2}{\sqrt{\lambda_{33}}} > \frac{\mu_2^2}{\sqrt{\lambda_{22}}}. \label{InertF0DM1}
\ee
Of course, if the relation in Eq.~(\ref{InertF0DM1}) does not hold but all masses-squared in Eqs.(\ref{massmh1}-\ref{massmhc2}) are positive, then \texttt{2-Inert} is a minimum and we face the possibility of a metastable vacuum. This is shown as a dark blue region in Figure \ref{fig-coex}. 
If the lifetime of this minimum is large enough, then the Universe may as well be in this exact state. However, establishing the validity of a metastable \texttt{2-Inert} state requires calculations of the  tunnelling time. For simplicity, we are limiting our studies only to a region of the parameter space where \texttt{2-Inert} is a global minimum, i.e. the relation in Eq.~(\ref{InertF0DM1}) must be respected. As we will show in Section \ref{sec-constraints}, this choice has a significant impact on the studied parameter space.

\paragraph{2-Inert and F0DM2} 
As in the case of \texttt{DM1} and \texttt{DM2}, in the \texttt{F0DM2} state $\phi_1$ and $\phi_2$ switch roles with respect to the \texttt{F0DM1} state. Therefore, the existence of a deeper \texttt{2-Inert} minimum is guaranteed by a similar condition:
\be
\frac{\mu_3^2}{\sqrt{\lambda_{33}}} > \frac{\mu_1^2}{\sqrt{\lambda_{11}}}. \label{InertF0DM2}
\ee

\begin{figure}[h]
\begin{center}
\includegraphics[scale=0.5]{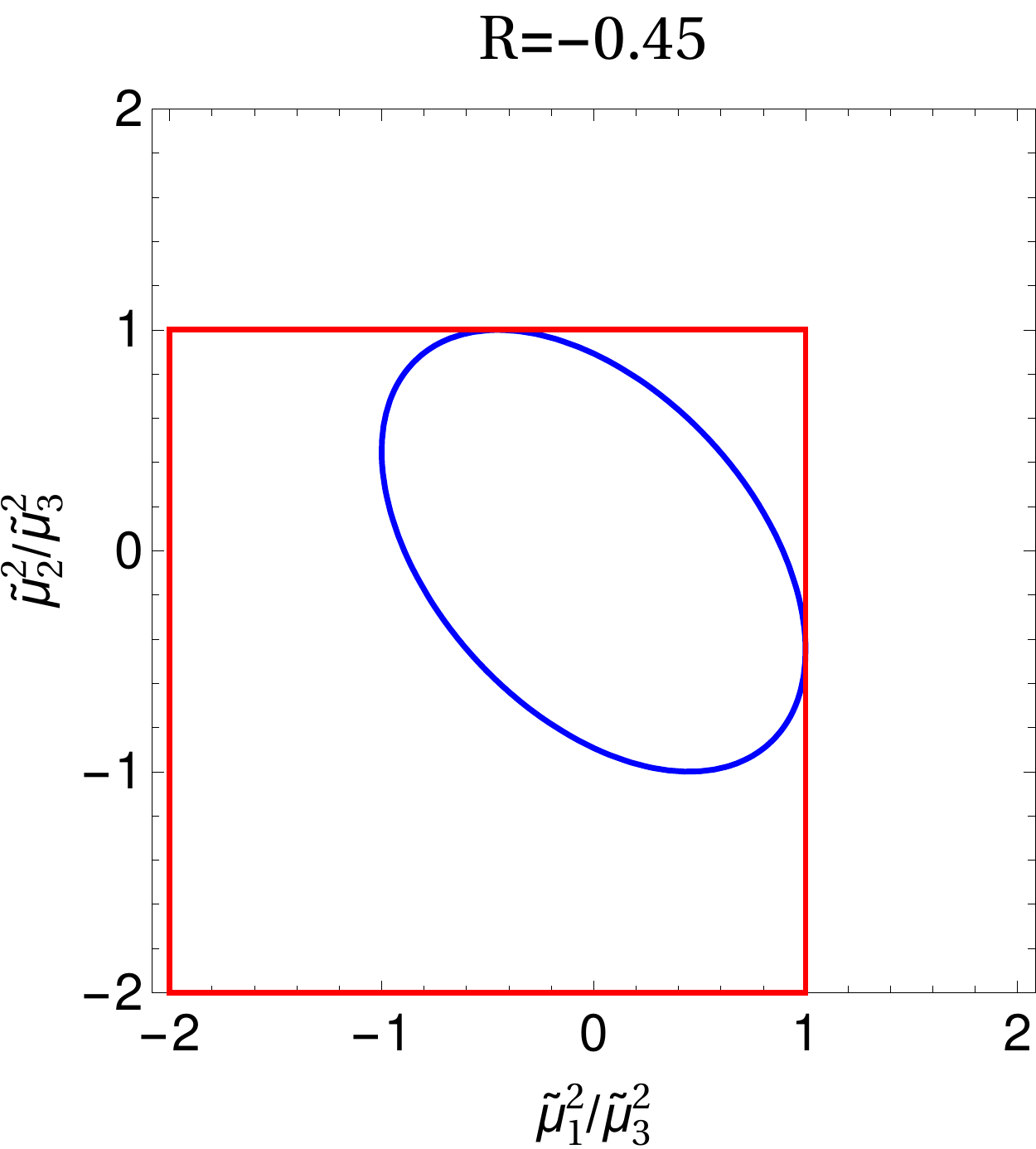} \quad \includegraphics[scale=0.5]{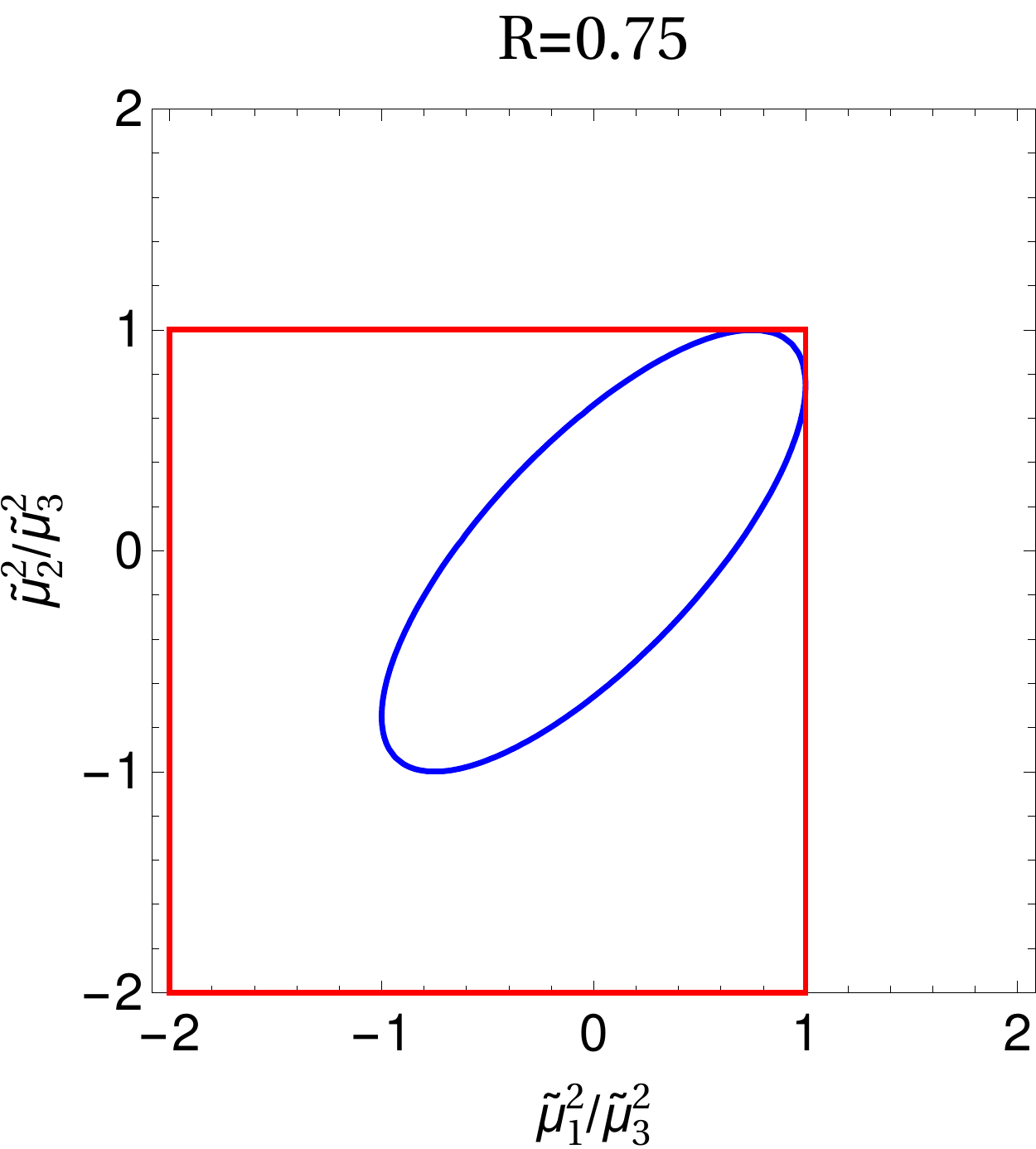}
\caption{Regions in the ${\tilde{\mu}^2_1}/{\tilde{\mu}^3_2}\,$--$\,{\tilde{\mu}^2_2}/{\tilde{\mu}^3_2}$ plane where the \texttt{2-Inert} state can be the global minimum for $R=-0.45$ on the left and  $R=0.75$ on the right. Inside the red box is the region where the \texttt{2-Inert} state is stable against the \texttt{F0DM1} and \texttt{F0DM2} states. Inside the blue ellipse the \texttt{2-Inert} state is stable against the \texttt{F0DM0} one.}
 \label{fig-coex2}
\end{center}
\end{figure}

\paragraph{2-Inert and  F0DM0} 
In this case, the energy difference between the two states is calculated to be
\bea
&&\mathcal{V}_{\texttt{F0DM0}} - \mathcal{V}_{\texttt{2-Inert}}  = \frac{1}{4} \left(-(v_1^2)^{\texttt{F0DM0}} \mu_1^2-(v_2^2)^{\texttt{F0DM0}} \mu_2^2 + (v_3^2)^{\texttt{2-Inert}} \mu_3^2 \right) 
\\
&&= \frac{1}{4} \left((v_1^2)^{\texttt{F0DM0}} (m_{H_1}^2)^{\texttt{2-Inert}} +(v_2^2)^{\texttt{F0DM0}} (m_{H_2}^2)^{\texttt{2-Inert}} - (v_3^2)^{\texttt{2-Inert}} (m_{H_3}^2)^{\texttt{F0DM0}} \right),\nonumber
\eea
where $(m_{H_3}^2)^{\texttt{F0DM0}}$ denotes the mass of the ``Higgs'' particle in \texttt{F0DM0} coming from $\phi_3$. This is yet another situation where the \texttt{2-Inert} state is not automatically stable. The energy difference can also be written in terms of the parameters of the potential as:
%\be
%\mathcal{V}_{\texttt{F0DM0}} - \mathcal{V}_{\texttt{2-Inert}}  = \frac{1}{4} \left(-\frac{\lambda_{22}\mu_1^4+\lambda_{11}\mu_2^4-2\Lambda_1\mu_1^2\mu_2^2}{\Lambda_1^2 -\lambda_{11}\lambda_{22}}  + \frac{\mu_3^4}{\lambda_{33}}\right),
%\ee
\be
\mathcal{V}_{\texttt{F0DM0}} - \mathcal{V}_{\texttt{2-Inert}}  = \frac{1}{4} \left(-\frac{\tilde{\mu}_1^4 + \tilde{\mu}_2^4-2\,R\,\tilde{\mu}_1^2\,\tilde{\mu}_2^2}{R^2-1}  + \tilde{\mu}^4_3\right), \label{InertF0DM0}
\ee
where we have introduced the notation of:
\be
\tilde{\mu}_i^2 = \mu_i^2/\sqrt{\lambda_{ii}},  \quad R = \Lambda_1/\sqrt{\lambda_{11} \lambda_{22}}\, .
\ee
The relation between energies in this case is much more complicated in comparison to the \texttt{F0DM1} or \texttt{F0DM2} states, and it strongly depends on the values of the dark couplings between doublets $\phi_1$ and $\phi_2$, i.e. self-interactions of inert particles in the \texttt{2-Inert} state, in particular the parameter $R$. 
This quantity is constrained by the positivity conditions (stability of the potential) and positivity of the masses in \texttt{F0DM0} (see Eq.~\ref{F0DM0min}):
\bea
&& R>-1, \quad R^2-1<0 \quad \Rightarrow \quad \vert R \vert < 1 \Rightarrow \textrm{ \texttt{F0DM0} is a minimum,}\\
&& R > 1 \quad \Rightarrow \textrm{ \texttt{F0DM0} is not a minimum.}
\eea
Therefore, one easy way to ensure that there is no possibility of tunnelling into an undesirable \texttt{F0DM0} minimum is to choose parts of parameter space where $R>1$. This can also be achieved by requiring $\lambda_1 >0$ or $\lambda'_{12}/2+\lambda_1 >0$ (for detailed calculations see Appendix \ref{ap-masses}). Otherwise, a numerical comparison has to be performed between different vacua. 
An example is shown in Figure \ref{fig-coex2} where inside the red box is the region where the \texttt{2-Inert} state is a deeper minimum compared to the \texttt{F0DM1} or \texttt{F0DM2} states while the region inside the blue ellipse is where the \texttt{2-Inert} state is deeper than the \texttt{F0DM0} state.
It is clear that stability against \texttt{F0DM1} or \texttt{F0DM2} does not guarantee stability against \texttt{F0DM0}.

\paragraph{2-Inert and  N} Finally, for the last remaining CP-conserving neutral stationary point, the energy difference between \texttt{2-Inert} and \texttt{N} is given by:
\be
\mathcal{V}_{\texttt{N}} - \mathcal{V}_{\texttt{2-Inert}} = \frac{1}{4} \left((v_1^2)^{\texttt{N}} (m_{H_1}^2)^{\texttt{2-Inert}} + (v_2^2)^{\texttt{N}} (m_{H_2}^2)^{\texttt{2-Inert}}\right).
\ee
From this we can conclude that if the \texttt{2-Inert} minimum exists, it is automatically a deeper minimum than the \texttt{N} one.

\paragraph{2-Inert and sCPv} The energy difference between the CP-violating extremum and the \texttt{2-Inert} state is:
\bea
\mathcal{V}_{\texttt{sCPv}} - \mathcal{V}_{\texttt{2-Inert}} = \frac{1}{4} \left((v_1^2 \cos^2\xi_1)^{\texttt{sCPv}} (m_{H_1}^2)^{\texttt{2-Inert}} + (v_2^2 \cos^2\xi_2)^{\texttt{sCPv}} (m_{H_2}^2)^{\texttt{2-Inert}}\right. \nonumber\\
\left.+(v_1^2 \sin^2\xi_1)^{\texttt{sCPv}} (m_{A_1}^2)^{\texttt{2-Inert}} + (v_2^2 \sin^2\xi_2)^{\texttt{sCPv}} (m_{A_2}^2)^{\texttt{2-Inert}}\right) >0.
\eea
Therefore, if \texttt{2-Inert} is a minimum, it will be deeper than a possible CP-violating extremum. Notice that by taking a CP-conserving limit, $\xi_1,\xi_2\to 0$, we restore the condition for the \texttt{N} state where only the masses of the CP-even particles appears in the energy difference expression. Whereas in the CP-violating state, the masses of the ``pseudoscalar'' particles also enters the equation.

\subsection{Stability against the charged minimum \label{ap-charged}}

Although the Higgs sector in the SM is safe against inducing a complete breaking of the EW symmetry, this is not necessarily the case in multi-doublet models. It is known that even in the 2HDM there exist the possibility of the so-called \textit{charge breaking} minimum, where at least one of the doublets develops a VEV in the upper component. This leads to the appearance of four Goldstone bosons, one of which is ``eaten'' by the photon giving it a mass. This is certainly not the vacuum that is realised in our Universe. We will now show that for the phenomenologically interesting region of the \texttt{2-Inert} configuration there is no possibility of tunnelling to the charge breaking minimum.

As we can always choose the so-called neutral direction in the isospin space, the most general charge breaking minimum VEV configuration, i.e. one that breaks EW symmetry completely is:
\be
\langle \phi_1 \rangle = \frac{1}{\sqrt{2}}\doublet{u_1}{c_1} ,\quad 
\langle \phi_2 \rangle = \frac{1}{\sqrt{2}}\doublet{u_2}{c_2} , \quad 
\langle \phi_3 \rangle = \frac{1}{\sqrt{2}}\doublet{0}{c_3}  .
\ee

For the vacuum to be charge breaking, at least one of the upper components $u_1, \, u_2$ has to be non-zero. Assuming first that all VEV entries are non-zero, $u_{1,2}, c_{1,2,3} \neq 0$, extremum conditions for this type of vacuum are calculated to be:

\begin{flalign}
&c_1 \mu_1^2 = c_1 \left( \left( c_2^2+\frac{c_2 u_1 u_2}{c_1} \right) \left( \lambda_{1} + \frac{\lambda'_{12}}{2} \right)  + (c_1^2 + u_1^2) \lambda_{11} + \frac{c_2^2 + u_2^2}{2} \lambda_{12} + c_3^2 \left(\frac{\lambda_{31}}{2} + \lambda_{3} +\frac{\lambda'_{31}}{2} \right) \right), \label{charmu11}\\
&u_1 \mu_1^2  =  u_1 \left( \left(\frac{c_1 c_2 u_2}{u_1} + u_2^2 \right) \left( \lambda_{1} + \frac{\lambda'_{12}}{2}  \right) + (c_1^2 + u_1^2) \lambda_{11} + \frac{c_2^2 + u_2^2}{2} \lambda_{12} + \frac{c_3^2}{2} \lambda_{31} \right),\label{charmu12} 
\end{flalign}
\begin{flalign}
&c_2 \mu_2^2 = c_2 \left( \left(c_1^2 + \frac{c_1 u_1 u_2 }{c_2} \right) \left(\lambda_{1} + \frac{\lambda'_{12}}{2} \right) + (c_2^2 + u_2^2) \lambda_{22} + \frac{c_1^2 +u_1^2}{2} \lambda_{12} +  c_3^2 \left(\frac{\lambda_{23}}{2}   + \lambda_{2}   + \frac{\lambda'_{23}}{2} \right) \right), \label{charmu21}\\ 
&u_2 \mu_2^2 = u_2 \left( \left(u_1^2 + \frac{c_1 u_1 u_1 }{u_2} \right) \left(\lambda_{1} + \frac{\lambda'_{12}}{2} \right) + (c_2^2 + u_2^2) \lambda_{22} + \frac{c_1^2 +u_1^2}{2} \lambda_{12} + \frac{c_3^2}{2} \lambda_{23} \right) , \label{charmu22}\\ 
&c_3 \mu_3^2 =c_3 \left(c_2^2 \lambda_{2} + c_1^2 \lambda_{3} + c_3^2 \lambda_{33} +  \frac{c_1^2 + u_1^2}{2} \lambda_{31} + \frac{c_2^2 + u_2^2}{2} \lambda_{23} +  \frac{c_1^2}{2} \lambda'_{31} + \frac{c_2^2}{2} \lambda'_{23}\right). \label{charmu33}
\end{flalign}
It is easy to see that the above equations cannot be fulfilled simultaneously, unless further relations on the parameters and/or VEVs are imposed. 

Let us first focus on the choice 
\be
\texttt{CB1}: \quad c_1, \;c_2, \;c_3, \;u_1, \;u_2 \neq 0. \label{generalCB}
\ee
We can simultaneously fulfil two pairs of conditions in Eqs.~(\ref{charmu11})-(\ref{charmu12}) and Eqs.~(\ref{charmu21})-(\ref{charmu22}) only if:
\be
2 \lambda_1 + \lambda'_{12} =0, \quad 2 \lambda_2 + \lambda'_{23} =0, \quad 2 \lambda_3 + \lambda'_{31} =0. \label{specialChar}
\ee
This not only seems like an unnatural choice, as there is no symmetry that imposes these kinds of relations, but also contradicts the assumption in Eq.~(\ref{Hlightest}) and results in a partially degenerate mass spectrum where $m_{H_i} = m_{H_i^\pm}$. 
If $\lambda_{2}$ or $\lambda_3$ are negative, then both $H_i$ and $H^\pm_i$ will be lighter than $A_i$. Therefore, there will be four different stable particles with two of them charged, which is not in agreement with cosmological observations. 
In this paper we only consider parts of the parameter space where $m_{H_{i}} < m_{H^\pm_i} \Leftrightarrow  2 \lambda_2 + \lambda'_{23} \neq 0, \,2 \lambda_3 + \lambda'_{31} \neq 0$ which forbids the appearance of this charge breaking extremum. Therefore, in our analysis the \texttt{2-Inert} state will be stable against the \texttt{CB1} state.

This argument is not sufficient if we were to assume that the lightest inert particles from two generations were $A_1$ and $A_2$. In such a case the degeneracy between $H_i$ and $H^\pm_i$ would not lead to any immediate catastrophic consequences for DM. However, looking at the relative depths of \texttt{CB1} and \texttt{2-Inert}, following the procedure described in Sec. \ref{sec-bilinear}, we arrive at the relation:
\be
\mathcal{V}_{\texttt{CB1}} - \mathcal{V}_{\texttt{2-Inert}} = \frac{1}{4}\left((u_1^2 + c_1^2) (m_{H^\pm_1}^2)^{\texttt{2-Inert}} + (u_2^2+c_2^2) (m_{H^\pm_2}^2)^{\texttt{2-Inert}} \right) >0, \label{CB1Inertdif}
\ee
i.e. \texttt{2-Inert} is automatically stable against the \texttt{CB1} minimum, regardless of the mass order. Notice that yet again, the relative depth of these two extrema is related to the masses of particles connected to the symmetries that are violated. Which in this case are both the discrete $Z_2$ symmetries and the $U(1)_{EM}$, hence the difference is related to the masses of the charged inert scalars.

Conditions in Eq.s~(\ref{charmu11}-\ref{charmu33}) can also be fulfilled if certain VEV entries are set to zero. 
In particular, consider the points:
\bea
&& \texttt{CB2}: \quad c_1 = 0, \; c_2, \;c_3, \;u_1, \;u_2 \neq 0,\\
&&\texttt{CB3}: \quad c_2 = 0, \; c_1, \;c_3, \;u_1, \;u_2 \neq 0,\\
&&\texttt{CB4}: \quad c_3 = 0, \; c_1, \;c_2, \;u_1, \;u_2 \neq 0,\\
&&\texttt{CB5}: \quad u_1 = 0, \; c_1, \;c_2, \;c_3, \;u_2 \neq 0,\\
&&\texttt{CB6}: \quad u_2 = 0, \; c_1, \;c_2, \;c_3, \;u_1 \neq 0.
\eea
In all these points, at least one of the following conditions has to be fulfilled:
\be
2 \lambda_2 + \lambda'_{23} =0, \quad 2 \lambda_3 + \lambda'_{31} =0,
\ee
which would lead to the same conclusions as in case of the \texttt{CB1} extremum, with the corresponding VEV set to zero in Eq.~(\ref{CB1Inertdif}). 

The first type of charge breaking extremum that does not require  any further relations on the parameters of the potential is:
\be
\texttt{CB7}: \quad c_1 = \;c_2 =0, \quad c_3, \;u_1, \;u_2 \neq 0.
\ee
The structure here resembles that of the charge breaking state in the $Z_2$ symmetric 2HDM, where the charge breaking doublet only has a non-zero VEV in the upper component. As before, we calculate the relevant $X_{\texttt{CB7}}$ and $\mathcal{V}'_{\texttt{CB7}}$:
\be 
X_{\texttt{CB7}} = \frac{1}{2} \left(\begin{array}{c}u_1^2\\ u_2^2\\ c_3^2\\ u_1 u_2 \\ 0 \\ 0\end{array}\right), \quad \mathcal{V}'_{\texttt{CB7}} = -\frac{2\lambda_1 + \lambda'_{12}}{2}\left(\begin{array}{c}u_2^2\\u_1^2\\0\\-2u_1 u_2\\0\\0\end{array}\right).
\ee
The relative depths of the \texttt{CB7} and \texttt{2-Inert} states is calculated to be:
\be
\mathcal{V}_{\texttt{CB7}} - \mathcal{V}_{\texttt{2-Inert}} = \frac{1}{4}\left(u_1^2 (m_{H^\pm_1}^2)^{\texttt{2-Inert}} + u_2^2 (m_{H^\pm_2}^2)^{\texttt{2-Inert}} \right) >0, \label{CB7Inertdif}
\ee
which guarantees that the \texttt{2-Inert} minimum is stable against the \texttt{CB7} state. 

We can also consider special cases of the above scenario, where one $u_i$ is also zero:
\bea
&& \texttt{CB8}: \quad c_1 = \;c_2 = \;u_2  = 0, \quad c_3, \;u_1 \neq 0, \\
&& \texttt{CB9}: \quad c_1 = \;c_2 = \;u_1 = 0, \quad c_3, \;u_2 \neq 0,
\eea
and the energy difference between \texttt{CB8} and \texttt{2-Inert} (\texttt{CB9} and \texttt{2-Inert}) is given by the first (second) term in Eq.~(\ref{CB7Inertdif}). Therefore the \texttt{2-Inert} minimum is stable against all possible charge breaking vacua.

\subsection{The 2-Inert state as the global minimum \label{sec-stabsummary}}

In general, our considered model exhibits a similar behaviour to other multi-scalar models:
\begin{enumerate}
\item The most symmetric EW breaking minimum, in our case the \texttt{2-Inert}, is {\sl automatically stable} against a number of other potential neutral minima with higher level of symmetry breaking, i.e. \texttt{DM1}, \texttt{DM2}, \texttt{N} and \texttt{sCPv}.  

\item The \texttt{2-Inert} state is {\sl not automatically stable} against a class of non-physical minima where fermions are massless, i.e. \texttt{F0DM1}, \texttt{F0DM2} and \texttt{F0DM0}. The global/local nature of the \texttt{2-Inert} state as a minimum depends on the relations between parameters of the potential and for phenomenological studies we have to ensure, that there is no possibility of tunnelling into any of these inert-like states.

\item The \texttt{2-Inert} state is {\sl automatically stable} against charge breaking scenarios.
\end{enumerate} 

Regarding the second point, we want to acknowledge that the presented analysis is done only at tree-level. The authors of \cite{Ferreira:2015pfi} have shown that the relative energy difference between inert and inert-like minima in the I(1+1)HDM can change once one-loop level corrections to the potential are included. In most cases this does not lead into any devastating effects -- the tree-level global minimum usually retains its status as global even when loop effects are included. However, the allowed parameter space is modified (e.g. there appear solutions for negative values of $\mu_i^2 <0$ whereas previously only the \texttt{EWs} vacuum could be realised). 
However, in rare cases, when the two minima are almost degenerate, the nature of the global minimum can change, rendering a tree-level global minimum metastable. This highlights the need for performing loop-level calculations in order to ensure the stability of a chosen vacuum fully. 

Furthermore, one can ask the question whether moving beyond the tree-level analysis can also change conclusions of point 1 -- i.e. is there  a part of the parameter space where the \texttt{2-Inert} minimum can coexist with, and in the worst scenario can be higher than, any of the \texttt{DM1}, \texttt{DM2} and \texttt{N} states? In this case, one is dealing with not only changing the global/local properties of a minimum but in fact, transforming a saddle point into a minimum. This question is not yet fully answered even in the 2HDM framework. In such cases, as one or more of the squared scalar masses at these saddle points will be negative, we may anticipate the one-loop effective potential having an imaginary component. As the full analysis is clearly beyond the scope of this work, we postpone further analysis of this question to a future publication.

\section{Theoretical and experimental constraints \label{sec-constraints}}

\begin{figure}[h]
\begin{center}
\includegraphics[scale=1]{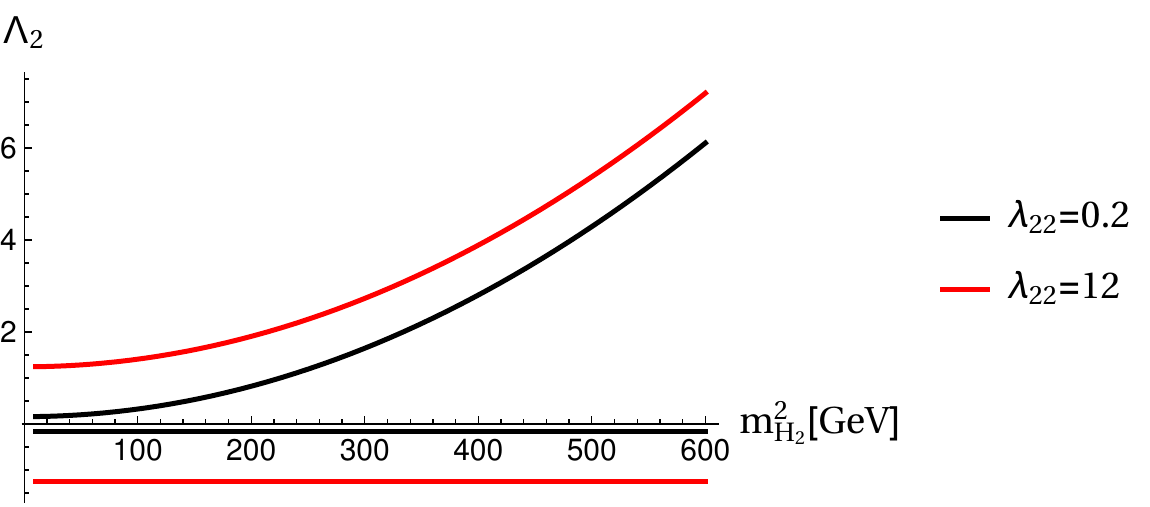}
\end{center}
\caption{Upper and lower limits on $\Lambda_2$ from stability of the vacuum condition for $\lambda_{22} = 0.2$ (black) and $\lambda_{22} = 12$ (red). In the region between the two corresponding curves the \texttt{2-Inert} vacuum is stable. \label{fig-L2limits}}
\end{figure}

\paragraph{Stability of the vacuum} We impose previously discussed positivity constraints, i.e. potential boundedness from below in Eqs.(\ref{positivity1}--\ref{positivity3}) and conditions for the \texttt{2-Inert} state to be the global minimum in Eqs.(\ref{InertF0DM1}--\ref{InertF0DM2}). 
Naively, the conditions for the global minimum seem to limit only quadratic and self-coupling parameters. However, it is clearly not the case, as we can rewrite Eqs.(\ref{InertF0DM1}--\ref{InertF0DM2}) in terms of masses of the inert particles to obtain:
\begin{flalign}
&-\sqrt{\lambda_{33} \lambda_{22}} < \Lambda_2 < \sqrt{\lambda_{33} \lambda_{22}} + \frac{m_{H_2}^2}{v^2},\qquad - \sqrt{\lambda_{33} \lambda_{11}} < \Lambda_3 < \sqrt{\lambda_{33} \lambda_{11}} + \frac{m_{H_1}^2}{v^2}, \label{Lam23stab}
\end{flalign}
where $\lambda_{33}$ is fixed by the Higgs mass and $v$. We therefore have an upper bound on parameters $\Lambda_2$ and $\Lambda_3$ which govern the interactions between DM candidates and the Higgs particle. The lower bounds in above equations come from the positivity constraints. 
%This can be particularly important in the low mass region, where $m_{H_i}^2 < v^2$, while in the heavy mass regime it is generally less so.
We can see once again the importance of establishing the conditions for the global minimum. These conditions depend on the values of self-coupling parameters $\lambda_{11}$ and $\lambda_{22}$ which are otherwise virtually unconstrained by experiments and have no impact on tree-level DM and collider phenomenology. However, they strongly influence the allowed parameter space, including crucial Higgs-DM couplings,  as shown in Figure~\ref{fig-L2limits}, where we plot the maximum and minimum values for $\Lambda_2$ as a function of the mass of DM candidate (a similar plot can be obtained for $\Lambda_3$ in terms of $m_{H_1}$ and $\lambda_{11}$). Especially in the low mass region, these conditions have a very strong impact on the range of physical parameters, and as we show later, they are stronger than standard limits arising from perturbative unitarity constraints. We would also like to stress again, that limits in Eq.~(\ref{Lam23stab}) are directly related to the \texttt{2-Inert} state being the \textit{global} minimum and would be relaxed if we were to consider cases of metastable vacua.

As discussed before, ensuring stability against the \texttt{F0DM0} state in Eq.(\ref{InertF0DM0}) results in strong conditions on values of all parameters of the potential. In the region of the parameter space where $\vert R \vert<1$, the \texttt{2-Inert} state can be the global minimum only if $\tilde{\mu}_1^2 $ and $\tilde{\mu}_2^2$ are of the same order. 
Unless there is a considerable discrepancy between $\lambda_{11}$ and $\lambda_{22}$, it will forbid large mass differences between masses of two families of dark particles. 
Furthermore, the existence of the \texttt{F0DM0} minimum and its depth are directly related to the values of dark self couplings. 
While $\lambda_{11}$ and $\lambda_{22}$ govern only undetectable scatterings\footnote{Assuming we are not in the region where DM self-interacts strongly.}
% \orange{(VK: maybe we should avoid showing very large dark couplings in Figure \ref{fig-L2limits} then.)}} 
of inert scalars within one family, $H_i H_i \to H_i H_i$, parameter $\Lambda_1$ (which is a  combination of $\lambda_1, \lambda_{12}$ and $\lambda'_{12}$) is responsible for the strength of the conversion processes, $H_2 H_2 \to H_1 H_1$, and will have a major influence on the thermal evolution of DM particles. 
 
\paragraph{Perturbative unitarity} We require that the scalar $2 \to 2$ scattering matrix is unitary, i.e. the absolute values of all eigenvalues of scattering matrices for scalars with specific hypercharge and isospin should be smaller than $8 \pi$. Furthermore, all quartic scalar couplings should be perturbative, i.e. $\lambda_i \leq 4 \pi$.
Following the analysis of \cite{Moretti:2015cwa} we define the relevant quantities as:
\begin{small}
\begin{flalign}
&X_1 = 
\begin{pmatrix} 
6\lambda_{33} &2\lambda_{31} + \lambda'_{31} & 2\lambda_{23} + \lambda'_{23}  \\
2\lambda_{31} + \lambda'_{31} & 6\lambda_{11} & 2\lambda_{12} + \lambda'_{12}\\
 2\lambda_{23} + \lambda'_{23} & 2\lambda_{12} + \lambda'_{12} & 6\lambda_{22} 
\end{pmatrix}, ~
X_2  =
\begin{pmatrix} 
2\lambda_{33} &\lambda'_{31} & \lambda'_{23}  \\
\lambda'_{31} & 2\lambda_{11} &  \lambda'_{12}\\
\lambda'_{23} & \lambda'_{12} & 2\lambda_{22} 
\end{pmatrix},~
X_3 = 2
\begin{pmatrix} 
\lambda_{33} &\lambda_{3} & \lambda_{2}  \\
\lambda_{3} & \lambda_{11} &  \lambda_{1}\\
\lambda_{2} & \lambda_{1} & \lambda_{22} 
\end{pmatrix}, \nonumber 
\end{flalign}
\end{small}
\begin{flalign}
&y_1^\pm = \lambda_{12} + 2\lambda'_{12} \pm 6\lambda_{1}, \qquad
y_4^\pm = \lambda_{12} \pm 2\lambda_{1}, \qquad
y_7^\pm  =  \lambda_{12} \pm \lambda'_{12} ,
\nonumber\\[1mm]
&y_2^\pm = \lambda_{31} + 2\lambda'_{31} \pm 6\lambda_{3}, \qquad
y_5^\pm  = \lambda_{31} \pm 2\lambda_{3}, \qquad
y_8^\pm =  \lambda_{31} \pm \lambda'_{31} , 
\label{y9}  \\[1mm]
&y_3^\pm = \lambda_{23} + 2\lambda'_{23} \pm 6\lambda_{2}, \qquad
y_6^\pm  = \lambda_{23} \pm 2\lambda_{2},  \qquad
y_9^\pm  =  \lambda_{23} \pm \lambda'_{23},
\nonumber
\end{flalign}
and require the following relations:
\be
\vert \hat{x}_i \vert \leq 8 \pi, \quad y_i^\pm \leq 8 \pi, \quad i=1,...,9,
\ee
where $\hat{x}_{1-3}$, $\hat{x}_{4-6}$ and $\hat{x}_{7-9}$ are the eigenvalues of matrices $X_1$, $X_2$ and $X_3$, respectively.

In general, the maximum values of $\lambda_i$ allowed by perturbative unitarity constraints can be found only numerically. However, for specific cases we can derive certain preliminary constraints to illustrate general ranges for these parameters:
\begin{enumerate}
\item 
From the eigenvalues of $X_1$ in the limit of $\lambda_{ij},\lambda_{ij}'\to 0$ we obtain:
\be
\lambda_{11} \leq \lambda_{11}^{max} = \frac{4}{3} \pi, \quad \lambda_{22} \leq \lambda_{22}^{max} = \frac{4}{3} \pi.
\ee
Notice that this condition together with Eq.(\ref{Lam23stab}), provides a lower bound for $\Lambda_2$ and $\Lambda_3$ to be:
\be
\Lambda_{2,3} > \Lambda_{2,3}^{min} \approx - 0.735.
\ee

\item 
In the limit of $\lambda_{22} \to 0$ and $\lambda_{2},\lambda_{23}' \to 0$ ($\lambda_{11} \to 0$ and $\lambda_{3},\lambda_{31}' \to 0$) from eigenvalues of $X_1,\,X_3$ and $y_{5},\,y_{6}$ we find the maximum value for $\Lambda_2$ ($\Lambda_3$) to be:
\begin{equation}
\Lambda_2 \leq \Lambda_2^{max} = \frac{1}{2}\lambda_{23}^{max} = 2 \pi, \quad \Lambda_3 \leq \Lambda_3^{max} = \frac{1}{2}\lambda_{31}^{max} = 2 \pi.
\end{equation}
These conditions can be weaker or stronger than conditions in Eq.~(\ref{Lam23stab}), depending on the value of $m_{H_1}$ ($m_{H_2}$). As shown in Figure~\ref{fig-L2limitsPert}, for lower masses the limits from ensuring the stability of the \texttt{2-Inert} vacuum are stronger than the bounds from perturbative unitarity, while the opposite is true for heavier masses.  
\begin{figure}[h]
\begin{center}
\includegraphics[scale=1]{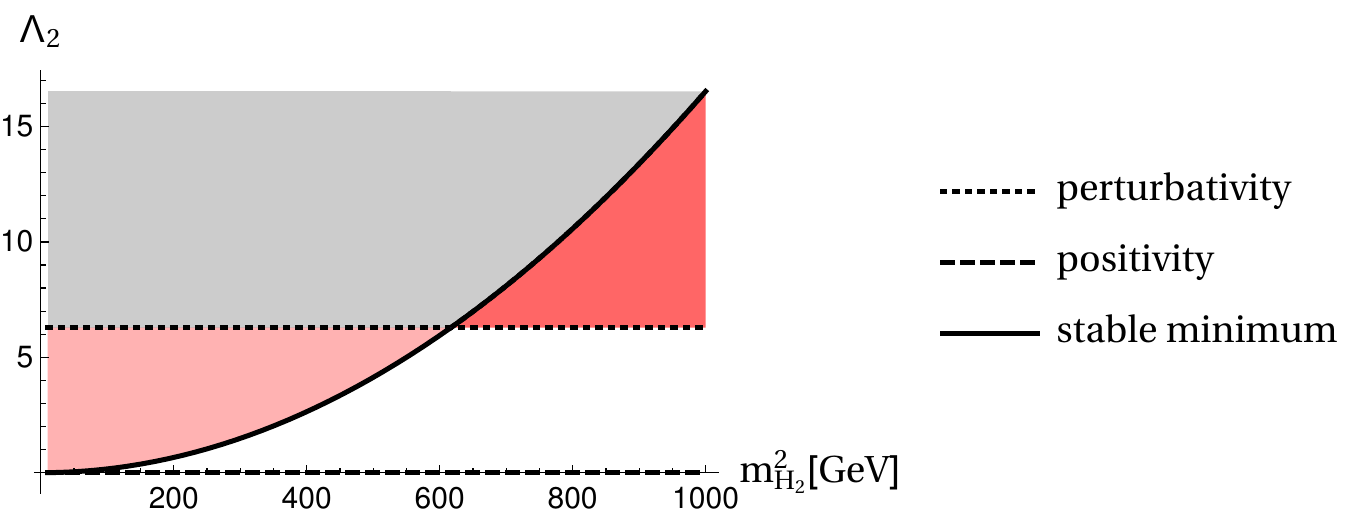}
\end{center}
\caption{The interplay between positivity (dashed), perturbative unitarity (dotted) and global minimum (solid) conditions on the value of $\Lambda_2$ in the $\lambda_{22} \to 0$ limit. The white region is the allowed region while the light red region is excluded by vacuum stability, the dark red region is excluded by perturbativity and the gray region is excluded by both. \label{fig-L2limitsPert}}
\end{figure}
\item Similarly, for $\Lambda_1$ the range is limited to:
\be
-\frac{4}{3}\pi < \Lambda_1 \leq 2 \pi,
\ee
where the lower limit comes from positivity constraints, while the perturbative unitarity condition imposes the upper limit.
\end{enumerate}

\paragraph{Electroweak precision observables (EWPO)} We demand a 2$\sigma$ (i.e. 95\% C.L.) agreement with electroweak precision observables, which are  parametrized through the electroweak oblique parameters $S,T,U$. When the $Z_2\times Z'_2$ symmetry is exact (there is no $\mu_{12}^2 \phi_1^\dagger \phi_2$ soft-breaking term), the two inert doublets do not mix, and as far as the gauge interactions are concerned, the model acts like a doubled I(1+1)HDM. Following \cite{Moretti:2015cwa,Barbieri:2006dq,Belyaev:2016lok} we may explicitly write the inert contributions to the oblique parameters as:
\begin{flalign}
&S = 
\frac{1}{72\pi}\frac{1}{(x_2^2-x_1^2)^3}
\left[ 
x_2^6 F_1(x_2) -x_1^6 F_1(x_1)
+ 9 x_2^2 x_1^2( x_2^2 F_2(x_2) - x_1^2 F_2(x_1)
\right]\nonumber\\
& \quad\;+ \frac{1}{72\pi}\frac{1}{(x_4^2-x_3^2)^3}
\left[ 
x_4^6 F_1(x_4) -x_3^6 F_1(x_3)
+ 9 x_4^2 x_3^2( x_4^2 F_2(x_4) - x_3^2 F_2(x_3)
\right], \label{Sparam}\\[1mm]
& F_1(x) = -5+12\log(x), \quad F_2(x)=3-4\log(x), \nonumber\\ 
& x_1=\frac{m_{H_1}}{m_{H_1^+}}, \quad  x_2=\frac{m_{A_1}}{m_{H^+_1}}, \quad  x_3=\frac{m_{H_2}}{m_{H_2^+}}, \quad  x_4=\frac{m_{A_2}}{m_{H^+_2}}, \nonumber
\end{flalign}
and
\begin{flalign}
&T = \frac{1}{32\pi^2\alpha v^2}\left(F_3(m_{H_1^{+}}^2,m_{A_1}^2) + F_3(m_{H_1^{+}}^2,m_{H_1}^2) - F_3(m_{A_1}^2,m_{H_1}^2)\right. \label{Tparam}\\
&\quad\quad\quad\quad\quad\quad\left.+F_3(m_{H_2^{+}}^2,m_{A_2}^2) + F_3(m_{H_2^{+}}^2,m_{H_2}^2) - F_3(m_{A_2}^2,m_{H_2}^2)\right),\nonumber\\
&F_3(x,y) = 
\begin{cases}
\frac{x+y}{2}-\frac{xy}{x-y}\log{\left(\frac{x}{y}\right)}, & x\neq y\\
0, & x = y.
\end{cases}\nonumber
\end{flalign}
From the form of these corrections we see that they are symmetric with respect to the change $m_{H_i} \leftrightarrow m_{A_i}$, which underlines the point that we cannot distinguish these physical states as far as their CP properties are concerned. 

\begin{figure}[h]
\includegraphics[scale=1]{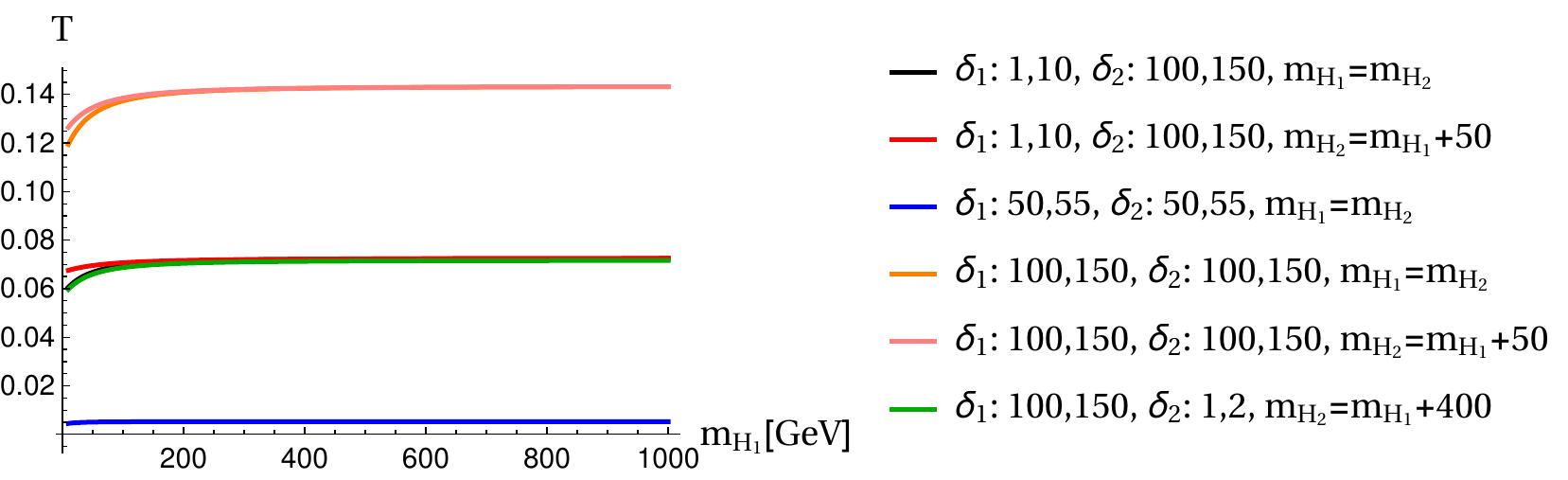}
\caption{Contribution to the $T$ parameter from the inert particles for different choices of mass splittings denoted by $\delta_i=(m_{A_i} - m_{H_i}, m_{H^\pm_i}-m_{H_i})$ [GeV]. \label{fig-T}}
\end{figure}

\begin{figure}[h]
\includegraphics[scale=1]{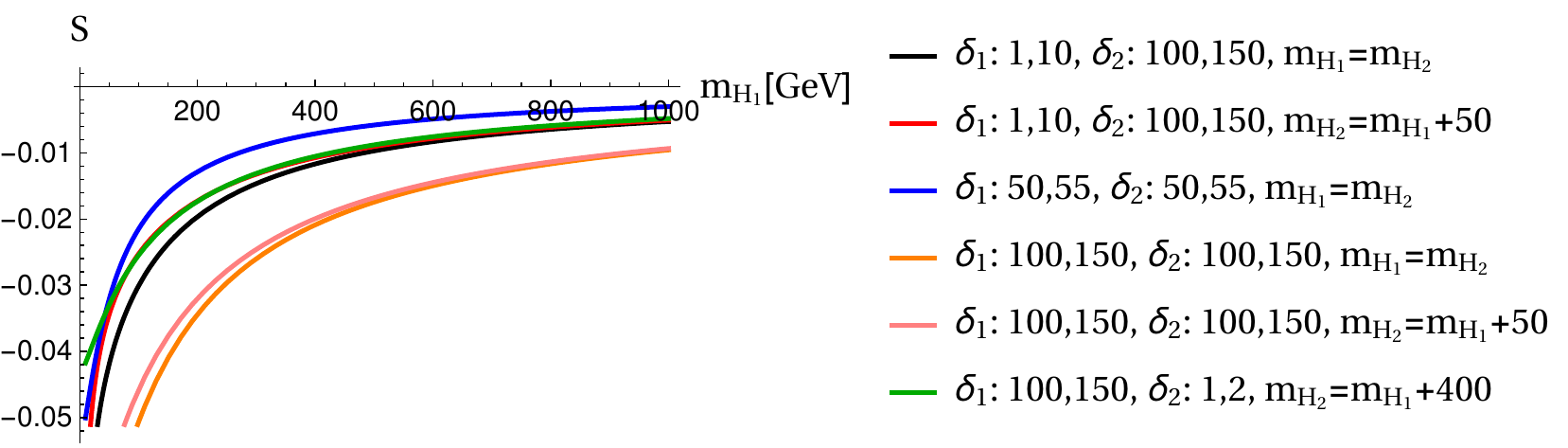}
\caption{Contribution to the $S$ parameter from the inert particles for different choices of mass splittings denoted by $\delta_i=(m_{A_i} - m_{H_i}, m_{H^\pm_i}-m_{H_i})$ [GeV]. \label{fig-S}}
\end{figure}

In general, the parameter $T$ is more sensitive to the mass splittings than the $S$ parameter, as illustrated in Figures~\ref{fig-T} and \ref{fig-S}. From Figure~\ref{fig-T} one can see also that the masses of the lightest stable particles (in this case $H_1$ and $H_2$) have a small impact on the value of the $T$ parameter and the main differences are related to the mass splitting between the heavier neutral scalar and its charged partner. On the other hand, in Figure~\ref{fig-S} we can observe the logarithmic dependence of the $S$ parameter on the values of $m_{H_1}$ and $m_{H_2}$, with the dominant effect visible for lower DM masses. 

In the I(1+1)HDM, EWPO constraints significanly limit the parameter space, effectively imposing a fixed order of masses, in which the charged inert scalar is heavier than both neutral inert scalars, and mass splitting between two heaviest particles is limited to roughly 50 GeV \cite{Belyaev:2016lok,Ilnicka:2015jba}. In the case of I(2+1)HDM, however, it does not have to be so. Notice that nothing forces the two generations of inert particles to have the same mass order, as these mass splittings are given by independent parameters: 
\be 
m_{A_1} - m_{H^\pm_1} \sim (\lambda_{31}-2\lambda_3) 
\quad \textrm{ and } \quad
m_{A_2}-m_{H_2^\pm} \sim (\lambda_{23}-2\lambda_2).
\ee
One may therefore consider a case where $m_{H_1} < m_{A_1} < m_{H^\pm_1}$ while $m_{H_2} < m_{H_2^\pm} < m_{A_2}$. 
In particular, we can expect either cancellation or amplification effects, depending on the relative signs of the contributions from two different doublets in Eqs.~(\ref{Sparam}-\ref{Tparam}).  

Assuming an SM Higgs boson mass of $m_h$ = 125 GeV, the central values of the oblique parameters are given by~\cite{Baak:2014ora}:
\be 
\hat{S} = 0.05 \pm 0.11 ,\qquad \hat{T} = 0.09 \pm 0.13, \qquad \hat{U}=0.01\pm 0.11.
\label{eq:ewpt}
\ee
with correlation coefficients $(S,T)= 0.9$, $(S,U)=-0.59$ and $(T,U)=-0.83$. As these observables are strongly correlated, it is not enough to separately test an agreement with observed values of $\hat{S},\hat{T},\hat{U}$. We study the correlation by calculating
\be
\chi^2 = \mathbf{x}^T \mathbf{C}^{-1} \mathbf{x},
\ee
where $\mathbf{x}^T = (S-\hat{S},T-\hat{T},U-\hat{U})$ and the covariance matrix is given by:
\be 
\mathbf{C} = \left(\begin{array}{ccc}
0.0121& 0.0129 & -0.0071\\
0.0129& 0.0169 & -0.0119\\
-0.0071 & -0.0119 & 0.0121
\end{array}\right)\,.
\ee
We then accept points with $\chi^2<8.025$, which corresponds to the 2$\sigma$ limit for $\chi^2$ with three degrees of freedom.

\begin{figure}[h!]
\begin{center}
\includegraphics[scale=0.9]{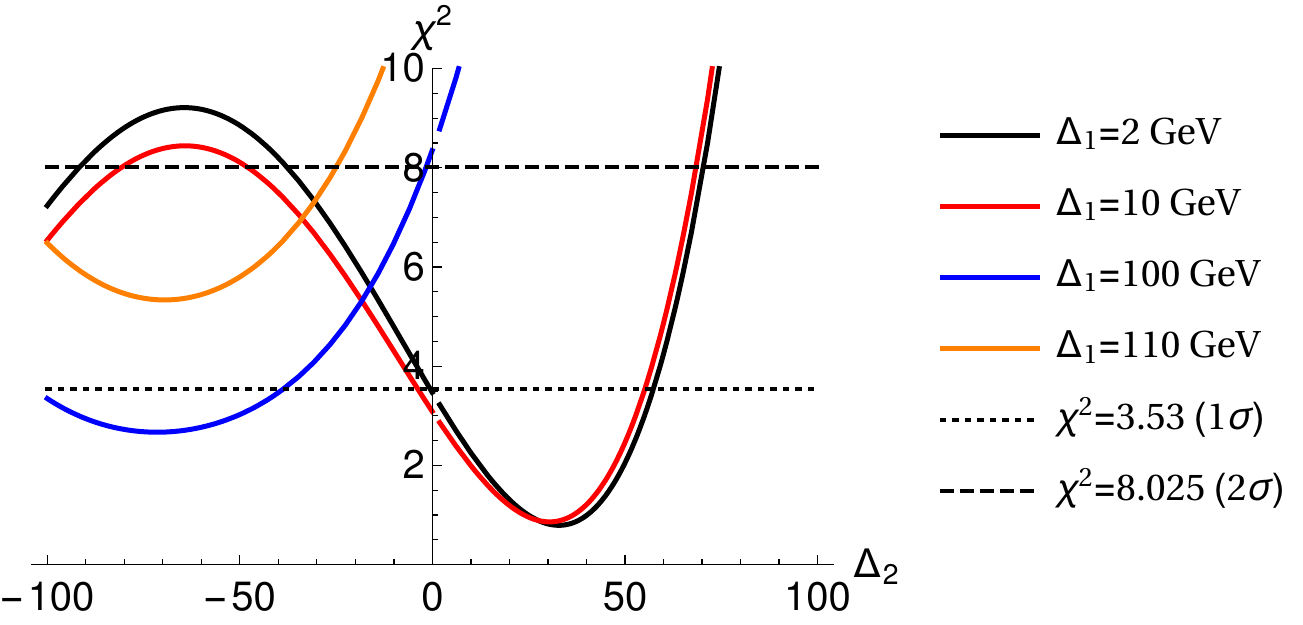}
\end{center}
\caption{Agreement with EWPO for: $m_{H_1} = 100$ GeV, $m_{A_1} = 101$ GeV, $m_{H_1^\pm} = 101 + \Delta_1$ GeV; $m_{H_2} = 100$ GeV, $m_{A_2} = 200$ GeV, $m_{H_2^\pm} = 200 + \Delta_2$ GeV. \label{fig-STUchi1}}
\end{figure}

\begin{figure}[h!]
\begin{center}
\includegraphics[scale=0.9]{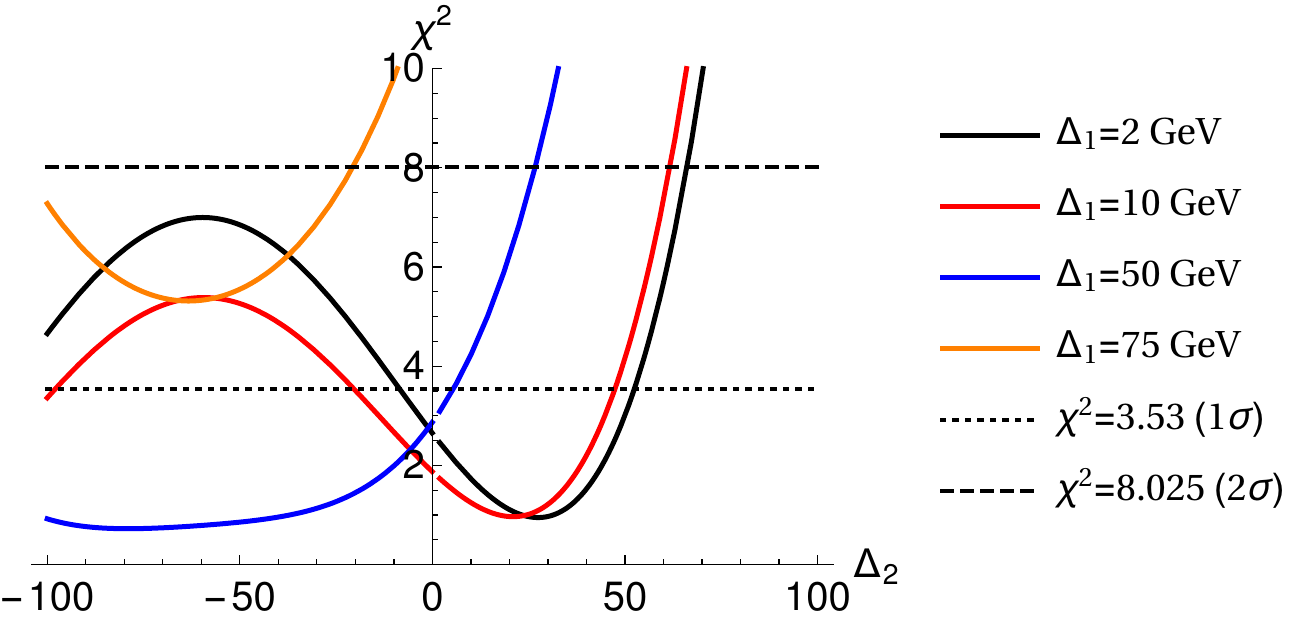}
\includegraphics[scale=0.9]{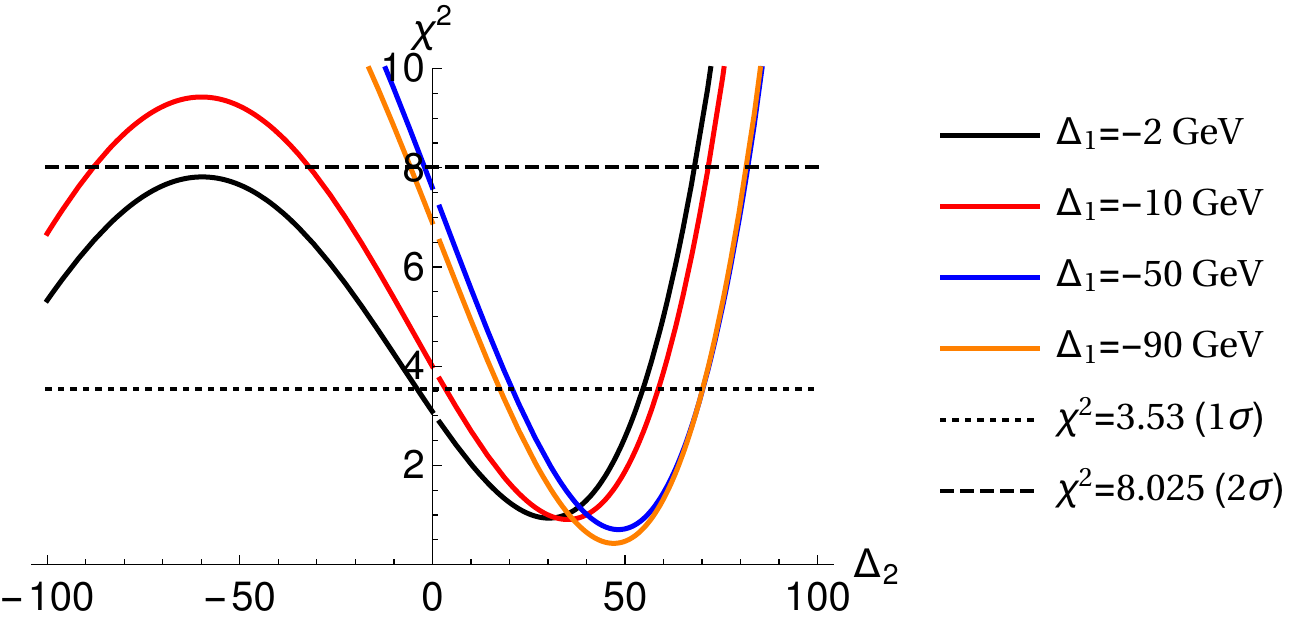}
\caption{Agreement with EWPO for: $m_{H_1} = 100$ GeV, $m_{A_1} = 200$ GeV, $m_{H_1^\pm} = 200 + \Delta_1$ GeV; $m_{H_2} = 150$ GeV, $m_{A_2} = 250$ GeV, $m_{H_2^\pm} = 250 + \Delta_2$ GeV. Upper plot: $m_{A_1} < m_{H_1^\pm}$, lower plot: $m_{A_1} > m_{H_1^\pm}$.}
 \label{fig-STUchi2}
\end{center}
\end{figure}

Figures \ref{fig-STUchi1} and \ref{fig-STUchi2} show the values of $\chi^2$ depending on the masses and mass splittings for two chosen benchmarks, where we define 
\be
\Delta_1 = m_{H_1^\pm} - m_{A_1}, \qquad 
\Delta_2 = m_{H_2^\pm} - m_{A_2}\,.
\ee
In Figure \ref{fig-STUchi1}, in the first family of inert particles, the two neutral scalars are almost degenerate in mass, $m_{H_1} \simeq m_{A_1}$. In the I(1+1)HDM this equality would automatically mean that the charged particle has to be relatively close in mass with its neutral partners, i.e. $\Delta_1$ would be small. 
However, this is not the case in the considered model. Due to the cancellation between contributions from the two separate dark families, we can expect significant (up to 100 GeV) difference between $m_{A_1}$ and $m_{H^\pm_1}$. Furthermore, the order of masses in the second generation of inert particles in such a case will be flipped, i.e. $m_{A_2} > m_{H^\pm_2}$.

The different mass orders can be clearly seen in Figure \ref{fig-STUchi2}. In the upper plot we consider cases where  $m_{H_1} < m_{A_1} < m_{H^\pm_1}$, i.e. $\Delta_1$ is positive. Depending on the value of $\Delta_1$ we see different cases of available $\Delta_2$, both positive and negative. Negative $\Delta_2$ (i.e. the flipped order) is needed if $\Delta_1$ is relatively large. In the lower plot, we fix the mass order to be $m_{H_1}  < m_{H^\pm_1} < m_{A_1} \Rightarrow \Delta_1 < 0$. This results in strong constraints on the allowed value of $\Delta_2$ to ensure proper cancellations.

In short, while in the I(1+1)HDM the preferred order of masses is for both neutral particles to be lighter than their charged partner, this is definitely not the case in the presence of a second inert doublet where we can expect to have different solutions, with different mass order and significantly larger mass splittings between neutral and charged dark particles.

%
%
%
%\begin{Figure}
%\includegraphics[scale=1]{STU3.pdf}
%\caption{Contribution to $T$ from two doublets, $\delta_i=(m_{A_i} - m_{H_i}, m_{H^\pm_i}-m_{H_i})$}
%\end{figure}
%%
%\begin{Figure}
%\includegraphics[scale=1]{S3.pdf}
%\caption{Contribution to $S$ from two doublets, $\delta_i=(m_{A_i} - m_{H_i}, m_{H^\pm_i}-m_{H_i})$}
%\end{figure}

\paragraph{Collider searches for new physics} Presence of additional scalars, especially if they are sufficiently lights, can influence properties of SM particles, e.g. their decay channels and widths.
We forbid decays of EW gauge bosons into new scalars by enforcing:
\be 
\label{eq:gwgz}
m_{H_i}+m_{H^\pm_i}\,\geq\,m_W^\pm,~~ m_{A_i}+m_{H^\pm_i}\,\geq\,m_W^\pm,~~
\,m_{H_i}+m_{A_i}\,\geq\,m_Z,\,~~
2\,m_{H_i^\pm}\,\geq\,m_Z.
\ee
Furthermore, we adopt LEP 2 searches for supersymmetric particles re-interpreted for the I(1+1)HDM in order to exclude the region of masses where the following conditions are simultaneously satisfied \cite{Lundstrom:2008ai} ($i=1,2$):
\be 
\label{eq:leprec}
m_{A_i}\,\leq\,100\,\GeV,\,~~
m_{H_i}\,\leq\,80\,\GeV,\,\, ~~
\Delta m {(A_i,H_i)}\,\geq\,8\,\GeV,
\ee
since this would lead to a visible di-jet or di-lepton signal.

The model also has to agree with null-results for additional neutral scalar searches at the LHC. 
As it is the case for other non-supersymmetric multi-scalar models, current searches at the LHC for multi-lepton final states with missing transverse energy are in general not sensitive enough. 
This is mainly due to a relatively large cut on missing transverse energy used in the experimental analysis. 
As it corresponds to a rather large mass splittings between scalars in the dark sector, the production cross-sections are significantly reduced and fall below current experimental sensitivity. 
On the other hand, for benchmarks with smaller mass splittings between scalars, the production cross-section is large enough for the scalars to be produced in abundance even at the current stage of the LHC; however, they require smaller cuts on missing energy to be detected. 

\paragraph{Charged scalar mass and lifetime}

We take a conservative model-independent lower estimate on the masses of charged scalars: $m_{H^\pm_i} > 70$ GeV ($i=1,2$) \cite{Pierce:2007ut}. Furthermore, in this work we will not consider scenarios with possibly long-lived charged particles, and following \cite{Heisig:2018kfq} we set the limit for the charged scalar lifetime to be $\tau\,\leq\,10^{-7}\,{\rm s}$ .
Notice that as new charged particles are inert and hence do not couple to fermions, they are not subject to many constraints that are present in the 2HDM framework. In particular, flavour bounds on the charged scalar mass, e.g. from $b\to s \gamma$, are not applicable here.

\paragraph{Higgs mass and signal strengths} 

The combined ATLAS and CMS Run-1 result for the Higgs mass
is~\cite{Aad:2015zhl}:
\be
m_h = 125.09\pm 0.21 \textrm{ (stat.)} \pm 0.11 \textrm{ (syst.)} \; \GeV.
\ee
The Higgs particle detected at the LHC is  in very good agreement with the SM predictions. 
By construction, $h$ in \texttt{2-Inert} vacuum in Eq.~(\ref{vac-inert}) is SM-like and its couplings to gluons, massive gauge bosons and fermions are identical to the SM values. However, inert scalars can significantly alter a number of Higgs observables:

\begin{enumerate}
\item 
The Higgs total width can be modified through additional decays into light inert scalars $S$ by contributing to the $h\to SS$ decay channel when $m_S \leq m_h/2$, and through changes to decay channels already present in the SM, in particular the $h \to \gamma \gamma$ decay. In this analysis we take the upper limit on the Higgs total decay width to be \cite{Sirunyan:2019twz}:
\be
\Gamma_{tot} \leq 9.1 \; \textrm{MeV}.
\ee
The latest constraints on the Higgs invisible decays from CMS and ATLAS are \cite{Sirunyan:2018owy, Aaboud:2019rtt}:
\be
\textrm{BR}(h \to \textrm{ inv.}) < 0.19 \; (\textrm{CMS}), ~~ 0.26 \; (\textrm{ATLAS}).
\ee
As it is common in models that exhibit Higgs portal-type of interaction, this constraint has a significant impact on the allowed values of Higgs-inert couplings for inert scalars which are lighter than $m_h/2$. 

\item 
The partial decay width $\Gamma(h\to \gamma\gamma)$ can receive significant contributions from the two charged inert scalars, and these corrections are formally of the same order as the SM process. In this work, we use the combined  ATLAS and CMS Run I limit for the signal strength \cite{Khachatryan:2016vau}:
\be
\mu_{\gamma \gamma} = 1.14^{+0.19}_{-0.018},
\ee
and ensure 2$\sigma$ agreement with the observation.
\end{enumerate}

\paragraph{DM constraints}
The total relic density is given by the sum of the contributions from both DM candidates $H_1$ and $H_2$
\be 
\label{planck-relic}
\Omega_{T}h^2 = \Omega_{{H_1}}h^2 + \Omega_{{H_2}}h^2 \, ,
\ee
and is constrained by Planck data \cite{Aghanim:2018eyx} to be:
\be
\Omega_{\text{\rm DM}}h^2 = 0.1200 \pm 0.0012.
\label{PLANCK_lim}
\ee
The current strongest upper limit on the spin independent (SI) scattering cross-section of DM particles on nuclei 
$\sigma_{DM-N}$ is provided by the XENON1T experiment relevant for all regions of DM mass \cite{Aprile:2018dbl}.

Regarding indirect detection searches, for light DM annihilating into $bb$ or $\tau\tau$, the strongest constraints come from the Fermi-LAT satellite, ruling out the canonical cross-section 
$\langle \sigma v\rangle \approx 3\times 10^{-26}~{\rm cm}^3/{\rm s}$ for $m_{\rm DM} \lesssim 100 \mbox{ GeV}$ 
\cite{Ackermann:2015zua}.
For heavier DM candidates the PAMELA and Fermi-LAT experiments provide similar limits of $ \langle \sigma v\rangle \approx 10^{-25}~{\rm cm}^3/{\rm s}$ for  $m_{\rm DM}=200 \mbox{ GeV} $ in the $bb,\tau\tau$ or $WW$ channels \cite{Cirelli:2013hv}. 
%HESS measurements of signal coming from the Galactic Centre set limits of $ \langle \sigma v\rangle \approx 10^{-25}-10^{-24}~{\rm cm}^3/{\rm s}$ for masses up to TeV scale \cite{Abramowski:2011hc}. 

\section{DM in $Z_2 \times Z'_2$ symmetric I(2+1)HDM \label{sec-DM}}

In our analysis, we use the \texttt{micrOMEGAs} package~\cite{Belanger:2006is} to calculate the relic density of two DM candidates. During the analysis we follow the standard assumptions included in the code, namely: 1) particles within the dark sector are in thermal equilibrium, 2) they have the same kinetic temperature as those of the SM, 3) the number densities  of DM particles can differ from the equilibrium values  once their number density multiplied by their annihilation cross-section becomes too small to compete with the expansion rate of the Universe.  

The lightest particle in each family is a viable DM candidate. As discussed previously, without loss of generality, we can take $H_1$ and $H_2$ from each family to be the respective DM candidates. However, other dark particles from both families have a significant impact on the final relic abundances of these two stable particles, as they influence the thermal evolution and decoupling rate of DM particles.

In what follows, $x_{a}$ denotes any relevant dark particle for a particular process from a respective dark sector. There are two\footnote{Due to imposed symmetry there are no processes that would be classified as semi-annihilation, i.e. processes of the form $x_a x_b \to \textrm{SM } x_d$. } distinct classes of processes that can influence number densities of dark particles in each sector. 
In the first class there are (co)annihilation processes in the form of:
\be
x_a x_a \to \textrm{ SM SM}.
\ee
In this class of processes we first include a standard DM annihilation, which in our case corresponds to $H_i H_i \to  \textrm{ SM SM}$ with  $i=1,2$. 
The product of this annihilation depends mostly on the mass of the DM particle: we observe mostly Higgs-mediated annihilation into fermions in case of relatively light DM, while heavier DM particles annihilate predominantly into gauge boson pairs, either directly or through Higgs $s$-channel. 
In our calculations, we also include annihilation into virtual gauge bosons, as these processes have a significant impact on DM annihilation rate for medium DM masses. 
Furthermore, if the mass difference between the DM candidate and other neutral or charged inert scalars from the same generation is small, then coannihilation channels, e.g. $H_i A_i \to Z \to \textrm{SM SM}$ play an essential role. 
This is, in fact, the dominant class of processes if DM is relatively heavy, exactly as it happens in the I(1+1)HDM. We would like to stress that due to the $Z_2 \times Z'_2$ imposed symmetry in this class of processes, two DM sectors are separated. 
There are no vertices that involve fields from two separate families, e.g. Higgs or gauge bosons couple only to a pair of particles from the same generation (see Feynman rules in Appendix \ref{ap-vertices} for details). 

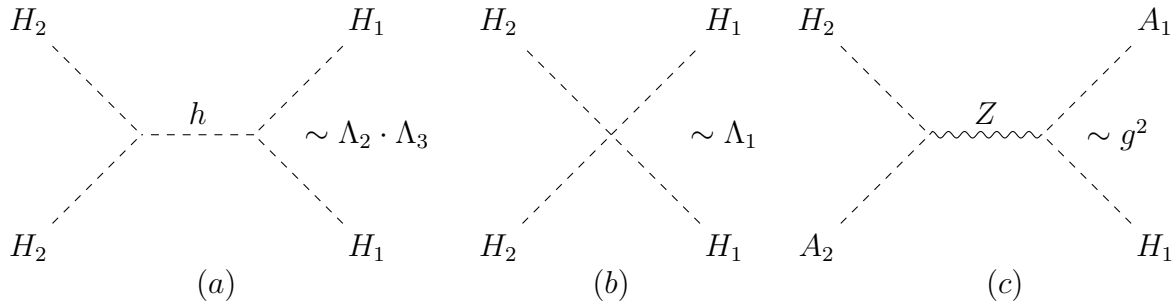
\begin{figure}[h!]
\begin{center}
\begin{tikzpicture}
    \begin{feynman}
        \vertex (m) at ( 0, 0);
        \vertex (n) at (1.5,0);
        \vertex (a) at (-1.5,-1.5) {\(H_2\)};
        \vertex (b) at ( 3,-1.5) {\(H_1\)};
        \vertex (c) at (-1.5, 1.5) {\(H_2\)};
        \vertex (d) at ( 3, 1.5) {\(H_1\)};
        \vertex (e) at ( 1, -2) {\((a)\)};  
		\vertex (f) at ( 3, 0) {\(\sim \Lambda_2 \cdot \Lambda_3\)};  
        \diagram* {
            (a) -- [scalar] (m),
            (b) -- [scalar] (n),
            (c) -- [scalar] (m),
            (d) -- [scalar] (n),
			(n) -- [scalar, edge label'=\(h\)] (m),
        };
    \end{feynman}
\end{tikzpicture} \;
\begin{tikzpicture}
    \begin{feynman}
        \vertex (m) at ( 0, 0);
        \vertex (a) at (-1.5,-1.5) {\(H_2\)};
        \vertex (b) at ( 1.5,-1.5) {\(H_1\)};
        \vertex (c) at (-1.5, 1.5) {\(H_2\)};
        \vertex (d) at ( 1.5, 1.5) {\(H_1\)};
        \vertex (e) at ( 0, -2) {\((b)\)};        
		\vertex (f) at ( 1.5, 0) {\(\sim \Lambda_1\)};  
        \diagram* {
            (a) -- [scalar] (m) -- [scalar] (c),
            (b) -- [scalar] (m) -- [scalar] (d),
        };
    \end{feynman}
\end{tikzpicture}\;
\begin{tikzpicture}
    \begin{feynman}
        \vertex (m) at ( 0, 0);
        \vertex (n) at (1.5,0);
        \vertex (a) at (-1.5,-1.5) {\(A_2\)};
        \vertex (b) at ( 3,-1.5) {\(H_1\)};
        \vertex (c) at (-1.5, 1.5) {\(H_2\)};
        \vertex (d) at ( 3, 1.5) {\(A_1\)};
        \vertex (e) at ( 1, -2) {\((c)\)};  
		\vertex (f) at ( 2.5, 0) {\(\sim g^2\)};  
        \diagram* {
            (a) -- [scalar] (m),
            (b) -- [scalar] (n),
            (c) -- [scalar] (m),
            (d) -- [scalar] (n),
			(n) -- [boson, edge label'=\(Z\)] (m),           };
    \end{feynman}
\end{tikzpicture} 
\end{center}
\caption{Example of DM conversion diagrams: (a) Higgs-mediated conversion of $H_2 H_2 \to H_1 H_1$, present always as long as $\Lambda_{2,3} \neq 0$, (b) direct DM conversion depending on the self-interaction parameter $\Lambda_1$, (c) $Z$-mediated conversion due to coannihilation channels.}
 \label{fig-conversion}
\end{figure}

The second class of processes is DM conversion. In this case, a pair of heavier scalars from one generation converts, either directly or through interaction with an SM particle, into a pair of dark particles from the other generation:
\be
x_a x_a \to x_b x_b \, .
\ee
Note that the conversion between two generations of DM particles occurs even if all self-interaction couplings are switched off. In Figure \ref{fig-conversion} diagrams (a) and (b) represent the same initial/final state in the process of a pair of $H_2$ particles converting to a pair of $H_1$ through either Higgs-mediated or direct annihilation. 
Even if $\Lambda_1$ were set to zero, there would still be a non-zero contribution coming from the diagram (a), as long as both particles couple to the Higgs boson ($\Lambda_{2,3}\neq 0$). 
We expect cancellations or enhancement depending on the relative sign of $\Lambda_2 \Lambda_3$ and $\Lambda_1$, in a similar vein to the effect observed in the standard annihilation into gauge boson pair. 
Furthermore, depending on the masses and couplings, we also need to take into account annihilation of heavier dark particles from the second generation directly into stable particles from the first generation, e.g. $A_2 A_2 \to h \to H_1 H_1$. All these processes are automatically included in our numerical analysis.

As discussed in Section \ref{sec-inertvac} in Eq.~(\ref{physpar}) we can parametrise the model by using masses of scalar particles and their couplings, namely:
\begin{enumerate}
\item 
masses of inert particles, $m^2_{H_1}, m^2_{H_2}, m^2_{A_1}, m^2_{A_2}, m^2_{H^\pm_1}, m^2_{H^\pm_2}$ dictate the annihilation patterns of DM particles. Depending on the absolute values of masses, but also on mass splittings between particles, we can expect different dominant channels of annihilation, coannihilation and conversion.

\item 
couplings of DM particles to the Higgs boson, $\Lambda_2$ and $\Lambda_3$, govern not only DM annihilation and DM conversion but also influence possible invisible decays of the Higgs particle, as well as direct and indirect detection of DM. 
In our numerical analysis we find that in particular the following vertices have significant impact on DM phenomenology:
\bea
g_{hH_1H_1} &=& 2\lambda_3 +\lambda_{31} + \lambda'_{31} = 2 \Lambda_3 \,, 
\\
g_{hH_2H_2} &=& 2\lambda_2 +\lambda_{23} + \lambda'_{23} = 2 \Lambda_2 \,, \\
g_{hA_1A_1} &=& -2\lambda_3 +\lambda_{31} + \lambda'_{31} = 2\Lambda_3 + 2(m^2_{A_1} - m^2_{H_1})/v^2  \,,  
\label{Eq:cubic-couplings}
\\
g_{hA_2A_2} &=&  -2\lambda_2 +\lambda_{23} + \lambda'_{23} = 2\Lambda_2 + 2(m^2_{A_2} - m^2_{H_2})/v^2 \,.
\eea

\item 
DM self-couplings, $\lambda_1, \lambda'_{12}, \lambda_{12}$ and $\lambda_{11}, \lambda_{22}$, play two different roles. The first set governs interactions between two different families, and will have observable impact on DM relic abundance through DM conversion processes. In particular, the following couplings have a crucial impact on DM phenomenology:
\bea
g_{H_1H_1H_2H_2} &=&  \phantom{-} 2\lambda_1 +\lambda_{12} + \lambda'_{12} =  4\Lambda_1 -(\lambda_{12} + \lambda'_{12}) \,,
\\
g_{A_1A_1H_2H_2}  = g_{A_2A_2H_1H_1} &=&  -4\lambda_1 +\lambda_{12} + \lambda'_{12} =-4\Lambda_1 + 2(\lambda_{12} + \lambda'_{12}) \,.
\eea
In turn, $\lambda_{11}$ and $\lambda_{22}$ do not directly contribute to any observable process and do not influence the DM abundance. However, as we discussed in Section \ref{sec-constraints} they have a fundamental impact on the range of other parameters through vacuum stability conditions.
\end{enumerate}

\begin{figure}[h!]
\begin{center}
\includegraphics[width=85mm]{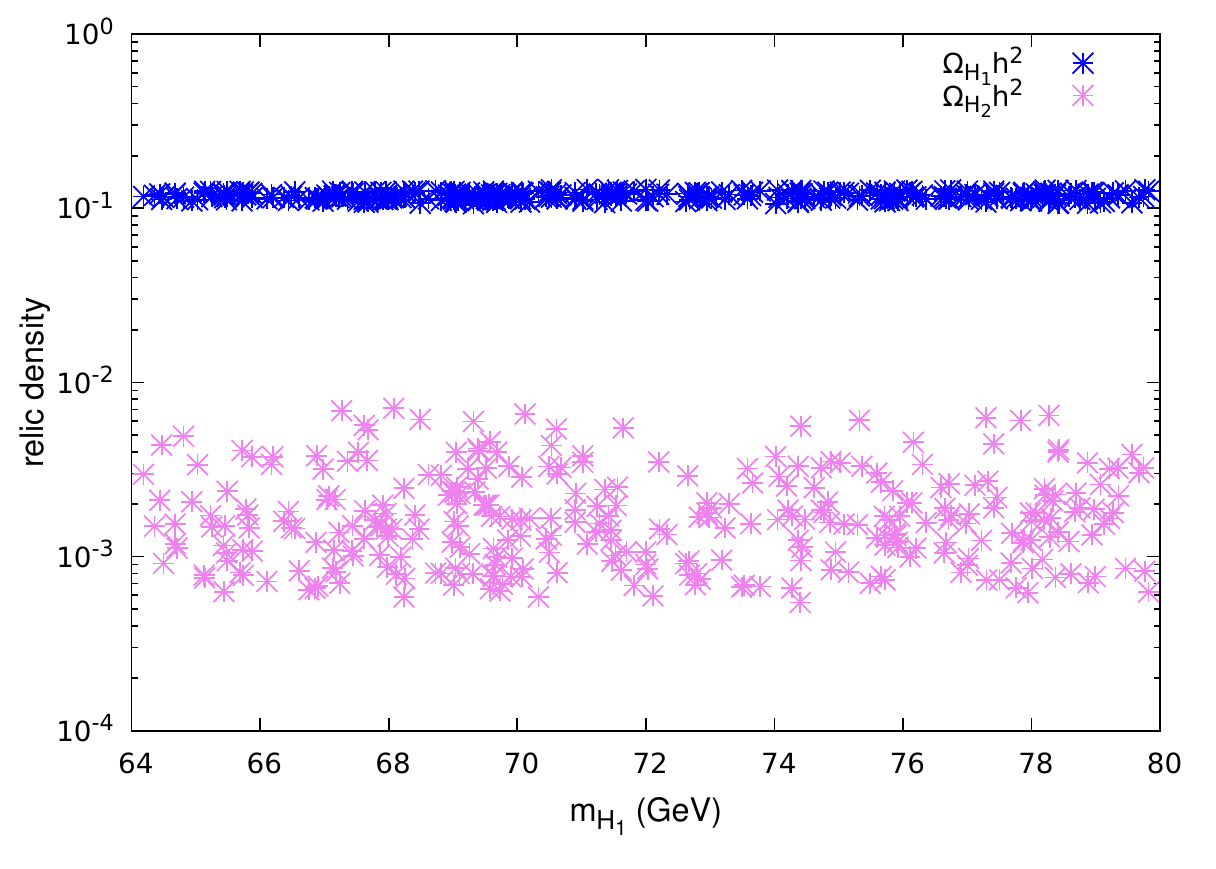}
\vspace{-5mm}
\caption{Relic density contribution of the light, $H_1$, and heavy, $H_2$, DM candidates. The total relic density is the sum of the two contributions $\Omega_{T}h^2=\Omega_{H_1}h^2+\Omega_{H_2}h^2$.}
\label{Fig:Relic}
\end{center}
\end{figure}

Our numerical analysis shows that due to the existence of the conversion processes, the total DM relic density receives its dominant contribution from $H_1$, while the contribution from $H_2$ is of a few per cent of the total DM relic density as illustrated in Figure \ref{Fig:Relic}.

\subsection{The anatomy of the parameter space}

\begin{figure}[h!]
\centering
\includegraphics[width=85mm]{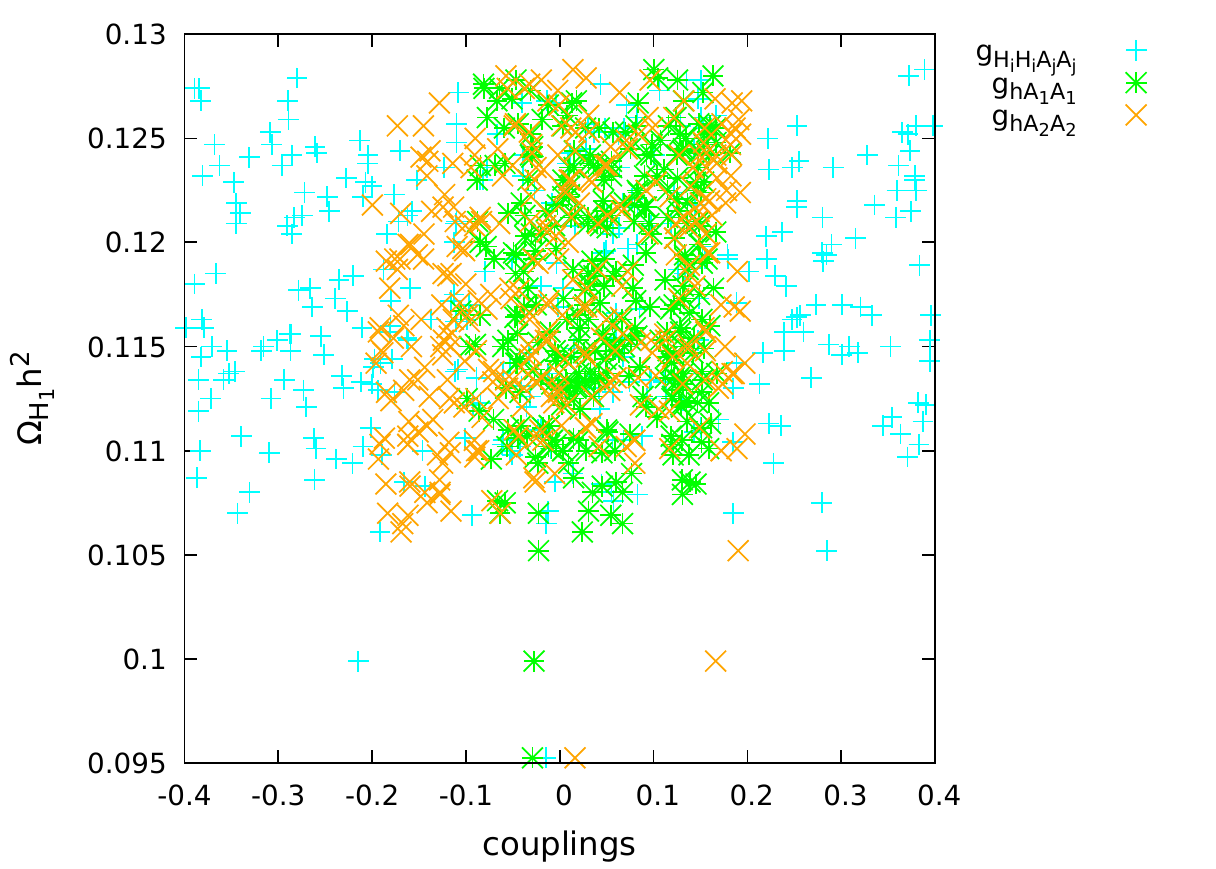}~
\includegraphics[width=85mm]{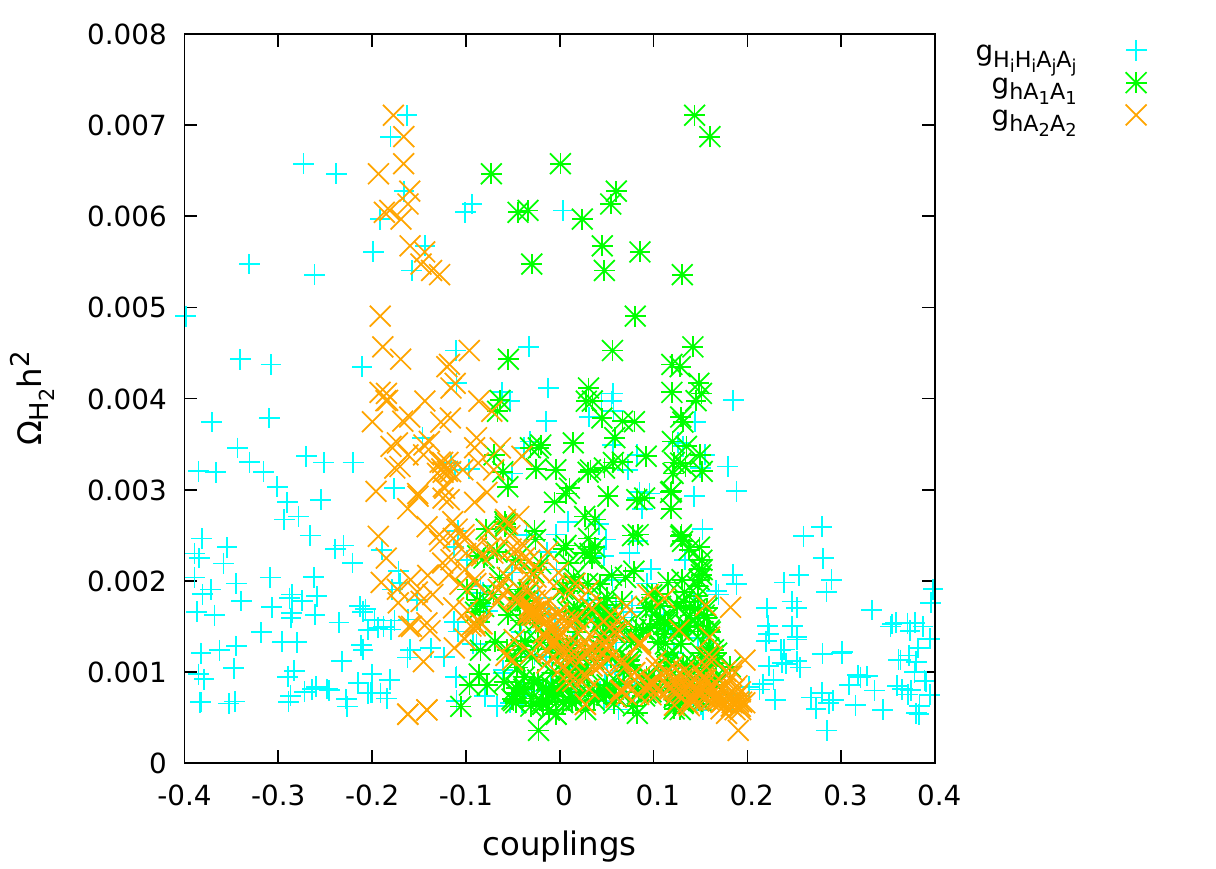}\\
\includegraphics[width=80mm]{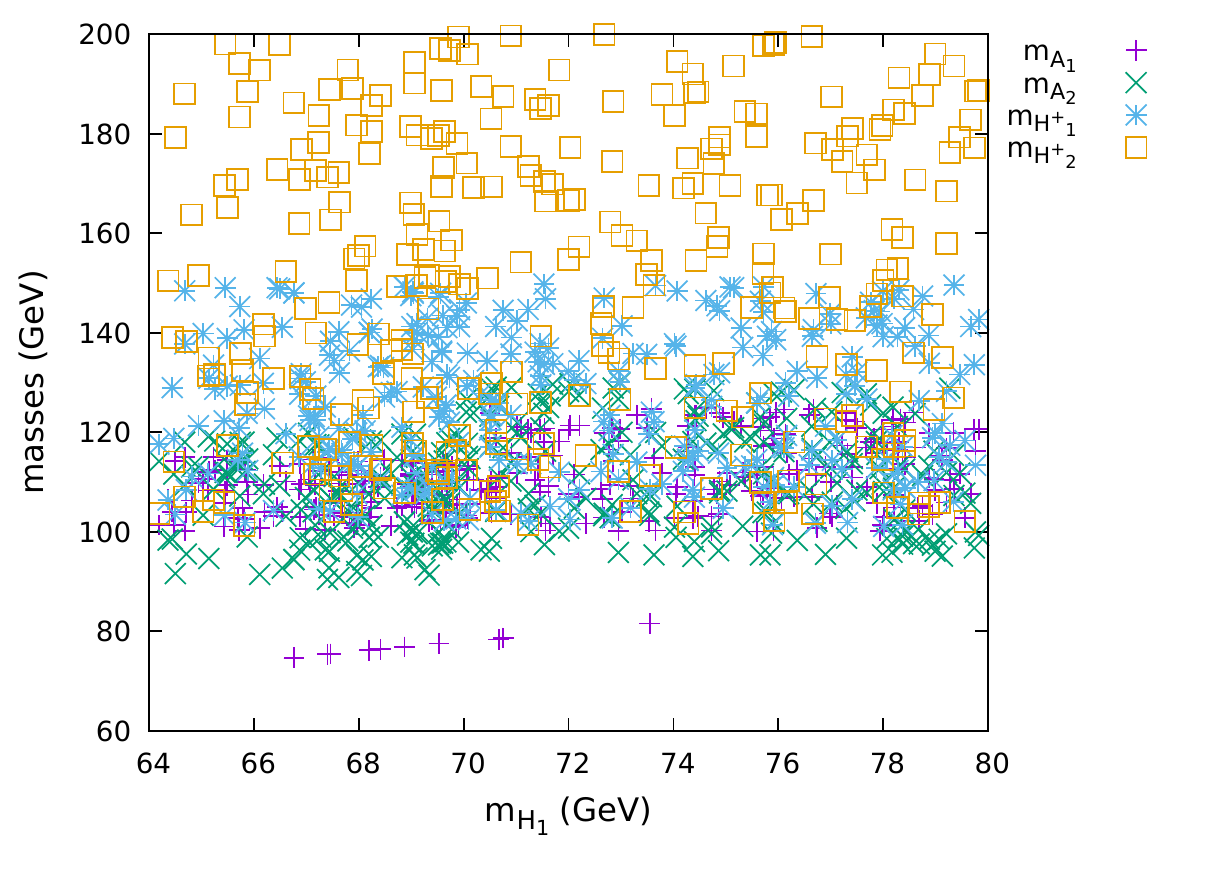}~
\includegraphics[width=80mm]{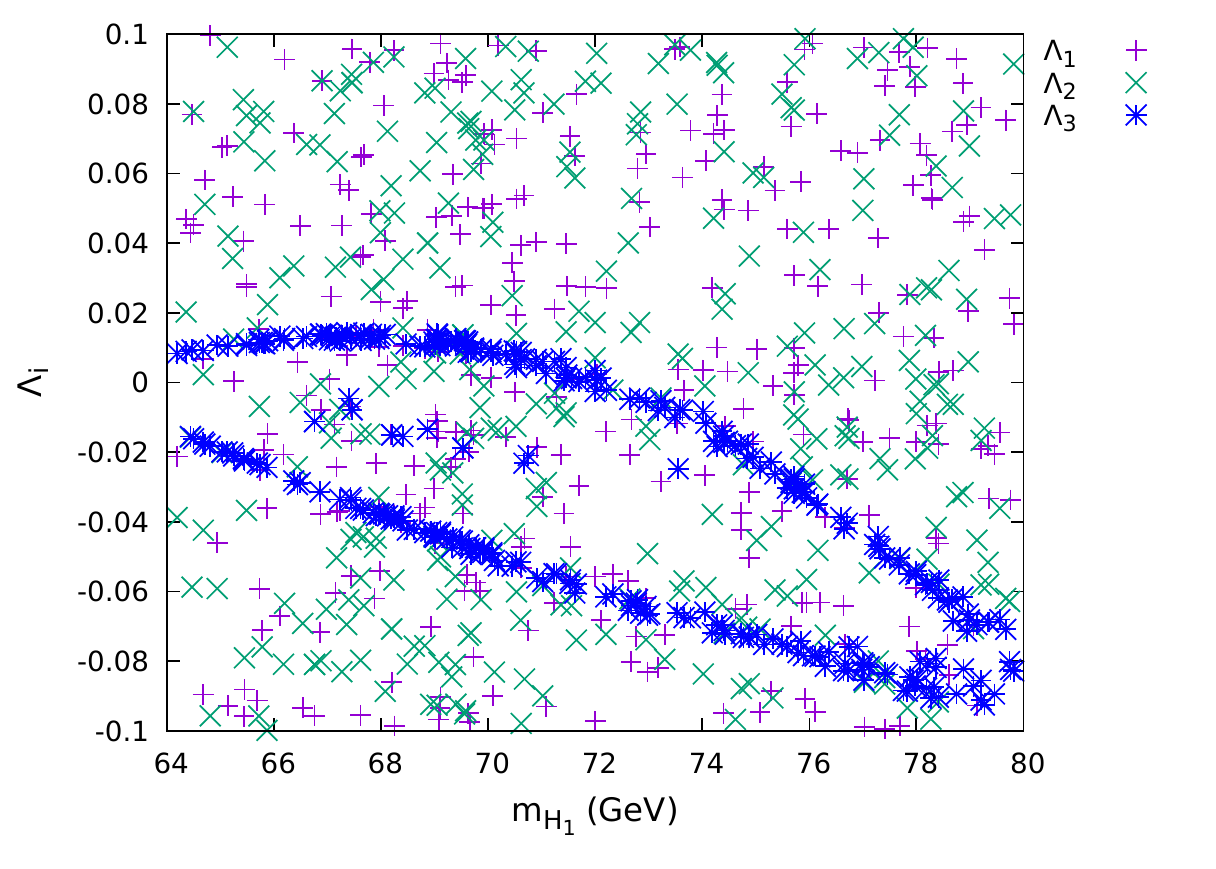}
\caption{The region of the parameters space relevant for our analysis, $ \frac{1}{2}m_h  < m_{H_1} <80$ GeV and $m_{H_2} \simeq 100$ GeV. The masses of other inert particles and couplings vary within the allowed ranges, with $\Omega_{H_1} h^2+ \Omega_{H_2} h^2= \Omega_{T} h^2$ within $3 \sigma$ deviation of the Planck experiment.}
\label{Fig:masses-couplings}
\end{figure}

In order to present the distinct and complementary signatures of the light and heavy components of DM in the direct and indirect detection experiments, it is instructive to divide the parameter space in the following way. As discussed before, the contribution of $H_2$ to the total DM relic density is negligible. Therefore, the characteristics of each benchmark scenario are dictated by the $H_1$ behaviour.
\begin{itemize}

\item 
{\sl The $46 \GeV < m_{H_1} < \frac{1}{2}m_h$ and $m_{A_1} <100$ GeV region}\\[1mm]
In this regime, the 
reinterpreted SUSY bounds in Eq.~(\ref{eq:leprec}) require
$m_{A_1} - m_{H_1} < 8$ GeV which means that the decay channel $h \to H_1H_1$ and possibly $h \to A_1A_1$ are kinematically allowed. Therefore, not just the Higgs-DM coupling $g_{hH_1H_1}$ but also the $g_{hA_1A_1}$ coupling has to be very small $\mathcal{O}(10^{-3})$ which also reinforces the need for a small $H_1$-$A_1$ mass splitting since $g_{hA_1A_1}$ depends on $m_{A_1}^2 - m_{H_1}^2$ as in Eq.~(\ref{Eq:cubic-couplings}).
As a result, the efficiency of the $H_1$-$A_1$ coannihilation is increased, leading to an under-production of the total DM relic density. 
Moreover, the small $g_{hH_1H_1}$ coupling, as discussed in detail in section \ref{sec-num}, leads to a small event rate in the recoil energy of the Xenon nucleus well below the sensitivity of the experiment. This region of the parameter space is therefore of little interest to us.

\item 
{\sl The $46 \GeV < m_{H_1} < \frac{1}{2}m_h$ and $m_{A_1} >100$ GeV region}\\[1mm]
In this regime, the reinterpreted SUSY bounds do not apply and the large $H_1$-$A_1$ mass splitting forbids the coannihilation leading to an over-production of the DM.
Moreover, the $h \to H_1H_1$ decay is kinematically allowed which severely constrains the $g_{hH_1H_1}$ coupling to very small values. As a result, not only the inefficient annihilation of $H_1$ reinforces the over-production of DM, but also the smallness of $g_{hH_1H_1}$ suppresses the event rate of nucleon recoil energy to undetectable values. Thus, we shall not focus on this region of the parameter space.

\item 
{\sl The $ \frac{1}{2}m_h \lesssim m_{H_1} <80$ GeV and $m_{A_1} <100$ GeV region}\\[1mm]
In this mass range, the reinterpreted SUSY bounds still apply and require $m_{A_1} - m_{H_1} < 8$ GeV. However, since the $h \to H_1 H_1$ channel is closed, in general, larger values of $\Lambda_3$ are allowed.
For $m_{H_1} \sim \frac{1}{2}m_h$, the annihilation of $H_1$ through Higgs is resonant leading to a very small $g_{hH_1H_1}$ coupling. 
As the DM mass grows, the point annihilation to gauge bosons, $H_1H_1 \to V V $ with $V=W^\pm,Z$, becomes the dominant process. The destructive interference with the Higgs mediated process $H_1H_1 \to h \to V V $ is needed to ensure the proper relic density which occurs for negative values of the $g_{hH_1H_1}$ coupling.
For larger DM masses, the point annihilation is more efficient, requiring larger values of $g_{hH_1H_1}$. This behaviour is manifest in the top right panel of Figure \ref{Fig:masses-couplings}.

\item 
{\sl The $ \frac{1}{2}m_h \lesssim  m_{H_1} <80$ GeV and $m_{A_1} >100$ GeV region}\\[1mm]
In this mass range, the reinterpreted SUSY bounds are not applicable, so any $H_1$-$A_1$ mass splitting is allowed. However, the behaviour of the model is almost identical to the previous case, since
the same dominant processes dictate the behaviour of the model.

\item
{\sl The $m_{H_1} >80$ GeV region}\\[1mm]
In this regime, 
any mass splitting between $A_1$ and $H_1$, and large values of $\Lambda_3$ are allowed. However, the light DM, $H_1$, is too heavy to have an observable effect in the nuclear recoil energy event rates in direct detection searches. Therefore, we do not discuss this region of the parameter space any further.
\end{itemize}

In Figure~\ref{Fig:masses-couplings}, we show the region of the parameters space relevant for our analysis, $ \frac{1}{2}m_h  < m_{H_1} <80$ GeV and $m_{H_2} \simeq 100$ GeV. We allow for the masses of other inert particles and couplings to vary within the allowed ranges discussed in section \ref{sec-constraints} while fixing the total relic abundance, $\Omega_{H_1} h^2+ \Omega_{H_2} h^2= \Omega_{T} h^2$, within $3 \sigma$ deviation of the Planck experiment observation in Eq.~(\ref{PLANCK_lim}).

For the numerical analysis in the following section we make use of the \texttt{MicrOMEGAs} routines \texttt{nuclearRecoil} \cite{Belanger:2013oya,Belanger:2014vza} and \texttt{PhotonFlux} \cite{Belanger:2010gh}. 

\subsection{Complementarity of the direct and indirect probes \label{sec-num}}

\begin{figure}[h!]
\centering
\includegraphics[width=130mm]{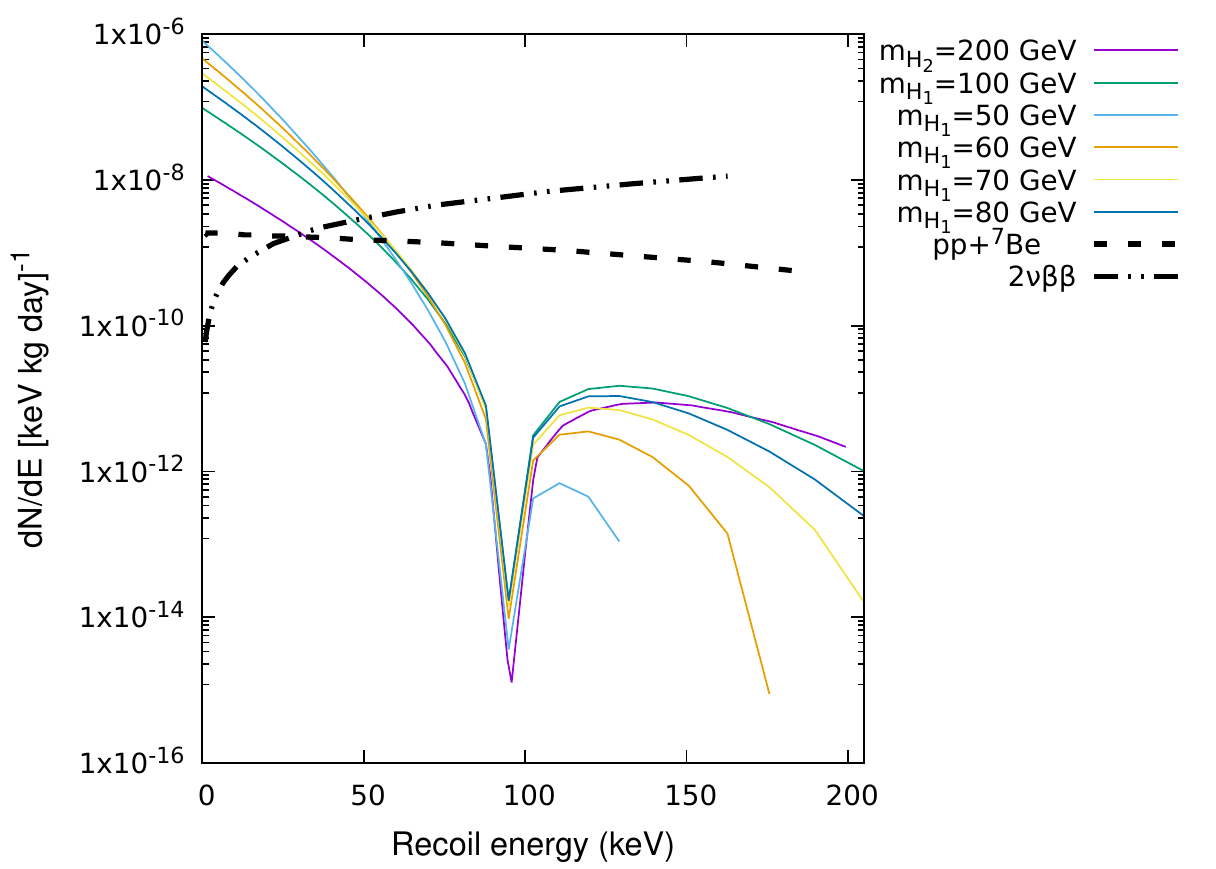}
\vspace{-3mm}
\caption{The complementary signals of our two-component DM model; the Figure show the Xe nuclear recoil energy event rate from the two DM components where the light DM, $H_1$, has the highest event rate in the low keV region within reach of XENONnT/LZ or DARWIN.}
\label{Fig:general-behaviour-DD}
\end{figure}

As discussed before, in our benchmark scenarios that follow, the light DM candidate, $H_1$, is lighter than 80 GeV while the second DM candidate, $H_2$, has a mass above 100 GeV.
We show that $H_1$ can be probed at future direct detection experiments such as XENONnT/LZ and DARWIN,  in the nuclear recoil energy event rate\footnote{
The electron recoil energy (momentum transfer) for a generic WIMP candidate is very small \cite{Belanger:2008sj} ($v_{DM}=0.0001 \, c$)
\be 
\sqrt{q^2}= 2 \, v_{DM} \frac{m_e m_{DM}}{m_e + m_{DM}} \simeq m_{e} \times 10^{-3} \,, \nonumber
\ee
and hard to access experimentaly.}.
The sensitivity of the XENONnT experiment in the nuclear recoil energy event rate is as low as $10^{-7}$ for the Xenon nucleon \cite{Aprile:2015uzo}. 
As shown in Figure \ref{Fig:general-behaviour-DD}, for fixed values of the couplings, the nuclear recoil event rate is higher for lower values of $m_{H_1}$ and decreases with increasing $H_1$ mass. The heavier DM candidate $H_2$, consequently, has a sub-dominant event rate which is well below the sensitivity of the experiment. 

In Figure \ref{Fig:general-behaviour-DD}, we also show the background \cite{Aalbers:2016jon,Baudis:2013qla}, which for liquid Xe experiments includes the electron recoil spectrum from the double beta decay of $^{136}$Xe $(2\nu\beta\beta)$, and the summed differential energy for $pp$ and $^7$Be neutrinos, $(pp + ^7$Be), undergoing neutrino-electron scattering. 
Any signal below these lines is in the so-called \emph{neutrino floor} region which is very challenging for any present or near future experiment to probe.

\begin{figure}[h!]
\begin{center}
\includegraphics[width=80mm]{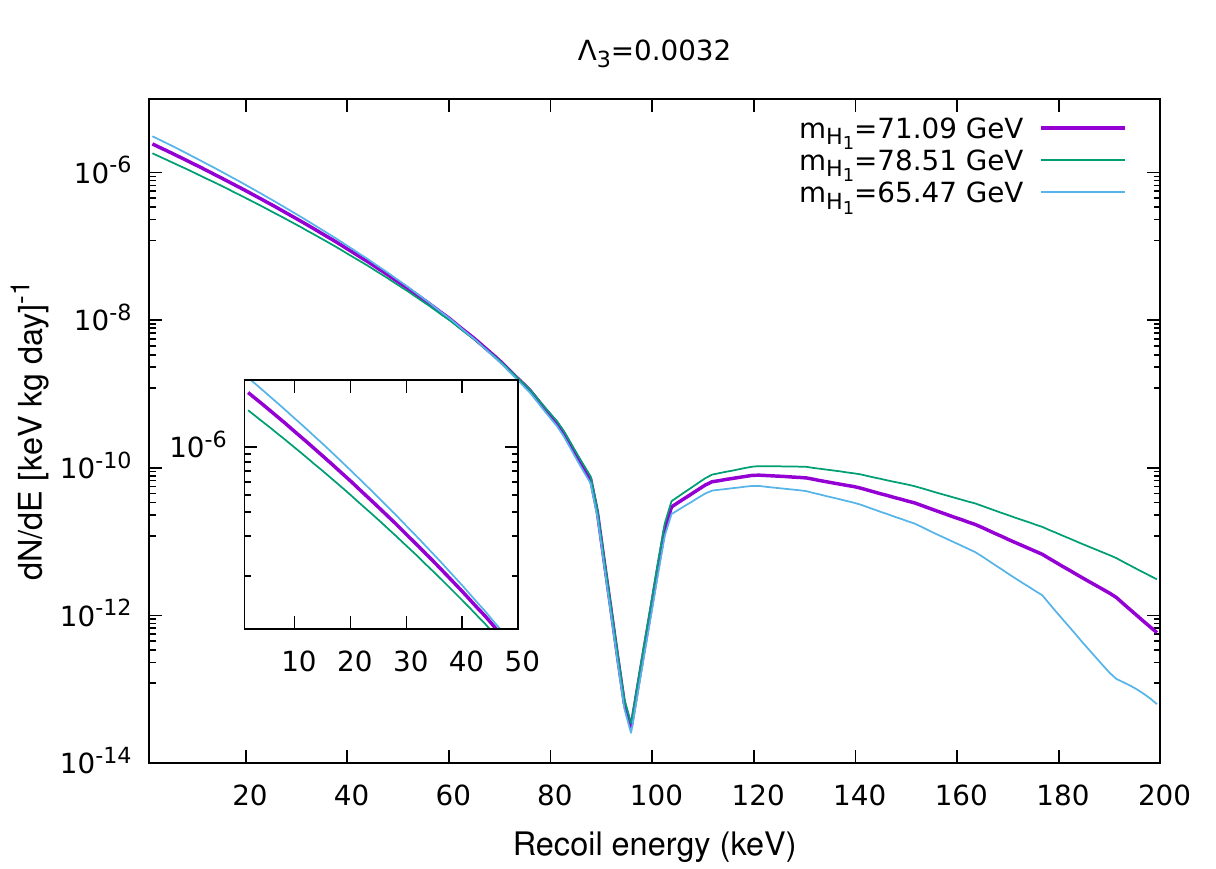}~~
\includegraphics[width=80mm]{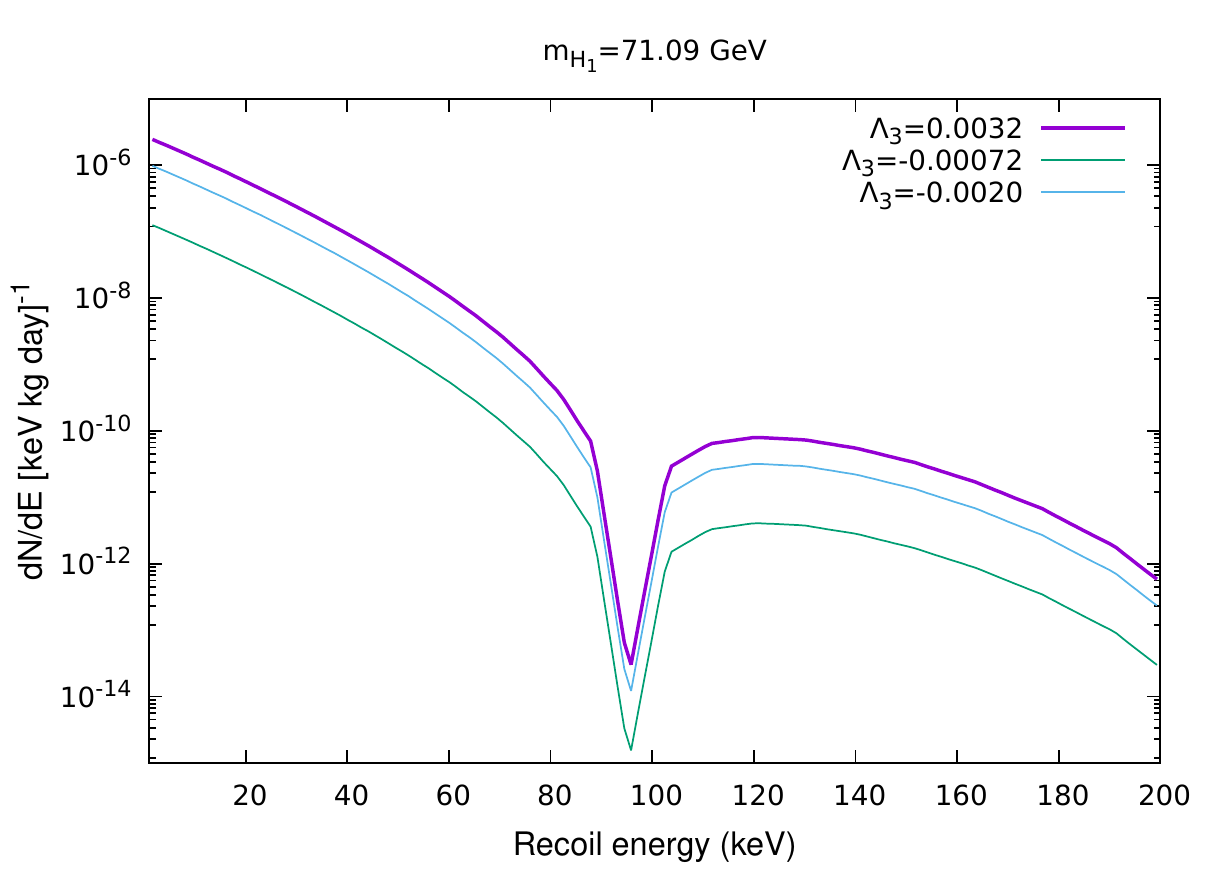}
\caption{The event rate versus recoil energy for a fixed value of 
$\Lambda_3$ while changing $m_{H_1}$ on the left and for a fixed $H_1$ mass while changing $\Lambda_3$ on the right.}
\label{Fig:recoil}
\end{center}
\end{figure}

To clarify the dependence of the direct probes of $H_1$ on the relevant parameters of the model, in Figure~\ref{Fig:recoil}
we show the event rate versus recoil energy for a fixed value of 
$\Lambda_3$ while changing $m_{H_1}$ on the left and for a fixed $H_1$ mass while changing $\Lambda_3$ on the right.

\begin{figure}[h!]
\centering
\includegraphics[width=130mm]{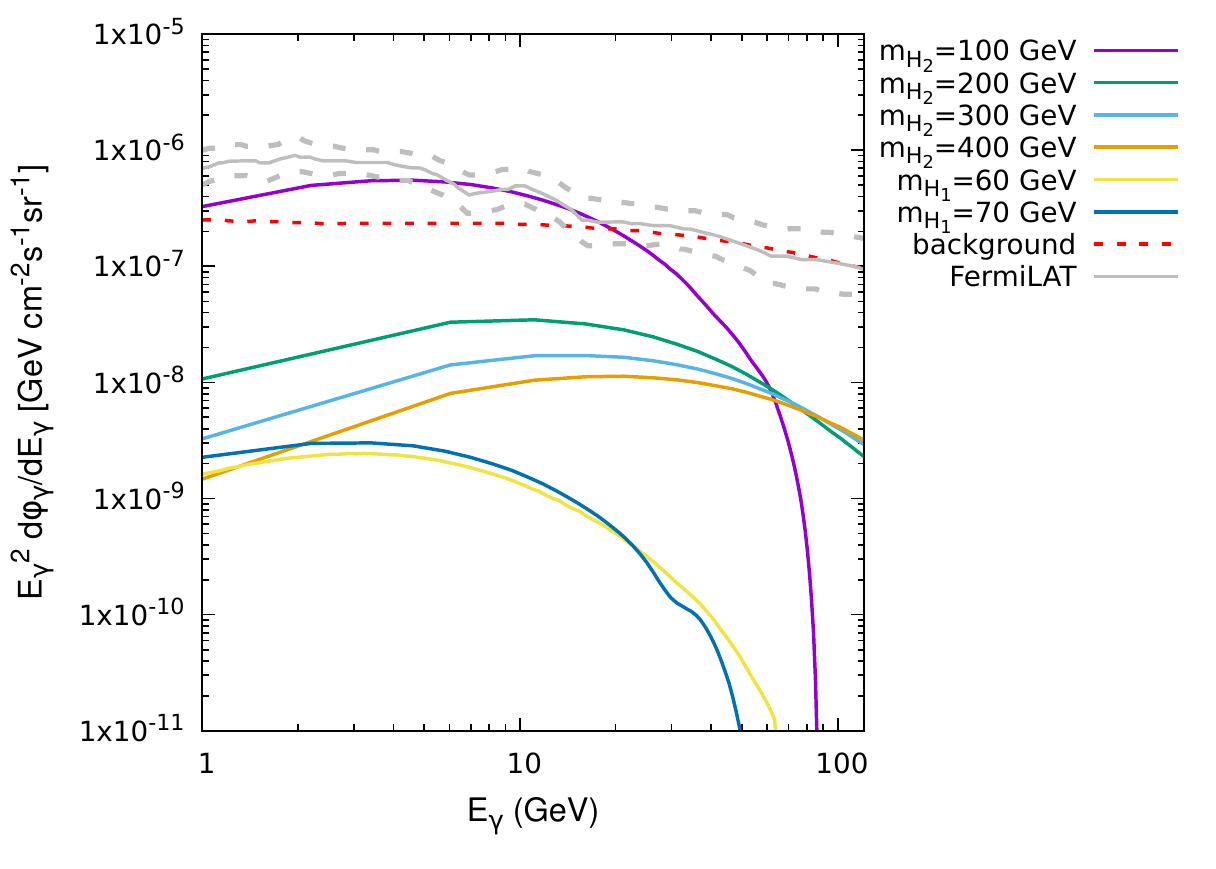}
\vspace{-5mm}
\caption{The complementary signals of our two-component DM model; the Figure shows that the heavy DM candidate $H_2$ with a mass of $\sim$100 GeV could explain the galactic centre photon flux excess measured by Fermi-LAT.}
\label{Fig:general-behaviour-ID}
\end{figure}

\begin{figure}[h!]
\begin{center}
\includegraphics[width=70mm]{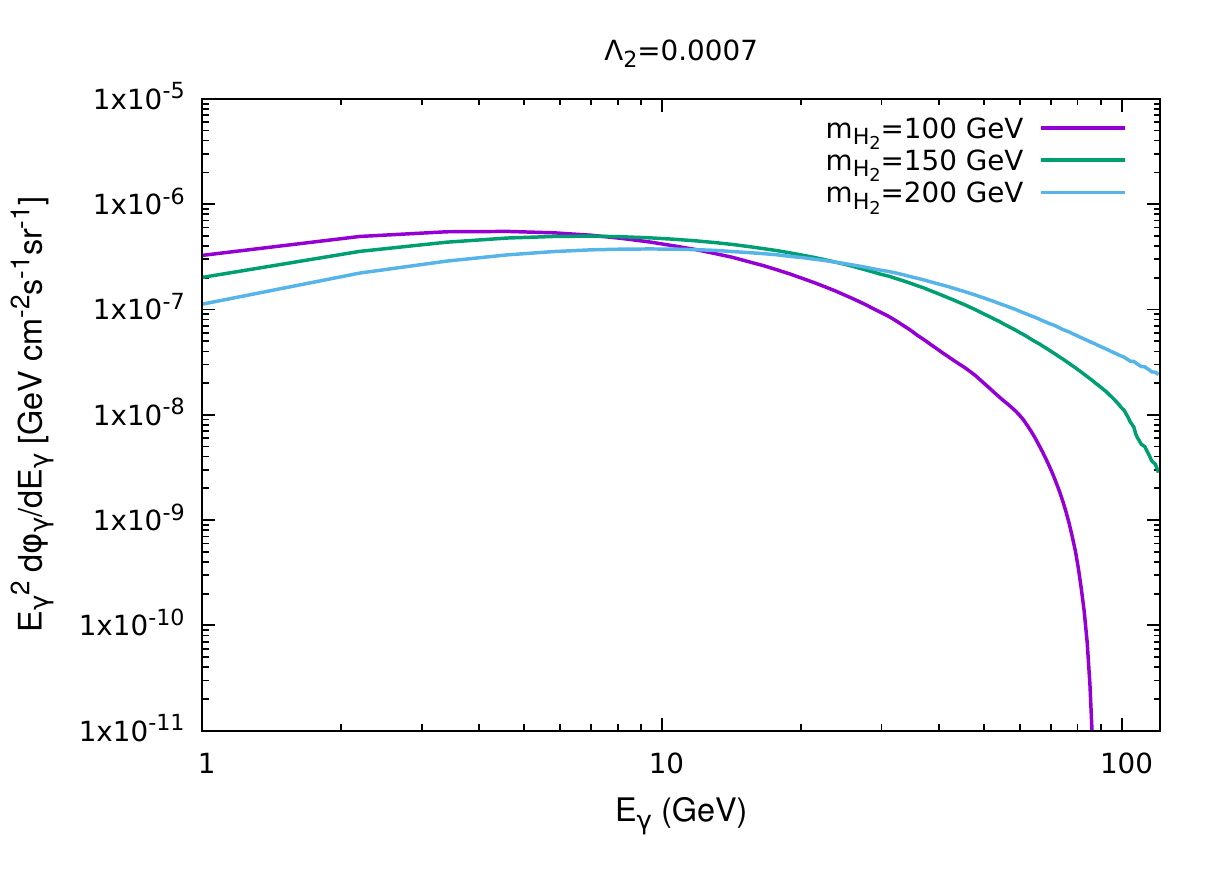}~~
\includegraphics[width=70mm]{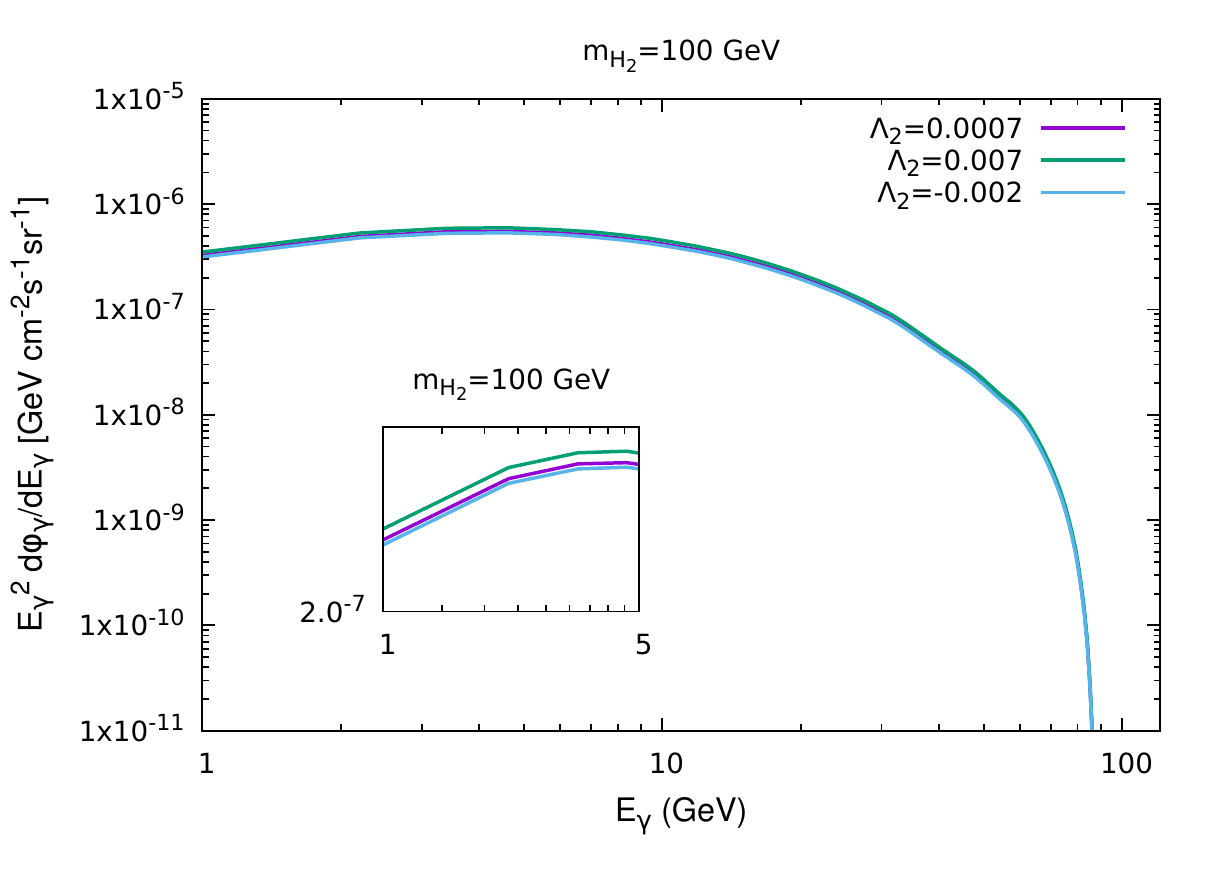}
\caption{The photon flux for a fixed value of 
$\Lambda_2$ while changing $m_{H_2}$ on the left and for a fixed $H_2$ mass while changing $\Lambda_2$ on the right.}
\label{Fig:photon-flux}
\end{center}
\end{figure}

To probe the heavy DM candidate $H_2$, we calculate its contribution to the photon
flux excess from the galactic centre in the indirect detection experiments such as Fermi-LAT \cite{Ackermann:2012qk}. 
This signature depends heavily on the mass of the DM candidate and peaks around $m_{H_2} \simeq 100$ GeV, where the contribution is of the precise order of magnitude of the excess observed by Fermi-LAT \cite{Ackermann:2012qk}.
As shown in Figure \ref{Fig:general-behaviour-ID}, for heavier $H_2$ masses, the photon flux is lower in the $E_\gamma <10$ GeV region,  where Fermi-LAT has observed an excess over the background.
The contribution from the light DM candidate $H_1$ is sub-dominant and has a negligible contribution to the photon flux.
To clarify the dependence of the indirect probes of $H_2$ on the relevant parameters of the model, in Figure~\ref{Fig:photon-flux}
we show the photon flux for a fixed value of 
$\Lambda_2$ while changing $m_{H_2}$ on the left and for a fixed $H_2$ mass while changing $\Lambda_2$ on the right.

\begin{figure}[h!]
\centering
\includegraphics[width=85mm]{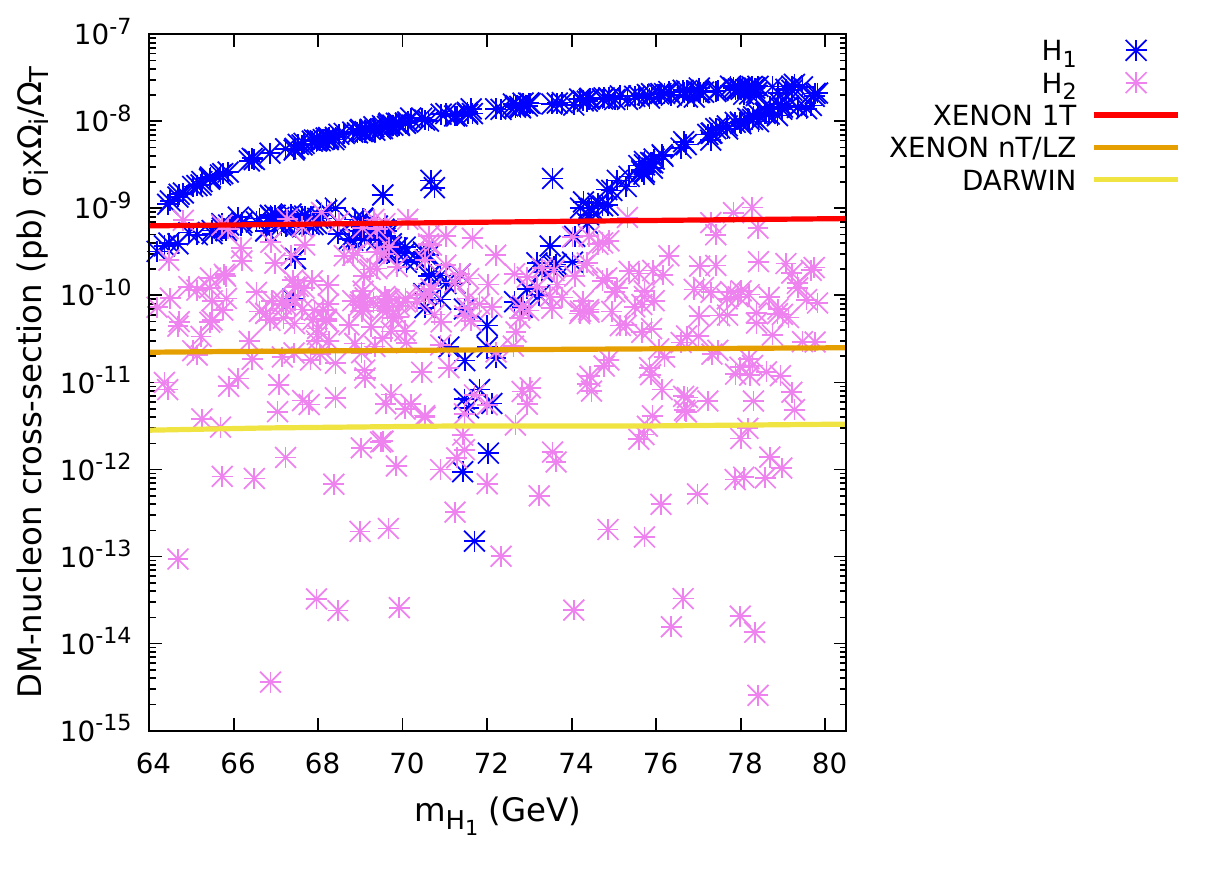}~
\includegraphics[width=85mm]{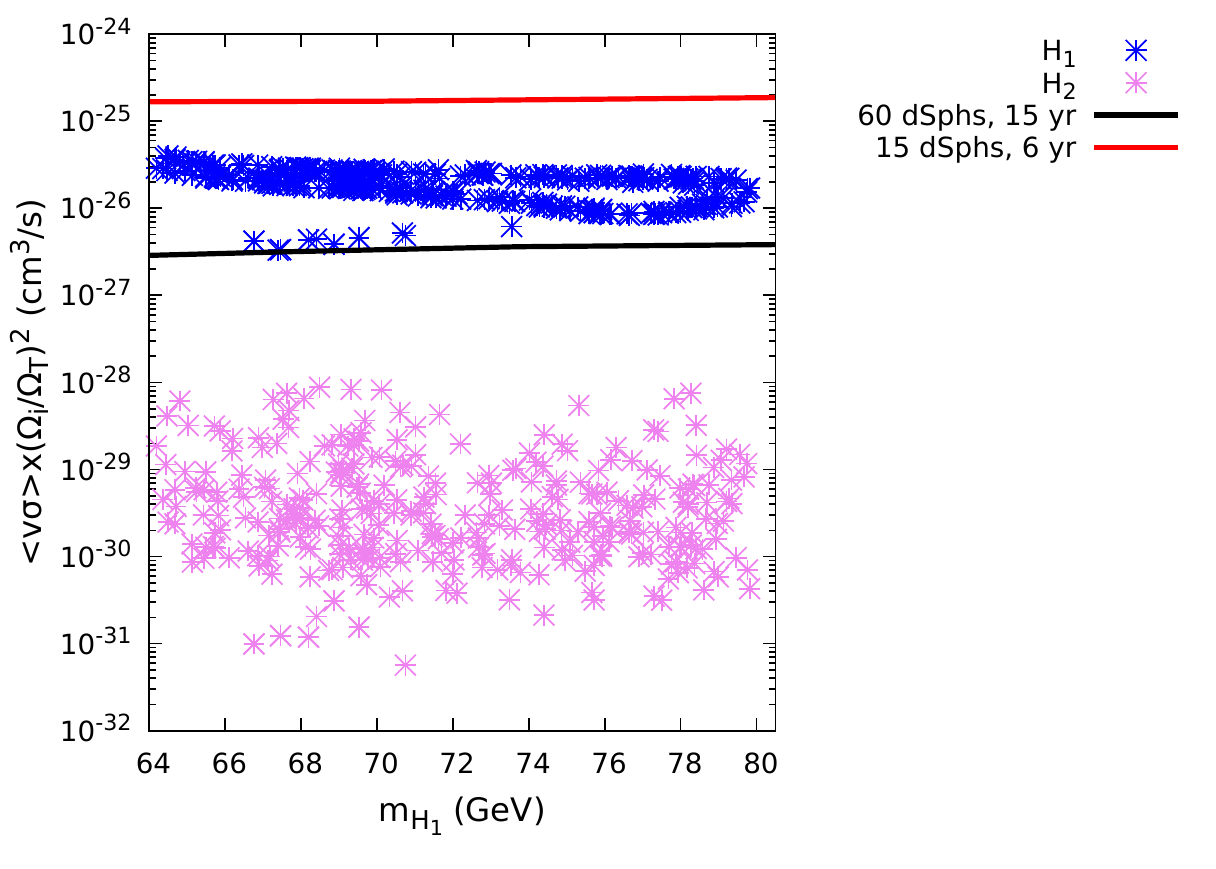}
\caption{Direct (left) and indirect (right) detection cross-section of the two DM components with the upper bounds from XENON1T and Fermi-LAT experiments in red.
Note that the cross-section values are re-scaled by the factor $\Omega_{H_i}/\Omega_T$, where $\Omega_T=\Omega_{H_1}+\Omega_{H_2}$.
We also show the projected sensitivity of future experiments for both direct and indirect searches.}
\label{Fig:future}
\end{figure}

To conclude this section, we show the projected sensitivity of current and future direct and indirect detection experiments.
In the left panel of Figure \ref{Fig:future}, we show the points that survive the XENON1T exclusion limits, i.e. the points below the red line. The projected sensitivity of XENONnT/LZ and DARWIN \cite{Aalbers:2016jon} is shown by the orange and yellow lines, respectively.
In the right panel, the red line represents the latest indirect detection cross-section limit from the FermiLAT experiment \cite{Boddy:2019qak} which set a limit on the DM annihilation into $b\bar{b}$ ($W^+W^-$).
The black line shows the projected sensitivity of 60 dwarf spheroidal galaxies (dSphs) after 15 years of observation, as reported by the Fermi collaboration \cite{Charles:2016pgz}.
In both plots, the pink and blue points show the contribution from the heavy DM candidate $H_2$ and the light DM candidate $H_1$, respectively. The two blue bands correspond to the two values of $\Lambda_3$ for a given mass in Figure \ref{Fig:masses-couplings}, leading to the correct relic density. 

One can see that most of the points are within the observation/exclusion reach of the XENONnT/LZ experiment. Note that the heavy DM candidate (represented by pink points) also has the potential to be probed in direct detection experiments even though its cross-section is scaled down due to its fractional relic density contribution.
In indirect detection searches, when the search target comes from dSphs, it is the lightest DM component that is potentially within reach of future Fermi-LAT sensitivity. This could also act as a complementary probe of the light DM candidate.
The heavier DM component, due to its low number density, annihilates very poorly and is not within reach of future observations. However, it fits well with the observed photon flux excess from the galactic centre.

\section{Conclusion and outlook\label{sec-conclusion}}

Our analysis showed that establishing the vacuum structure and determining the nature of the global minimum is a complicated but necessary task while dealing with multi-scalar models as well as that extrapolating from properties of other models may not be correct. For example, 
in our limited 3HDM analysis, we have identified situations that do not occur in the 2HDM framework. Several possible minima can coexist, and the stability of the chosen vacuum is not guaranteed even at tree-level. 
For example, two distinct minima of the I(1+2)HDM type considered here can exist at the same time. This highlights possible problems with studies of multi-scalar models as, due to the number of parameters, often numerical analysis is employed. Thus, random scans over parameter space can easily cover regions where the chosen minimum is not a global one.

In this work, we have derived conditions that guarantee tree-level stability of the most symmetric vacuum, the \texttt{2-Inert} configuration, where two DM candidates exist. However, the possibility of coexistence between other types of minima was left mostly unexplored, so it has to be done in the future if one wants to do any phenomenological studies. 
One possible situation, when knowing the full vacuum structure is a necessity, is the  study of the thermal history of the Universe, where thermal corrections to the potential could lead to undesirable minima appearing at a given temperature. Furthermore, as we have  pointed out, one-loop corrections to the potential even at $T=0$ could influence the stability of the vacuum. These important questions are to be explored in our future work.

We would also like to stress that demanding the \texttt{2-Inert} state to be the \textit{global} minimum certainly limits the parameter space considered in this work. In the same vein as imposing too strong positivity conditions may rob us of exploring potentially different phenomenology of the model, we may be neglecting an interesting part of the parameter space where the \texttt{2-Inert} state is a sufficiently long-lived metastable vacuum. However, 
conditions for the global minimum heavily depend on the self-couplings of DM particles which govern only DM scattering and have no impact on any cross-sections that influence relic density or DM detection at experiments.

Therefore, within this model configuration, we have proceeded to establish the conditions which are necessary for the I(2+1)HDM parameter space to onset a DM sector which can accommodate two (viable) DM candidates. As a precondition, two discrete symmetries are needed, $Z_2$ and $Z'_2$, which we have implemented onto the two inert doublets, each of which is odd under one of the parities, while all SM states (thus including the active Higgs doublet) are even under both. Following these theoretical conditions, we have proceeded to implement current experimental constraints from both collider and DM data. The key constraint mostly responsible for shaping the mass and coupling structure of the inert sector of the I(2+1)HDM is relic density, which imposes the contribution to   the corresponding abundance due to the heaviest of the two DM candidate to be generally sub-leading with  respect to the contribution of the lightest one. However, the mass difference allowed between the two DM candidates needs not to be large,  in fact, with both being in the EW regime, we have proceeded to test whether there could be the possibility of experimentally accessing both DM states, by exploiting a complementary approach to available DM data. We have found that this can be the case if one pursues detection of the lightest DM state in 
direct direction experiments (e.g., XENONnT/LZ or DARWIN), specifically, by studying the spectrum of the nuclear recoil energy, and that of the heaviest DM state in indirect detection experiments (e.g., FermiLAT), chiefly, by studying the photon flux from the galactic centre. 

As an outlook of our study, we highlight the fact that another consequence of the masses of the two DM candidates being in the EW regime is that they could afford one with the possibility of exploring their signals at the LHC. We are currently addressing this in a forthcoming publication, aimed at establishing Higgs cascade decays into the two different DM candidates, wherein their mass difference (hence their simultaneous presence in data)  leads to notably different shapes of the ensuing missing transverse energy distributions. This would therefore correspond to another independent evidence of the two-component DM nature that may exist in the I(2+1)HDM.

Finally, in order to favour experimental investigations of both EDM and collider data towards establishing sensitivity to our theoretical construct, we have discussed benchmark configurations of the latter, in terms of masses and couplings of the two DM candidates,   that can be adopted for this purpose.

\subsection*{Acknowledgements}
SM acknowledges support from the STFC Consolidated Grant ST/L000296/1 and is financed in part through the NExT Institute. SM, VK, and DR-C acknowledge the H2020-MSCA-RISE-2014 Grant No. 645722 (NonMinimalHiggs).
DS is supported in part by the National Science Center, Poland, through the HARMONIA project under contract UMO-2015/18/M/ST2/00518.
DR-C is supported by the Royal Society Newton International Fellowship NIF/R1/180813 and by the National Science Centre (Poland) under the research Grant No.
2017/26/E/ST2/00470.
VK acknowledges financial support from Academy of Finland projects ``Particle cosmology and gravitational waves'' No. 320123
and ``Particle cosmology beyond the Standard Model'' No. 310130.
J H-S acknowledges support from SNI-CONACYT, PRODEP-SEP and VIEP-BUAP.
The authors acknowledge the use of the IRIDIS High Performance Computing Facility, and associated support services at the University of Southampton, in the completion of this work.

\appendix

\section{Feynman rules in the 2-Inert vacuum \label{ap-vertices}}

In Table \ref{vertex-table}, we recap the Feynman rules of the model considered here in its general set-up.

\begin{table}[t!]
\begin{center}
\begin{tabular}{|l|c|}\hline
vertex & vertex coefficient \\ \hline
\hline		
$hhh$ 			 
&
$ \lambda_{33}v$ 
\\ 	[0.5mm]
$H^+_{1,2} H^-_{1,2} h$ &
$ \lambda_{23}v$ 
\\ 	[0.5mm]
$H_1 H_1  h $ & 
$(\lambda_{13}+ \lambda'_{13}+2\lambda_{3})v$ 
\\ 	[0.5mm]
$H_2 H_2  h$ 		 
&
$(\lambda_{23}+ \lambda'_{23}+2\lambda_{2})v$   
\\ 	[0.5mm]
$A_1 A_1  h$ 	 
&
$(\lambda_{13}+ \lambda'_{13}-2\lambda_{3})v$ 
\\	[0.5mm]
$A_2 A_2  h$ 		 
&
$(\lambda_{23}+ \lambda'_{23}-2\lambda_{2})v$ 
\\	[0.5mm]
\hline	
\hline	
$W^+ W^- h$ 			 
&
$ \frac{1}{2} g^2v$ 
\\ 		[0.5mm]
$Z Z h$ &
$ \frac{1}{8}(g\cos\theta_W + g'\sin\theta_W)^2 v$ 
\\ 	[0.5mm]
$H^+_{1,2} H^-_{1,2} \gamma $ & 
$\frac{i}{2}(g\sin\theta_W + g'\cos\theta_W)$ 
\\ 	[0.5mm]
$H^+_{1,2} H^-_{1,2} Z$ 		 
&
$ \frac{i}{2}(g\cos\theta_W - g'\sin\theta_W)$   
\\ 	[0.5mm]
$H^\pm_{1,2} H_{1,2} W^\mp$ 	 
&
$\frac{i}{2}g$ 
\\	[0.5mm]
$H^\pm_{1,2} A_{1,2} W^\mp$ 	 
&
$\frac{1}{2}g$ 
\\	[0.5mm]
$H_{1,2} A_{1,2}  Z$ 		 
&
$ \frac{1}{2}(g\cos\theta_W + g'\sin\theta_W)$ 
\\	[0.5mm]
\hline
\end{tabular}
\caption{\small Scalar and gauge couplings in the $Z_2 \times Z'_2$ symmetric I(2+1)HDM. The Yukawa couplings are identical to the SM ones with the active doublet $\phi_3$ playing the role of the SM-Higgs doublet.}
\label{vertex-table}
\end{center}
\end{table}

\section{Mass matrices for neutral minima \label{ap-masses}}

Below we present formulae for masses, extremum conditions and VEVs used in establishing the (co)existence of different minima in section~\ref{sec-coexistence}. In the interest of keeping all relations easier to read, we omit the vacuum index (e.g. \texttt{DM1}) from all relevant quantities, but we remind the reader that all given relations for masses, VEVs and energies are derived within the context of considered minimum e.g. Eq.(\ref{DM1_Hc1}) for the mass of $H_1^\pm$ particle in the \texttt{DM1} vacuum should in fact be read as:
\[ \left( m_{H^\pm_1}^2 \right)^{\texttt{DM1}} = -\mu_1^2 +\frac{1}{2} ((v^2_2)^{\texttt{DM1}} \lambda_{12} + (v_3^2)^{\texttt{DM1}} \lambda_{31}). \]

\subsection{DM1}

This extremum is realised when:
\bea
&& \mu_2^2 = v_2^2 \lambda_{22} + \frac{1}{2}v_3^2 (2\lambda_2 + \lambda_{23} + \lambda_{23}'),\\
&& \mu_3^2 = v_3^2 \lambda_{33} + \frac{1}{2}v_2^2 (2\lambda_2 + \lambda_{23} + \lambda_{23}').
\eea
The first doublet, $\phi_1$ contains inert scalars with masses:
\bea
&& m_{H_1}^2 = -\mu_1^2 +(v^2_2 \Lambda_{1} + v_3^2 \Lambda_{3}), \\
&& m_{A_1}^2 = -\mu_1^2 +(v^2_2 \bar{\Lambda}_{1} + v_3^2 \bar{\Lambda}_{3}),\\
&& m_{H^\pm_1}^2 = -\mu_1^2 +\frac{1}{2} (v^2_2 \lambda_{12} + v_3^2 \lambda_{31})\, ,
\label{DM1_Hc1}
\eea
with the $\Lambda_i$s defined in section~\ref{sec-inertvac}.
The two active doublets mix and the masses are calculated to be:
\bea
&& m_{h}^2 = v_2^2 \lambda_{22} + v_3^2 \lambda_{33} - \sqrt{4 v_2^2 v_3^2 \Lambda_2^2 + (v_2^2 \lambda_{22} - v_3^2 \lambda_{33})^2},\\
&& m_{H}^2 = v_2^2 \lambda_{22} + v_3^2 \lambda_{33} + \sqrt{4 v_2^2 v_3^2 \Lambda_2^2 + (v_2^2 \lambda_{22} - v_3^2 \lambda_{33})^2},\\
&& m_{A}^2 = -2 (v_2^2 +v_3^2) \lambda_2,\\
&& m_{H^\pm}^2 = -(v_2^2 +v_3^2) \left(\lambda_2 + \frac{\lambda'_{23}}{2} \right).
\eea

For our purposes of establishing the coexistence with the \texttt{2-Inert} vacuum, it is convenient to look at the squared-mass matrix for the CP-even Higgs particles:
\be
M_{\textrm{CP-even}}^2 = 2 \left(\begin{array}{cc}
v_2^2 \lambda_{22} & v_2 v_3 \Lambda_2 \\
v_2 v_3 \Lambda_2 & v_3^2 \lambda_{33}
\end{array}\right).
\ee
This squared-mass matrix is positive definite if both diagonal elements and the determinant are greater than zero. Therefore, the  set of conditions that have to be realised for the \texttt{DM1} state to be a minimum is:
\bea
&& v_2^2 = \frac{\lambda_{33} \mu_2^2 - \Lambda_2 \mu_3^2}{\lambda_{22} \lambda_{33} - \Lambda_2^2} >0, 
\qquad  
v_3^2 = \frac{\lambda_{22} \mu_3^2 - \Lambda_2 \mu_2^2}{\lambda_{22} \lambda_{33} - \Lambda_2^2} >0, 
 \label{DM1global1}\\[2mm]
&& \lambda_{22} \lambda_{33} - \Lambda_2^2 > 0, 
\qquad
\left(\lambda_2 + \frac{\lambda'_{23}}{2} \right) <0, 
\qquad
\lambda_2  <0.
\eea
The vacuum energy in terms of parameters of the potential is:
\be
\mathcal{V}_{\texttt{DM1}} = - \frac{\lambda_{33} \mu_2^4+\lambda_{22
}\mu_3^4-2\Lambda_2\mu_2^2\mu_3^2}{4(\lambda_{22} \lambda_{33} - \Lambda_2^2)} \label{DM1energy}.
\ee

\subsection{DM2}

This extremum is realised when:
\bea
&& \mu_2^2 = v_1^2 \lambda_{11} + \frac{1}{2}v_3^2 (2\lambda_3 + \lambda_{31} + \lambda_{31}'),\\
&& \mu_3^2 = v_3^2 \lambda_{33} + \frac{1}{2}v_1^2 (2\lambda_3 + \lambda_{31} + \lambda_{31}').
\eea
The second doublet, $\phi_2$ contains inert scalars with masses:
\bea
&& m_{H_2}^2 = -\mu_2^2 + (v^2_2 \Lambda_{1} + v_3^2 \Lambda_{2}), \\
&& m_{A_2}^2 = -\mu_2^2 + (v^2_2 \bar{\Lambda}_{1} + v_3^2 \bar{\Lambda}_{2}),\\
&& m_{H^\pm_2}^2 = -\mu_2^2 +\frac{1}{2} (v^2_2 \lambda_{12} + v_3^2 \lambda_{23}). \label{DM2_Hc2}
\eea
The two active doublets mix and their masses are defined as follows:
\bea
&& m_{h}^2 = v_1^2 \lambda_{11} + v_3^2 \lambda_{33} - \sqrt{4 v_1^2 v_3^2 \Lambda_3^2 + (v_1^2 \lambda_{11} - v_3^2 \lambda_{33})^2},\\
&& m_{H}^2 = v_1^2 \lambda_{11} + v_3^2 \lambda_{33} + \sqrt{4 v_1^2 v_3^2 \Lambda_3^2 + (v_1^2 \lambda_{11} - v_3^2 \lambda_{33})^2},\\
&& m_{A}^2 = -2 (v_1^2 +v_3^2) \lambda_3,\\
&& m_{H^\pm}^2 = -(v_1^2 +v_3^2) \left(\lambda_3 + \frac{\lambda'_{31}}{2} \right).
\eea

The squared-mass matrix for CP-even Higgs particles is defined as:
\be
M_{\textrm{CP-even}}^2 = 2 \left(\begin{array}{cc}
v_1^2 \lambda_{11} & v_1 v_3 \Lambda_3 \\
v_1 v_3 \Lambda_3 & v_3^2 \lambda_{33}
\end{array}\right),
\ee
and the \texttt{DM2} state is a minimum if:
\bea
&& v_1^2 = \frac{\lambda_{33} \mu_1^2 - \Lambda_3 \mu_3^2}{\lambda_{11} \lambda_{33} - \Lambda_3^2} >0, 
\qquad  
v_3^2 = \frac{\lambda_{11} \mu_3^2 - \Lambda_3 \mu_1^2}{\lambda_{11} \lambda_{33} - \Lambda_3^2} >0, \label{DM2global1}\\[1mm]
&& \lambda_{11} \lambda_{33} - \Lambda_3^2 > 0, 
\qquad
\left(\lambda_3 + \frac{\lambda'_{31}}{2}  \right) <0, 
\qquad\lambda_3  <0.
\eea
The energy of the vacuum in terms of parameters of the potential is:
\be
\mathcal{V}_{\texttt{DM2}} = - \frac{\lambda_{33} \mu_1^4+\lambda_{11
}\mu_3^4-2\Lambda_3\mu_1^2\mu_3^2}{4(\lambda_{11} \lambda_{33} - \Lambda_3^2)}.
\ee

%
% we arrive at the following set of conditions for \texttt{DM2} being a minimum:
%
%\bea
%&& v_1^2 = \frac{\lambda_{33} \mu_1^2 - \Lambda_3 \mu_3^2}{\lambda_{11} \lambda_{33} - \Lambda_3^2} >0, \quad  v_3^2 = \frac{\lambda_{11} \mu_3^2 - \Lambda_3 \mu_1^2}{\lambda_{11} \lambda_{33} - \Lambda_3^2} >0, \label{DM2global1}\\
%&& \lambda_{11} \lambda_{33} - \Lambda_2^2 > 0.
%\eea
%%The energy is:
%%
%\be
%\mathcal{V}_{\texttt{DM2}} = - \frac{\lambda_{33} \mu_1^4+\lambda_{11
%}\mu_3^4-2\Lambda_3\mu_1^2\mu_3^2}{4(\lambda_{11} \lambda_{33} - \Lambda_3^2)}
%\ee

\subsection{F0DM1}

The extremum condition for the \texttt{F0DM1} state is:
\be
\mu_2^2 = v_2^2 \lambda_{22}.
\ee 

Physical scalar states have masses in the following form:
\bea
&& m_{H_1}^2 = -\mu_1 + v_2^2 \Lambda_1,\\
&& m_{A_1}^2 = -\mu_1 + v_2^2 \bar{\Lambda}_1,\\
&& m_{H^\pm_1}^2 = -\mu_1 + v_2^2 \lambda_{12}/2,\\
&& m_{H_2}^2 = 2 v_2^2 \lambda_{22}, \label{F0DM1H2}\\
&& m_{H_3}^2 = -\mu_3 + v_2^2 \Lambda_2,\\
&& m_{A_3}^2 = -\mu_3 + v_2^2 \bar{\Lambda}_2,\\
&& m_{H^\pm_3}^2 = -\mu_3 + v_2^2 \lambda_{23}/2.
\eea
The energy of the \texttt{F0DM1} stationary point is:
\be
\mathcal{V}_{\texttt{F0DM1}}= -\frac{\mu_2^4}{4\lambda_{22}}.
\ee

\subsection{FODM2}

The extremum condition for the \texttt{F0DM1} state is:
\be
\mu_1^2 = v_1^2 \lambda_{11}.
\ee 

Physical scalar states have masses in the following form:
\bea
&& m_{H_2}^2 = -\mu_2 + v_2^2 \Lambda_1,\\
&& m_{A_2}^2 = -\mu_2 + v_2^2 \bar{\Lambda}_1,\\
&& m_{H^\pm_2}^2 = -\mu_2 + v_2^2 \lambda_{12}/2,\\
&& m_{H_1}^2 = 2 v_1^2 \lambda_{11}, \label{F0DM2H1}\\
&& m_{H_3}^2 = -\mu_3 + v_2^2 \Lambda_3,\\
&& m_{A_3}^2 = -\mu_3 + v_2^2 \bar{\Lambda}_3,\\
&& m_{H^\pm_3}^2 = -\mu_3 + v_2^2 \lambda_{23}/2.
\eea
The energy of the \texttt{F0DM1} stationary point is:
\be
\mathcal{V}_{\texttt{F0DM1}}= -\frac{\mu_1^4}{4\lambda_{11}}.
\ee

\subsection{F0DM0}

This extremum is realised when:
\bea
&& \mu_1^2 = v_1^2 \lambda_{11} + \frac{1}{2}v_2^2 (2\lambda_1 + \lambda_{12} + \lambda_{12}'),\\
&& \mu_2^2 = v_2^2 \lambda_{22} + \frac{1}{2}v_1^2 (2\lambda_1 + \lambda_{12} + \lambda_{12}').
\eea

The masses of the physical particles are given by:
\bea
&& m_{h_1}^2 = v_1^2 \lambda_{11} + v_2^2 \lambda_{22} - \sqrt{4 v_1^2 v_2^2 \Lambda_1^2 + v_1^4 \lambda_{11}^2 - 2 v_1^2 v_2^2 \lambda_{11} \lambda_{22} + v_2^4 \lambda_{22}^2},\\
&& m_{h_2}^2 = v_1^2 \lambda_{11} + v_2^2 \lambda_{22} + \sqrt{4 v_1^2 v_2^2 \Lambda_1^2 + v_1^4 \lambda_{11}^2 - 2 v_1^2 v_2^2 \lambda_{11} \lambda_{22} + v_2^4 \lambda_{22}^2},\\ 
&& m_{h_3}^2 =   \frac{1}{2} v_1^2 (\lambda_{31} + 2 \lambda_{3} + \lambda'_{31}) + 
    \frac{1}{2} v_2^2 (2 \lambda_{2} + \lambda_{23} + \lambda'_{23}) - \mu_3^2,\\ 
&& m_{a}^2 =  -2 (v_1^2 +v_2^2) \lambda_{1}, \\
&& m_{h^\pm}^2 =  -(v_1^2 +  v_2^2) \left(\lambda_{1} + \frac{\lambda'_{12}}{2}\right).
\eea

As is the case with \texttt{DM1} and \texttt{DM2}, conditions for positivity of the masses (for states that are linear combinations of the fields inside the doublets $\phi_1$ and $\phi_2$ require that:
\bea
&& \lambda_{11} \lambda_{22} - \Lambda_1^2 > 0, \quad \lambda_1 <0, \quad \lambda_1 + \frac{\lambda'_{12}}{2} < 0,\\
&& v_1^2 = \frac{\Lambda_1 \mu_2^2 - \lambda_{22} \mu_1^2}{\Lambda_1^2 - \lambda_{11}\lambda_{22}} >0, \quad v_2^2 = \frac{\lambda_{11} \mu_2^2 - \Lambda_{1} \mu_1^2}{\Lambda_1^2 - \lambda_{11}\lambda_{22}} >0. \label{F0DM0min}
\eea

\subsection{N}

Extremum conditions for the \texttt{N} extremum are as follows:
\bea
&& \mu_1^2 =  \frac{1}{2} \left(2 v_1^2 \lambda_{11} + 
    v_2^2 (2 \lambda_{1} + \lambda_{12} + \lambda'_{12}) + 
    v_3^2 (\lambda_{31} + 2 \lambda_{3} + \lambda'_{31}) \right),\\
&& \mu_2^2 = \frac{1}{2} \left(2 v_2^2 \lambda_{22} + 
    v_1^2 (2 \lambda_{1} + \lambda_{12} + \lambda'_{12}) + 
    v_3^2 (2 \lambda_{2} + \lambda_{23} + \lambda'_{23}) \right),\\
&& \mu_3^2 = \frac{1}{2} \left(2 v_3^2 \lambda_{33} + 
    v_1^2 (\lambda_{31} + 2 \lambda_{3} + \lambda'_{31}) + 
    v_2^2 (2 \lambda_{2} + \lambda_{23} + \lambda'_{23}) \right).
\eea

All three doublets mix and the masses of physical states are the eigenvalues of the following mass-squared matrices:

\be
M_{\textrm{CP-even}}^2 = 2 \left(\begin{array}{ccc}
v_1^2 \lambda_{11} & v_1 v_2 \Lambda_1& v_1 v_3 \Lambda_3\\
v_1 v_2 \Lambda_1& v_2^2 \lambda_{22} & v_2 v_3 \Lambda_2\\
v_1 v_3 \Lambda_3 & v_2 v_3 \Lambda_2 & v_3^2 \lambda_{33}
\end{array}\right)\,,
\ee

\be
M_{\textrm{CP-odd}}^2 = 2 \left(\begin{array}{ccc}
-v_2^2 \lambda_{1} - v_3^2 \lambda_3 & v_1 v_2 \lambda_1& v_1 v_3 \lambda_3\\
v_1 v_2 \lambda_1& - v_1^2 \lambda_1 - v_2^2 \lambda_{2} & v_2 v_3 \lambda_2\\
v_1 v_3 \lambda_3 & v_2 v_3 \lambda_2 & -v_2^2 \lambda_2 - v_1^2 \lambda_{3}
\end{array}\right)\,,
\ee

%\be
%{\scriptsize M_{\textrm{charged}}= \frac{1}{2} \left(\begin{array}{ccc}
%-v_2^2 (2\lambda_{1} +\lambda'_{12}) - v_3^2 (2\lambda_3 +\lambda'_{31}) & v_1 v_2 (2\lambda_1 +\lambda'_{12}) & v_1 v_3 (2\lambda_3 +\lambda'_{31})\\
%v_1 v_2 (2\lambda_1 +\lambda'_{12}) & - v_1^2 (2\lambda_1 +\lambda'_{12}) - v_2^2 (2\lambda_{2}+\lambda'_{23}) & v_2 v_3 (2\lambda_{2}+\lambda'_{23})\\
%v_1 v_3 (2\lambda_3 +\lambda'_{31}) & v_2 v_3 (2\lambda_{2}+\lambda'_{23}) & -v_2^2 (2\lambda_{2}+\lambda'_{23}) - v_1^2 (2\lambda_3 +\lambda'_{31})
%\end{array}\right)}
%\ee

\be
M^2_{\textrm{charged}}= \frac{1}{2} \left(\begin{array}{ccc}
m_{11} & m_{12} & m_{13}\\
m_{12} & m_{22} & m_{23}\\
m_{13} & m_{23} & m_{33}
\end{array}\right),
\ee
where
\bea
&& m_{11} = -v_2^2 (2\lambda_{1} +\lambda'_{12}) - v_3^2 (2\lambda_3 +\lambda'_{31}), 
\qquad 
m_{12} = v_1 v_2 (2\lambda_1 +\lambda'_{12}), \\
&& m_{22} = - v_1^2 (2\lambda_1 +\lambda'_{12}) - v_2^2 (2\lambda_{2}+\lambda'_{23}),
\qquad 
m_{13} =  v_1 v_3 (2\lambda_3 +\lambda'_{31}), \\
&&
m_{33} = -v_2^2 (2\lambda_{2}+\lambda'_{23}) - v_1^2 (2\lambda_3 +\lambda'_{31}),
\qquad 
m_{23} =  v_2 v_3 (2\lambda_{2}+\lambda'_{23})\,.
\eea

\section{Towards the analysis of the full phase space \label{ap-DM1DM2}}
Here we discuss the two minima which exhibit the same level of symmetry breaking but with a different remnant symmetry, can in principle coexist and therefore are not equivalent through a basis change. 

Let us first focus on the comparison between the \texttt{DM1} and \texttt{DM2} states. We follow the bilinear formalism and define the vectors $X^T = (x_1,x_2,x_3,x_4,x_5,x_6)$ at a given minimum as\footnote{Both minima are CP-conserving, and therefore we do not explicitly write entries for $x_7,x_8,x_9$, as they are zero in both cases.} :
\bea
X^T_{\texttt{DM1}} = \frac{1}{2} \left( 0, (v_2^2)^\texttt{DM1}, (v_3^2)^\texttt{DM1}, 0,  0,  (v_2 v_3)^\texttt{DM1} \right), 
\qquad
X^T_{\texttt{DM2}} = \frac{1}{2} \left((v_1^2)^\texttt{DM2}, 0, (v_3^2)^\texttt{DM2},  0, (v_1 v_3)^\texttt{DM2}, 0\right).\nonumber
\eea
The values of $\mathcal{V}'$ are given by:
\be 
\mathcal{V}'_{\texttt{DM1}} = \left(\begin{array}{c}(m_{H_1^\pm}^2)^{\texttt{DM1}} \\ 0 \\0 \\ 0 \\0 \\ 0\end{array}\right)+ \left(-\frac{V_6'}{2 v_3 v_2} \right)^{\texttt{DM1}} \left(\begin{array}{c} 0 \\ (v_3^2)^{\texttt{DM1}}\\ 
(v_2^2)^{\texttt{DM1}}\\ 0\\ 0\\ 
-2(v_3 v_2)^{\texttt{DM1}}\end{array}\right),
\ee
and
\be 
\mathcal{V}'_{\texttt{DM2}} = \left(\begin{array}{c}0\\(m_{H_2^\pm}^2)^{\texttt{DM2}} \\0 \\ 0 \\0 \\ 0\end{array}\right)+ \left(-\frac{V_5'}{2 v_3 v_1} \right)^{\texttt{DM2}} \left(\begin{array}{c} (v_3^2)^{\texttt{DM2}} \\ 0\\ (v_1^2)^{\texttt{DM2}}\\ 
0\\ 
-2(v_3 v_1)^{\texttt{DM2}}\\0\end{array}\right).
\ee

In the above equation we show the $\mathcal{V}'$ expressions separately to clarify the influence of inert and active doublets. The inert parts are proportional, in both cases, to the masses of charged \textit{inert} scalars, $(m_{H_1^\pm}^2)^{\texttt{DM1}}$ and $(m_{H_2^\pm}^2)^{\texttt{DM2}}$, respectively. In the active parts, non-zero entries share a common factor. 
It can be shown easily by using extremum conditions, that e.g. in case of the \texttt{DM1} minimum we get:
\bea
&& 0 = \frac{\partial V}{\partial v_2} = \frac{\partial V}{\partial x_2} \frac{\partial x_2}{\partial v_2} + \frac{\partial V}{\partial x_6} \frac{\partial x_6}{\partial v_2} \Rightarrow \frac{\partial V}{\partial x_2} = - \frac{v_3}{2 v_2} \frac{\partial V}{\partial x_6},\\
&& 0 = \frac{\partial V}{\partial v_3} = \frac{\partial V}{\partial x_3} \frac{\partial x_3}{\partial v_3} + \frac{\partial V}{\partial x_6} \frac{\partial x_6}{\partial v_3} \Rightarrow \frac{\partial V}{\partial x_3} = - \frac{v_2}{2 v_3} \frac{\partial V}{\partial x_6},
\eea
and
\be
-V'_6 = -\frac{\partial V}{\partial x_6} = - (v_2 v_3)^{\texttt{DM1}} (2 \lambda_{2} + \lambda'_{23}) = 2 (m_{H^\pm}^2)^{\texttt{DM1}} \left( \frac{v_2 v_3}{v_2^2+v_3^2} \right)^{\texttt{DM1}},
\ee
i.e. $V'_6$ is proportional to the mass of the charged \textit{active} scalar in \texttt{DM1}. A similar condition for $V'_5$ can be obtained in the \texttt{DM2} state.

The active parts of $\mathcal{V}'$ show similar behaviour to the one found in the 2HDM when deriving formulas for the normal minimum. However, including the inert doublet fundamentally changes the picture and clearly shows that 2HDM studies can only be used as guidance, and cannot be simply extrapolated to 3HDMs.

Following Eq.~(\ref{energydiff}) for quantities defined above we arrive at the result:
\bea
 \mathcal{V}_{ \texttt{DM1}} - \mathcal{V}_{ \texttt{DM2}} &=& \frac{1}{4} \left[
(v_2^2)^\texttt{DM1} \left( (m_{H^\pm_2}^2)^{\texttt{DM2}}  - (m_{H^\pm}^2)^{\texttt{DM1}} \frac{(v_3^2)^{\texttt{DM2}}}{(v_2^2+v_3^2)^\texttt{DM1}}\right) \right. \nonumber\\
&& \left. \;\;\; - (v_1^2)^\texttt{DM2} \left( (m_{H^\pm_1}^2)^{\texttt{DM1}}  - (m_{H^\pm}^2)^{\texttt{DM2}} \frac{(v_3^2)^{\texttt{DM1}}}{(v_1^2+v_3^2)^\texttt{DM2}}\right) \right],
\eea
which depends on the masses of charged particles (both inert and active) and does not have a fixed sign. In short, the relative position of \texttt{DM1} and \texttt{DM2} is not determined, and either can be the deeper one.
Furthermore, numerical studies show that both of them can be minima.

The relative position of these two minima will be related to the masses of scalar particles. These, in turn, are heavily constrained by collider experiments, as the active sectors for \texttt{DM1} and \texttt{DM2} states are very much like the Type-I CP-conserving 2HDM. Therefore, it may be so that although coexistence of minima is in principle possible, experimental limits actually exclude this possibility. However, until this is proven to be so, in analysing the phase structure of the 3HDMs one has to consider the potential coexistence and metastability of neutral minima with different symmetry structures. 

Coexistence is also possible between the two inert-like minima, \texttt{F0DM1} and \texttt{F0DM2}, as the relative depths of these two possible solutions is given by:
\be 
\mathcal{V}_{ \texttt{F0DM1}} - \mathcal{V}_{ \texttt{F0DM2}} = \frac{1}{4} \left(
(v_2^2)^\texttt{F0DM1} (m_{H^\pm_2}^2)^{\texttt{F0DM2}}  - (v_1^2)^\texttt{F0DM2}  (m_{H^\pm_3}^2)^{\texttt{F0DM1}} \right).
\ee
Again, the remnant symmetries of the potential are different, however, in each case one $Z_2$ symmetry remains unbroken. The phase diagram is equivalent to the one presented in Figure~\ref{fig-coex}, with $\tilde{\mu}_3^2 \to \tilde{\mu}_1^2, \lambda_{33} \to \lambda_{11}, \Lambda_2 \to \Lambda_1$.  
As these two minima can coexist and in principle have different depths, we cannot dismiss one of them as equivalent to the other through a basis change. They both have to be taken into account when establishing the global minimum.

%%%%%%%%%%%%%%%%%%%%%%%%%%%%%%%%%%%%%%%%%%%%%%%%%%%%%%%%%%%%%%%%%%%%%%%%%%%%%%%%%%%%%%%%%%%%

\end{document}